\documentclass[final, 12pt]{elsarticle}

\usepackage{amssymb}

\usepackage{xcolor}
\usepackage{hyperref}       
\usepackage{url}            
\usepackage{booktabs}       
\usepackage{amsfonts}       
\usepackage{nicefrac} 
\usepackage{amsmath}
\usepackage{enumitem}
\usepackage{threeparttable}
\usepackage{multicol, multirow}
\usepackage{tcolorbox}

\usepackage{colortbl}
\usepackage{caption}
\usepackage{subcaption}

\definecolor{mySUred}{RGB}{140, 21, 21}

\usepackage{amsmath,amsfonts,bm}

\def\Figref#1{Figure~\ref{#1}}

\def\secref#1{section~\ref{#1}}

\def\eqref#1{equation~\ref{#1}}

\def\eqrefp#1{equation~(\ref{#1})} 

\def\tableref#1{Table~\ref{#1}}

\def\1{\bm{1}}

\def\eps{{\epsilon}}

\def\va{{\bm{a}}}
\def\vb{{\bm{b}}}

\def\vx{{\bm{x}}}

\DeclareMathAlphabet{\mathsfit}{\encodingdefault}{\sfdefault}{m}{sl}
\SetMathAlphabet{\mathsfit}{bold}{\encodingdefault}{\sfdefault}{bx}{n}

\newcommand{\R}{\mathbb{R}}

\newcommand{\normlone}{L^1}
\newcommand{\normltwo}{L^2}

\DeclareMathOperator*{\argmin}{arg\,min}

\usepackage{geometry}
\newgeometry{
    top=1in,
    bottom=0.99in,
    outer=0.99in,
    inner=1.2in,
}

\journal{Applied Energy}

\begin{document}

\begin{frontmatter}



\title{Constructing dynamic residential energy lifestyles using Latent Dirichlet Allocation}


\author[inst1]{Xiao Chen\fnref{label2}} 
\fntext[label2]{These authors contributed equally to this work.}
\ead{markcx@stanford.edu}
\affiliation[inst1]{organization={Civil \& Environmental Engineering},
            addressline={Stanford University}, 
            city={Stanford},
            postcode={94305}, 
            state={CA},
            country={U.S.}}

\author[inst1]{Chad Zanocco\fnref{label2}}

\ead{czanocco@stanford.edu}
\author[inst1]{June Flora}
\ead{jflora@stanford.edu}
\author[inst1,inst2]{Ram Rajagopal}
\ead{ramr@stanford.edu}

\affiliation[inst2]{organization={Electrical Engineering},
            addressline={Stanford University}, 
            city={Stanford},
            postcode={94305}, 
            state={CA},
            country={U.S.}}

\begin{abstract}
The rapid expansion of Advanced Meter Infrastructure (AMI) has dramatically altered the energy information landscape. However, our ability to use this information to generate actionable insights about residential electricity demand remains limited. In this research, we propose and test a new framework for understanding residential electricity demand by using a dynamic energy lifestyles approach that is iterative and highly extensible. To obtain energy lifestyles, we develop a novel approach that applies Latent Dirichlet Allocation (LDA), a method commonly used for inferring the latent topical structure of text data, to extract a series of latent household energy attributes. By doing so, we provide a new perspective on household electricity consumption where each household is characterized by a mixture of energy attributes that form the building blocks for identifying a sparse collection of energy lifestyles. We examine this approach by running experiments on one year of hourly smart meter data from 60,000 households and we extract six energy attributes that describe general daily use patterns. We then use clustering techniques to derive six distinct energy lifestyle profiles from energy attribute proportions. Our lifestyle approach is also flexible to varying time interval lengths, and we test our lifestyle approach  seasonally (Autumn, Winter, Spring, and Summer) to track energy lifestyle dynamics within and across households and find that around 73\% of households manifest multiple lifestyles across a year. These energy lifestyles are then compared to different energy use characteristics, and we discuss their practical applications for demand response program design and lifestyle change analysis.
\end{abstract}




\end{frontmatter}


\section{Introduction}
The growth of advanced metering infrastructure (AMI) has greatly expanded our potential to analyze household electricity usage. To date, AMI infrastructure provides hourly and sub-hourly electricity usage information via smart meter technology for tens of millions of households in the United States alone, with the deployment of residential smart meters increasing yearly by millions~\cite{Ghosal2019Key}. Prior analysis of smart meter data has provided insights about household daily load shapes and the variation of electricity use patterns both within and across households. While such information about household energy use patterns is being applied toward forecasting residential demand~\cite{Fahiman2017Improving,Barbour2018clustering,Yildiz2018household}, other nascent applications of smart meter data are of increasing interest to energy providers, including targeting households for demand response (DR)~\cite{Kwac2016DDtargetTransSG, wong2012simple, todd2020winners}, tailoring information to differing user segments about energy efficiency (EE) programs~\cite{kavousian2013determinants}, and making recommendations to customers for enrollment in variable rate pricing programs, such as time-of-use rate plans~\cite{di2017nudging, afzalan2019residential}. Moreover, in many extant applications, household energy use pattern information is typically treated as static, without consideration into how patterns may change across time, either cyclically (e.g., seasons, school calendar year, etc.) or as structural household shifts (e.g., new household members, change in work hours, etc.), potentially missing opportunities for more refined targeting, tailoring, and other program design considerations. 

As smart meters become more ubiquitous in households across the world, new advances are needed to efficiently process the deluge of data streams produced from these devices and then generate actionable insights---especially in a way that has low computational overhead and does not require continuous human-in-the loop interactions~\cite{dusparic2017residential}. In particular, using smart meter data to understand household level energy use is an on-going challenge, and with it comes the difficulties in developing meaningful interventions that can ease burdens on the grid while maintaining customer engagement and satisfaction~\cite{boudetpublic}. For example, one such motivation for household-level energy interventions is to reduce greenhouse gas emissions from nonrenewable generation sources (e.g., natural gas ``peaker'' power plants) during periods of high demand, while also expanding the potential for customers to change their energy behaviors and appliance purchases to save money on their energy bills~\cite{werner2021pricing}.

While existing work on customer identification and segmentation has been explored in the literature~\cite{ wong2012simple, Kwac2013DRtargeting, Kwac2016DDtargetTransSG,afzalan2019residential}, insights about customer segmentation for households do not demonstrate strong linkages to a variety of common occupant behaviors with few exceptions~\cite{Kwac2018LifestyleSeg} (see  \secref{lda:sec2:relatedwork:ls:desc} for examples). Additionally, while many methods have been proposed for gaining broad insights into customers and groups of customers' electricity consumption, these methods are often too complex to easily scale for use by utility companies and require higher-resolution data than what is typically available via smart meters~\cite{motlagh2019clustering, abubakar2017application}.
 
To address these challenges, we use Latent Dirichlet Allocation (LDA) to analyze daily energy consumption patterns of households. 
While LDA is most commonly associated with Natural Language Processing (NLP) tasks such as extracting latent topics from text documents~\cite{blei2003latent, hoffman2010online}, it is now increasingly applied in other domains, including remote sensing~\cite{Pinoli2014LDAGibbs, Lienou2010SemanticLDA} and biology and genomics~\cite{van2016topic, Pinoli2014LDAGibbs}. 

We propose a new application for LDA previously unexplored: the classification and interpretation of energy use patterns within the home. In doing so, we seek to identify latent patterns of daily energy consumption and then use these latent constructs to build residential energy lifestyle profiles. While there are many ways to describe a lifestyle, we adopt the definition that a lifestyle is the consolidation of a persistent set of patterns of behavior that occur within the home environment~\cite{socolow1978twin}. We propose that energy consumption is best understood as a consequence of lifestyles that reflect the organization, sequencing, synchronicity, habitualness, and are contingent or interdependent on the timing of the activities of daily life within a day and over weeks, months and years. 

Our approach does not directly characterize residential energy activities and behaviors through observational or self-reported methods~\cite{grunewald2018electricity, aydin2018information} or real time data disaggregation of household energy consumption--all of which can introduce complexity that makes it challenging to generalize across households. Instead, our conceptualization of energy lifestyles are more broadly construed, with the potential to generate meaningful insights without having to resort to finer grained, more nuanced understandings of energy use and energy--related activities within the home. Such an understanding of energy lifestyles could have applications for energy practitioners, such as electricity service providers, policymakers, and the research community for tasks including identifying energy use patterns, targeting customers, and understanding household demand flexibility and response of residential users.

Our approach toward developing energy lifestyles also affords us new opportunities in examining the temporal dimensions of lifestyles, or how these energy lifestyles may change across time intervals of varying length. Previous research has considered a lifestyle as a static attribute of a household, with the lifestyle referring to a component that does not change across time. However, research suggests that lifestyles can indeed have dynamic components~\cite{gladhart1987energy,lutzenhiser2009behavioral, torriti2017understanding,anderson2016laundry,Warde2007ActaSocial,southerton2006analysing}, even though much of this literature is limited to within--day time organization as opposed to across days, weeks, months, etc. On a global scale, we have recently experienced dramatic disruptive influences that has changed the nature, organization, and amount of electricity consumed--lifestyle changes that have occurred during the COVID-19 pandemic~\cite{memmott2021sociodemographic,giurge2021multicountry, zanocco2020exploring, buechler2022global}. While the measurement of lifestyle change through electricity use may only serve as an approximation for a variety of conditions and activity patterns that occur within a household across time, we postulate that such a lifestyle approach could provide a signal for when large changes related to energy use occurs in the home. Such changes could include anything from a  change in the number of household occupants (e.g., a child being born or leaving for college) to a change in the patterns of occupancy (e.g., new employment or retirement) to broader ``shocks'' such as  COVID-19 related restrictions. On the other hand, some households may experience little to no change in energy lifestyles across time, also providing important insights into the stability of energy use patterns and their associated household activities. Understanding these characteristics of lifestyles could bring new opportunities for energy providers to dynamically target energy programs during certain times throughout the year and allow the iterative identification of lifestyle patterns based on constantly updating data streams from AMI infrastructure. While this understanding has the potential to improve recruitment and engagement in both efficiency and demand response programs~\cite{todd2019spillover, todd2020winners}, we may also find that this more ``real'' life understanding of residential consumption leads to the development of new policies and programs.

In this research we break new ground in constructing temporally dynamic energy lifestyles using a novel application of LDA. Given this focus on generating and gaining insights from temporally dynamic energy lifestyles, our research seeks to answer the following research questions: (1) What residential energy lifestyle profiles emerge from empirical data and what are their prominent characteristics? and (2) What patterns of change, or stability, is observed in households' lifestyle profiles across time and what is related to these temporal dynamics?

We address these research questions in the following sections. First, we describe our approach for generating energy lifestyles by introducing our conceptual framework, method, and residential electricity dataset. Next, we describe our experimental setting, derive energy lifestyles, and then provide insights about their prominent features and patterns of change across time. Lastly, we discuss applications of this lifestyle approach and provide recommendations for future research.   

\section{Framework, methods, and data}\label{lda:sec2:relatedwork:ls:desc} %

\subsection{Conceptual framework}\label{lda:sec2:relatedwork:ls:concept}

To capture the temporal patterns in energy use, our conceptual approach toward energy lifestyles is built around the daily load shape as a core feature of household consumptive patterns, which imparts information about the relative magnitude, duration, and timing of energy use throughout a day (24 hour period). Embedded within this daily load shape representation is information about energy use related to the timing of household activities (e.g., cooking, cleaning, entertainment), appliance characteristics (e.g., heating/cooling technologies), household characteristics (e.g., number of occupants, age of occupants, etc.) and contextual and environmental characteristics (e.g., weather, climate, etc.). Features of the daily load shape, including the timing of peak (i.e., maximum hourly energy use in a day), base (i.e., minimum hourly energy use throughout a day), and the ratio of peak/base, contribute to insights about the relation between activities and electricity use. Finally, the variation of load shape patterns (i.e., entropy) imparts information about consistency or inconsistency of energy use patterns across time. The load shape itself, therefore, contains rich information about a household's energy use and everything within the household related to this use.

To encompass this broad representation of household energy use with daily load shape as a focus, we envision a framework that applies Latent Dirichlet Allocation (LDA). LDA is a generative statistical model that allows unobserved groups to be explained by a set of observations that have related  characteristics. The canonical application of LDA is identifying topics in text analysis~\cite{gerlach2018network, blei2012probabilistic,van2016topic}, where words are observations that are collected from documents, such as a newspaper article, and each document is some mixture of topics that can later be assigned  meaning   (e.g., politics, sports, etc.). In the text example, the process of generating a document is described by a sampling of topics from a mixture of topics, and a sampling of corresponding words according to those topics, and then repeating this process to generate all words in the document. Topics are initialized randomly and then updated through iterations using Variational Bayesian Inference~\cite{blei2003latent, hoffman2013stochastic} or Markov Chain Monte Carlo~\cite{salimans2015markov} approaches until a convergence criteria is met.
\begin{table}[!hbpt]
\centering
\begin{threeparttable}
    \centering
    \caption{Conceptual relationship between text analysis and energy analysis in a LDA setting}
    \label{lda:tab:term:relation}
    \begin{tabular}{ccc}
    \toprule
         Text and language & & Residential energy consumption  \\
         \midrule
         documents & $\longleftrightarrow$ & households \\
         words & $\longleftrightarrow$ & load shapes\\ 
         topics &$\longleftrightarrow$ & energy attributes \\
        \bottomrule
    \end{tabular}
\end{threeparttable}
\end{table}

Applying LDA to the context of analyzing energy demand, we develop a novel application that extracts latent patterns of energy consumption by considering households as documents and load shapes as words. A conceptual relationship of terms in the domain of text analysis to our proposed energy analysis is displayed in \tableref{lda:tab:term:relation}. In this comparison example, latent energy patterns, which we have named an energy attribute, is synonymous with a topic in the text analysis example.

This framework is expected to generate two nested components. The first component is the aforementioned energy attribute, a latent characterization of daily energy use patterns. These energy attributes are derived from daily load shapes and form the most common patterns that households consume energy on a daily basis. Energy attributes are therefore the building blocks of dominant daily energy use patterns in a household, and each household can contain different proportions of these latent attributes. When proportions of these energy attributes are aggregated across a large pool of households, their mixtures among certain household-types can be used to generate a second layer of abstraction which we refer to as an ``energy lifestyle". Energy lifestyles, therefore, are an expression of collective daily energy use patterns across groups of households, and can be applied to any temporally consumptive data stream (e.g., electricity, natural gas, water, etc.).

\subsection{Overview of methods}\label{lda:sec3:relatedwork:ls}

\begin{figure}[!hbpt]
\centering
\begin{subfigure}[t]{0.65\textwidth}
\includegraphics[width=0.99\columnwidth]{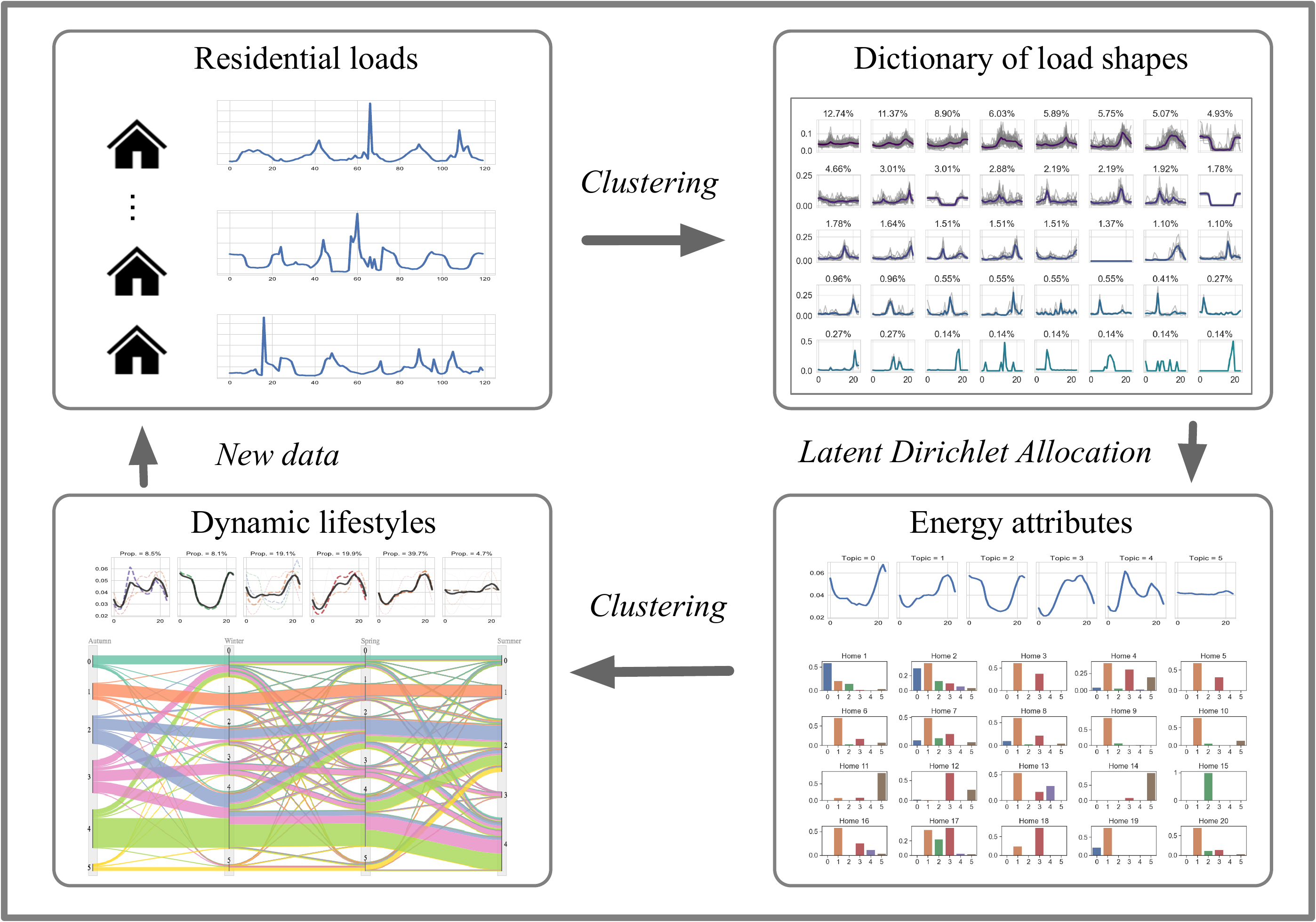}
\caption{Schematic flow of generating dynamic lifestyles.}
\label{lda:fig:schematic:flowchart}
\end{subfigure}
\hfill
\begin{subfigure}[t]{0.34\textwidth}
\includegraphics[width=0.99\columnwidth]{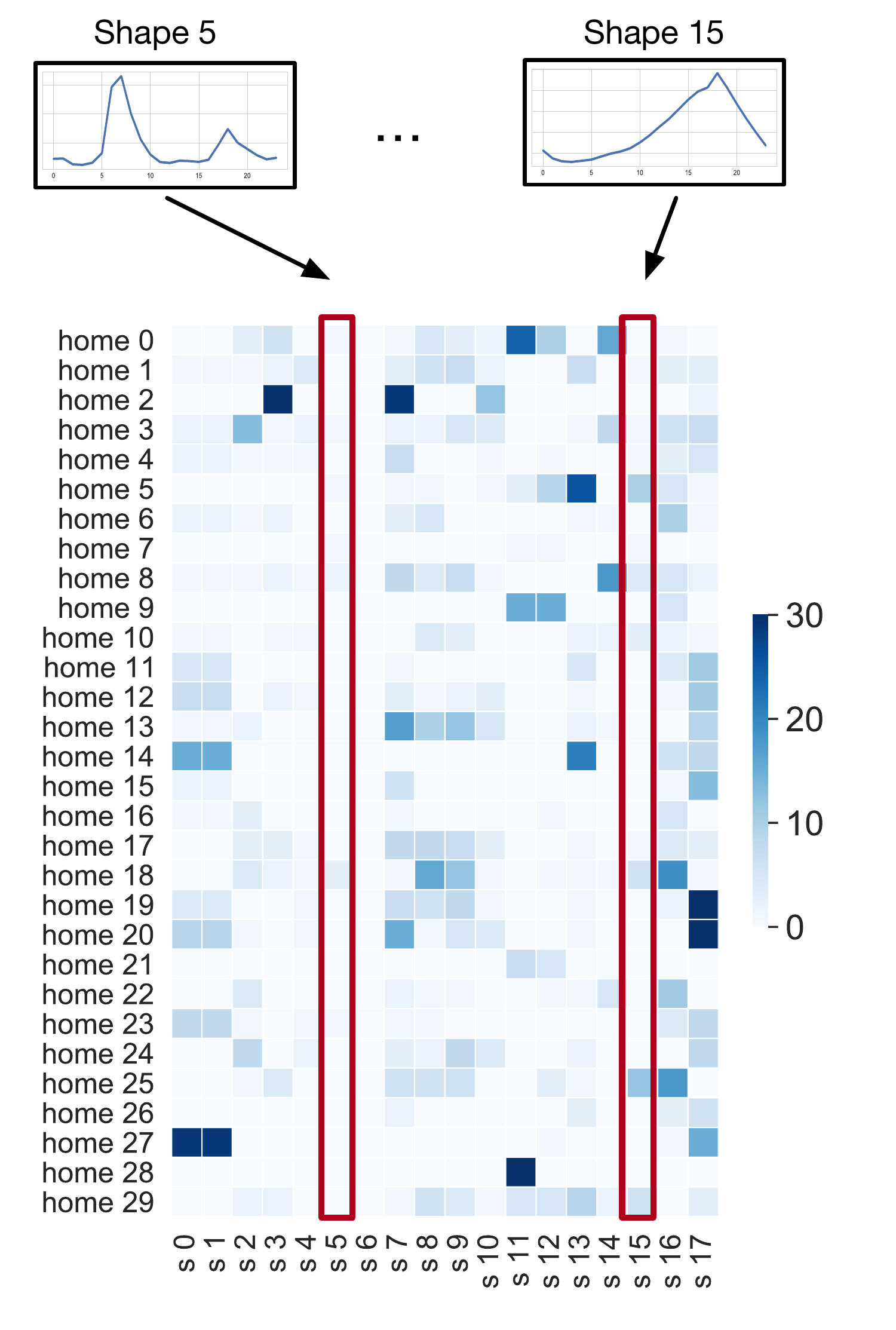}
\caption{Frequency counts of a few load shapes}
\label{lda:fig:schematic:cnt}
\end{subfigure}
\caption{(\subref{lda:fig:schematic:flowchart}) The process of generating dynamic lifestyles. We first  cluster raw smart meter data to create a dictionary of load shapes. Next, we apply LDA to identify representative energy attributes. We then cluster energy attributes to generate lifestyles across time. (\subref{lda:fig:schematic:cnt}) We present the frequency counts of 18 load shapes in a random subset (denoted as $s$ 0 to $s$ 17 on $x$-axis) of 30 homes ($home$ 0 to $home$ 29 on $y$-axis) over the Autumn season (Sept. -- Nov.). }
\end{figure}

In the context of analyzing energy demand, we develop a method to extract latent patterns using LDA. We choose LDA because it is a Bayesian approach and has a better generalization in topic modeling compared with other methods such as Latent Semantic Analysis (LSA)~\cite{landauer1998introduction} and Probabilistic Latent Semantic Analysis (pLSA)~\cite{hofmann2001unsupervised} (see \ref{lda:appx:sec:related:topic:models} for more details). Analogous to the text example where each document contains a mixture of topics, we assume that each household contains a mixture of energy attributes. Therefore, an objective is to identify latent attributes of energy consumption across many households and construct load shapes denoted as $s$. 
Specifically, for a $j$-th home having a mixture of $K$ attributes, the household's attribute mixture weights ${\theta}_j$ is a probability distribution drawn from a Dirichlet prior with parameter ${\alpha}$ and the $k$-th attribute is a multinomial distribution ${\psi}_k$ over a $S$-shape vocabulary (or dictionary). For $i$-th shape $s_{ji}$ in home $j$, a topic $z_{ji} = k$ is sampled from $\theta_j$ and $s_{ji}$ is drawn from ${\psi}_k$. The generative model can therefore be expressed as 
\begin{align}
    {\theta}_j \sim Dir({\alpha}), \psi_k \sim Dir({\beta}), \{z_{ji} = k\} \sim {\theta}_j, s_{ji} \sim {\psi}_k \quad . 
\end{align}
We briefly describe the LDA method here to build intuition about its application and then we expand upon this by providing a more detailed description in \ref{lda:appx:deferred:LDA:detail:explain}. Once energy attributes are finalized, we then apply the $k$-means clustering method on the energy attribute space of households in the second stage to generate a sparse representation of energy usage patterns over days (characterized by cluster centers), which we refer to as energy lifestyles because they contain latent patterns of energy usage generated across households. To provide an overview of the entire process of generating energy lifestyles, we constructed a simplified workflow displayed in \Figref{lda:fig:schematic:flowchart}, described in detail below.

To dynamically create and analyze energy lifestyles, we generate a collection of daily load shapes that covers a majority of household consumption patterns by clustering over raw meter data (\Figref{lda:fig:schematic:flowchart}). Such a collection forms a load shape dictionary that allows us to identify the frequency of generalized load shapes for each household (\Figref{lda:fig:schematic:cnt}). After obtaining the frequency counts of shapes for households, we use the LDA method to yield representative latent energy attributes. Correspondingly, the households can also be viewed as mixtures of attributes (\Figref{lda:fig:schematic:flowchart}). Because many households share similar energy attribute distributions, we use a clustering method to group similar households in their energy attribute space, yielding distinct clusters (i.e., lifestyles). Each cluster represents an energy lifestyle group that can be further interpreted using the proportions of attributes of energy use patterns within the cluster.

We next explore temporal dynamics by applying the LDA method on seasonal intervals (Autumn, Winter, Spring, Summer) and assign each household to its nearest computed lifestyle, which we display as seasonal transitions of lifestyles for households (\Figref{lda:fig:schematic:flowchart}). We can then iterate on these steps and re-generate insights about household consumption patterns as data streams are updated. While in this research we explore temporal dynamics seasonally, our method can be applied across other time intervals (e.g., monthly, bi-annually, annually, etc.).  

We implement our analysis using Python and Python-based tools such as Scikit-learn~\cite{scikit-learn} for LDA computation, Pandas~\cite{mckinney2010data} for data processing, Numpy~\cite{harris2020array} for matrix calculations, and Matplotlib~\cite{Hunter2007Matplotlib} and Seaborn~\cite{Waskom2021} for data visualization. We have three steps that involve large-scale computations in creating our energy lifestyles: (i) generating a dictionary of daily load shapes; (ii) running Latent Dirichlet Allocation to obtain energy attributes; (iii) using clustering to construct energy lifestyles. We implemented data parallelism in clustering for step (i) and step (iii), and applied LDA from Scikit-learn~\cite{scikit-learn} for step (ii). The major computational burden is creating a dictionary of load shapes but all computational workflows took less than three hours using a computer with a 2.6GHz 8‑core Intel Core i7 processor and 16GB RAM. See the Appendix (\ref{lda:sec:appx:compute:time}) for more details on computation time.

\subsection{Description of residential electricity data}
Our work utilizes a large dataset of residential electricity consumption from Pacific Gas \& Electricity (PG\&E) spanning from August 1, 2010 to August 1, 2011, which contains more than one hundred thousand customers. This data is confidential and cannot be shared publicly. For this analysis, we randomly selected 60,000 households from the dataset, each having complete electricity consumption data in hourly time intervals collected via smart meters. Spatially, these households are located within 436 ZIP codes in California, U.S.A, covering  eight different climate zones (\ref{lda:appx:sec:des:dataset}). Such a sample population, which is larger than many previous studies \cite{jain2017data, beckel2014revealing}, is appropriate for capturing heterogeneity in residential energy consumption patterns. From this data, we then convert each household's load shape pattern into the format of daily (24-hour) sequences over the course of a year for our analysis. This yields a complete dataset with the dimensions of 8,760 (24h $\times$ 365 days) by 60,000, or 525.6 million unique household -- time observations of hourly usage (kWh).

While we do not have information about the electricity pricing rate structure for households in our dataset, very few households with smart meters in California ($\sim 1$\%) were subject to time based variable pricing rates (e.g., time-of-use, critical peak, etc.) during this time period (2010-2011)~\cite{PGandE2012TariffReport}. Additionally, we do not have specific information about residential solar adoption or plug-in electric vehicle (e.g., BEV or PHEV) ownership for the households in our dataset, but during this time period there was relatively low adoption of these technologies in California overall, with less than 1\% of households having rooftop solar~\cite{CEC2019SolarAdoptionReport, Census2010CaliforniaGeo} and even fewer with electric vehicles~\cite{CEC2022EVsalesDashboard, DoT2010HighwayStat}. Lastly, we do not have information about hot water heating system technology deployment in our data, but a large proportion of California households (88\%) were estimated to use natural gas water heaters, and not electric water heaters, during this time period~\cite{CEC2010pubreport}.

\section{Results} 
Our framework is heavily driven by empirical data from actual residential households using several contemporary machine learning methods. Specifically, we first run an experiment for generating a representative load shape dictionary by clustering raw residential smart meter data. In our next series of experiments, we apply this load shape dictionary and then synthesize typical lifestyles by using LDA. When summarizing the lifestyle profiles, we assign them names according to a composite shape formed via reconstructing the weighted sum of load shapes. Once these lifestyle profiles are obtained, we run a series of experiments to validate the profiles by examining the electricity consumption features and show how these features (e.g., the ratio of morning to whole day energy use) support temporal characteristics of these lifestyles. Finally, we run a series of experiments to identify households that change lifestyles across seasons (Changer) and those that do not (No Changer). 

\subsection{Dictionary of load shapes}
Since households display a variety of load shapes across time, and that the mixture of these load shapes is associated with the lifestyle that households may possess, we first learn a dictionary of daily load shapes that is the foundation of our energy lifestyle approach. To generate a robust dictionary of load shapes, we utilize clustering methods with a careful selection of distance metrics (\ref{lda:appx:sec:cluster:detail:distance}).

Given a set $\mathcal{X}$ that includes all daily electricity loads and a data point $\vx \in \mathcal{X}$, we would like to find a number of representative points of clusters, denoted as a set $C \supseteq \mathcal{X} $, that can summarize a massive dataset into several typical patterns. To accomplish this, we minimize the distance between points and cluster sets $\min_{\vx\in \mathcal{X}} d(\vx, C)$ in metric $d$, where $d(\vx, C)=\min_{c\in C} d(\vx, c)$ is the minimum distance from $\vx$ to a center $c$. Taking the standard $k$-means as an example, we have an assignment $\phi: \mathcal{X} \rightarrow C$ of points to clusters so as to $\min_{\vx\in C} d(\vx, \phi(\vx, C)) = \min_C \sum_{i=1}^k \min_{\vx\in C_i}|| \vx - c_i||_2^2 $, where $d$ is the Euclidean distance between two points. In addition to the Euclidean distance, we also apply the cosine distance, the $\normlone$ distance, and dynamic time warping (DTW) distance~\cite{teichgraeber2019clustering} to perform the clustering for the load shape dictionary. We also test $k$-medians~\cite{bradley1997clustering}, hierarchical clustering~\cite{johnson1967hierarchical}, and DBSCAN~\cite{ester1996density} clustering methods for comparison (\ref{lda:appx:sec:cluster:detail:meth}). We set the load shape dictionary of size 200 using the $k$-medians method with a hybrid of DTW and Euclidean distances, because this setting yields a good coverage of profiled shapes with the highest score on the Calinski-Harabaz Index~\cite{calinski1974dendrite}. Further technical detail about creation of the load shape dictionary is provided in \ref{lda:appx:sec:eval:clustr:perform} and \ref{lda:appx:sec:cluster:detail:dic:size}.

\subsection{Energy lifestyles composition}
Once we have derived a dictionary of 200 load shapes, we use the clustered labels (i.e., shape indices) to represent each household's load shape pattern. Specifically, we calculate the frequency of the load shapes for a household and represent them as a 200-dimensional vector. For example, during a calendar year, if the home $i$ repeated ``shape 1'' for 23 days, ``shape 2'' for 17 days, ``shape 200'' for 325 days, then we have the vector $r_{i}=[23, 17, \hdots, 325]$ that describes the one-year pattern of home $i$, where $r_i\in \R_{+}^{200}$. Referring to \Figref{lda:fig:schematic:cnt}, we stack all households' load patterns into an $n$-by-200 matrix $M_r$ where $n$ is the number of homes.

\begin{figure}[!hbpt]
    \centering
    \includegraphics[width=0.99\textwidth]{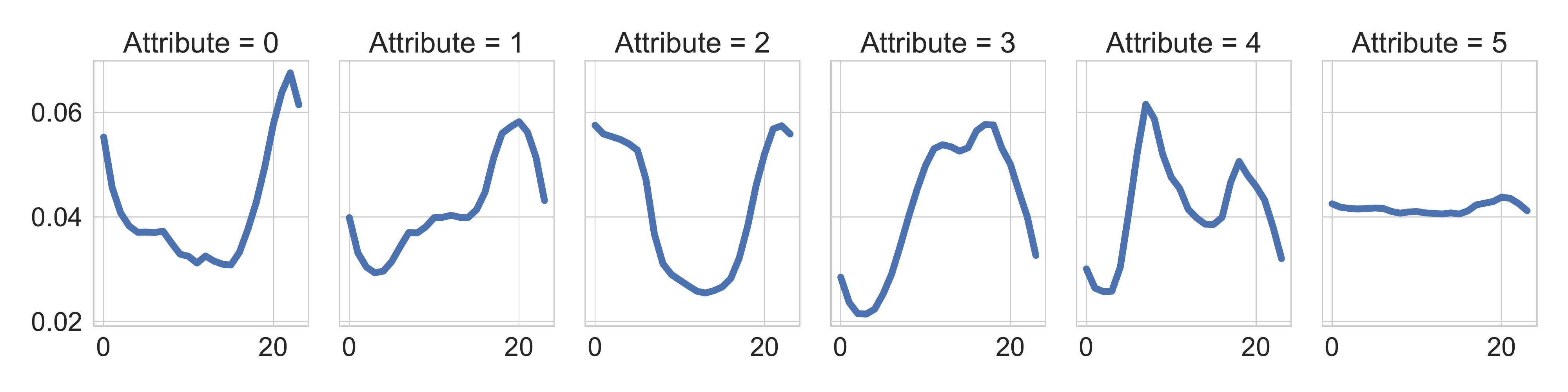}
    \caption{Energy attributes. Each attribute shape is a weighted sum of 200 dictionary shapes, where the weights are the normalized probabilities of each shape's occurrence.}
    \label{lda:fig:6topicshapes}
\end{figure}

We apply the LDA method to extract a few distinct and representative energy attributes of load shapes.  To determine how many attributes are appropriate to both capture all consumption patterns while also being sufficiently representative, we prescribed 10 energy attributes and then merge the neighboring attributes together using a bottom-up approach, i.e., by calculating correlations of attributes and projecting them down to lower dimensions (\ref{lda:sec:appx:gen:ene:attr}).  After consolidating similar attributes, this ultimately yields six representative energy attributes that are quantitatively and descriptively distinct in terms of daily consumption patterns (see ~\Figref{lda:fig:6topicshapes}). These six attributes are further verified via a pairwise distance matrix. More details, including why we chose six attributes, are provided in \ref{lda:sec:appx:gen:ene:attr}. We observe that \emph{attribute 0} has the peak consumption around 10pm--11pm with lower energy use during the daytime. A similar pattern is also observed in \emph{attribute 2} but with a longer time span of late-night consumption, whereas \emph{attribute 1} has a peak consumption around 6pm--7pm with lowest energy use around 2am--3am. Distinct from other attributes, \emph{attribute 3} has the highest energy usage in the afternoon from 12pm--5pm and \emph{attribute 4} displays peak usage around 7am--8am in the morning. Finally, we observe that \emph{attribute 5} has comparatively low variation in daily usage.

\begin{figure}[!hbpt]
    \centering
    \includegraphics[width=0.99\textwidth]{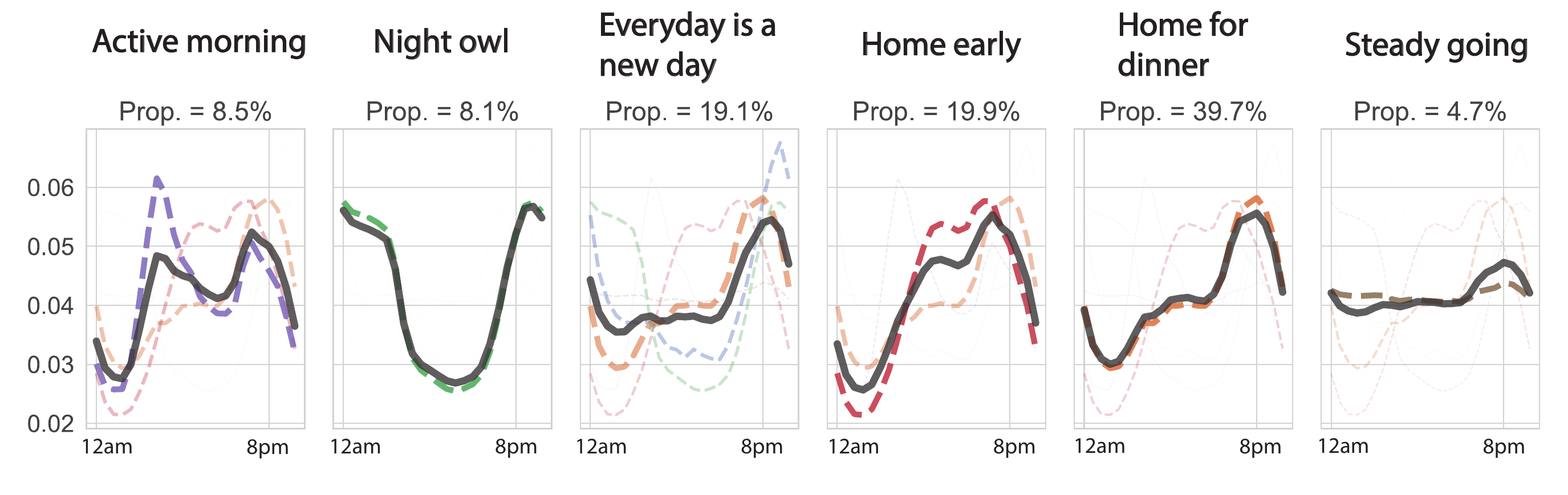}
    \caption{Energy lifestyles. From left to right, they are Active morning, Night owl, Everyday is a new day, Home early, Home for dinner, and Steady going. The different thickness of the dashed lines indicate different composition weights for corresponding attributes, depicted as different colors. The dark grey solid line represents the weighted average of all attributes. }
    \label{lda:fig:ls_mix_6topics}
\end{figure}

With these six summarized attributes, each home is then characterized by assigning a 6-dimensional vector where the value at each entry represents the corresponding attribute weight. The attribute weight at the $k$-th entry indicates how likely a home possesses \emph{attribute}-$k$. We found that six lifestyles were 
sufficient to cover the heterogeneity of daily lifestyle patterns according to the inertia heuristic and distinct mixture of probability mass of energy attributes (shown in \ref{lda:appx:sec:n:ls}).
In \Figref{lda:fig:ls_mix_6topics}, we plot each lifestyle as a dark grey line that represents a weighted average of different attribute weights depicted by the thickness of the dashed lines. Given their load shape characteristics, for ease of reference we assigned names to each of the lifestyles from left to right as Active morning, Night owl, Everyday is a new day, Home early, Home for dinner, and Steady going. In naming these lifestyles, we use the following as descriptive justification: Active morning has a distinguishing characteristic of energy use in the morning time period; Night owl has energy use in the very late night and very early morning with little morning through evening usage across days; Everyday is a new day displays substantial heterogeneity in daily energy use patterns across different days; Home early is distinguished by its late afternoon use; Home for dinner has energy use concentrated in the evening; and the Steady going lifestyle has use that remains relatively stable throughout the day. We have no additional, non electricity-use information about these households to verify or justify these lifestyle names, a challenge confronted by other  ``unsupervised'' learning applications~\cite{van2016topic,beckel2014revealing}.

The Home for dinner lifestyle is the most frequently occurring lifestyle among our sampled residential households, accounting for 39.7\% of the households in our dataset. The next most frequently occurring lifestyles are Home early and  Everyday is a new day, which account for 19.1\% and 19.9\% respectively, followed by Active morning and Night owl, both of which account for approximately 8\% of the sample. At 4.7\%, the least frequently occurring lifestyle is Steady going. Although the mean representations of Everyday is a new day and Home for dinner are similar, they are different in that the mixture weights of the attributes are evenly distributed for Everyday is a new day but the weights for Home for dinner are concentrated on \emph{attribute 1} (\ref{lda:appx:sec:n:ls}).            

\begin{figure}[!bpht]
    \centering
    \includegraphics[width=0.99\textwidth]{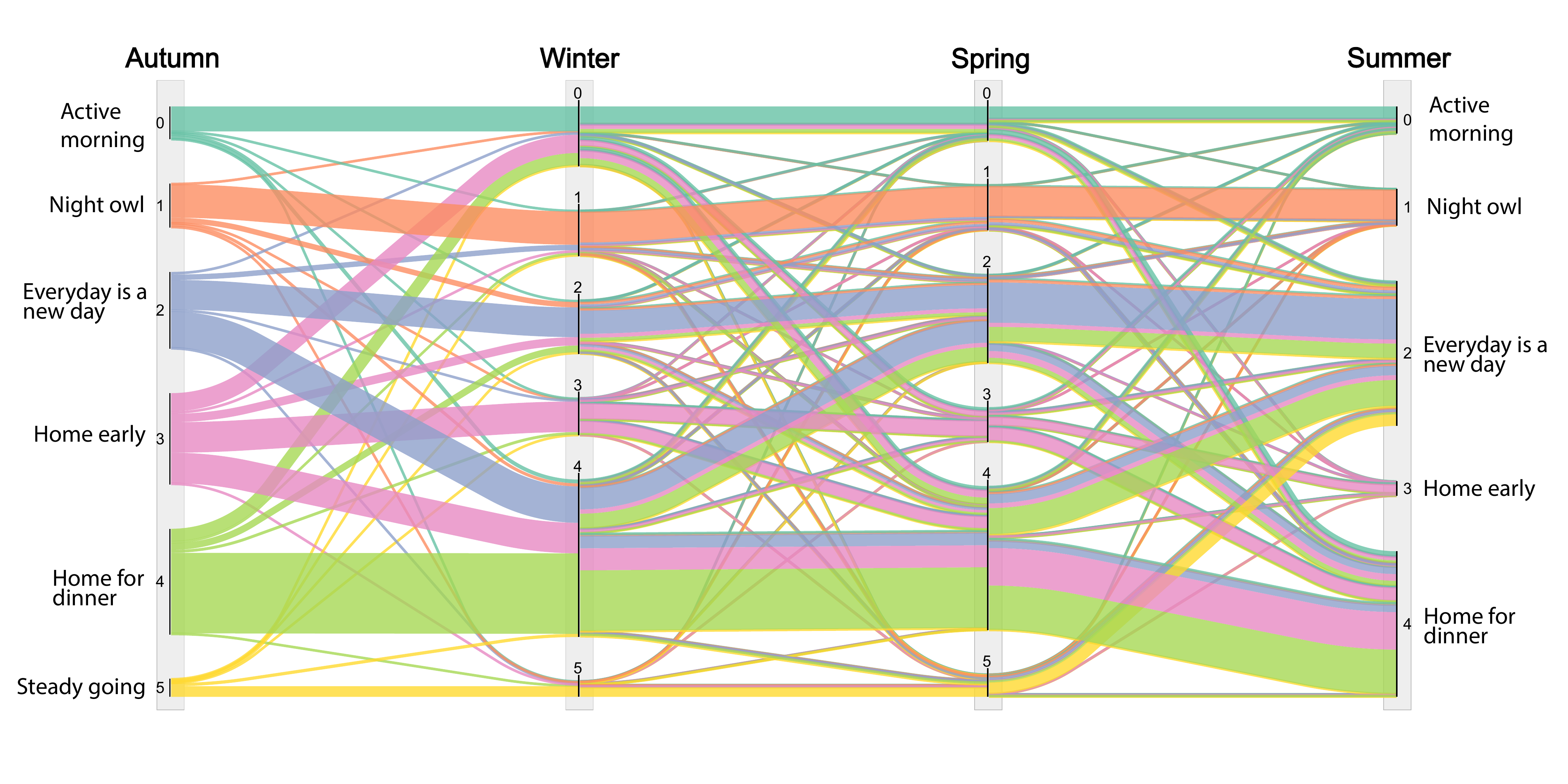}
    \caption{Seasonal transition of lifestyles spanning from Autumn 2010 through Summer 2011. The color of the lines represent the lifestyle designation in Autumn and tracks groups of households across time. The thickness of the lines represent the proportion of total households in each lifestyle at each seasonal interval, with wider lines indicating more households and thinner lines fewer households. }
    \label{lda:fig:sanky_diagram:seasons}
\end{figure}

While different households have different lifestyles, we also observe that a single household can display multiple lifestyles over the course of a year. For example, one set of lifestyle patterns could be related to the presence of children in the home, such as when children are on break during the summer months and in school during the fall. Change of lifestyle could be associated with a household members' behaviors (e.g., occupancy) under different time horizons (e.g., months, seasons, years, etc.). To this end, we next examine how these energy use behaviors change across time by choosing season as a convenient unit of measurement. Therefore, we partition our one year's worth of data into four seasons: Autumn (Sept. -- Nov.), Winter (Dec. -- Feb.), Spring (Mar. -- May), and Summer (June -- Aug.) and run lifestyle analysis for seasonal time intervals. The lifestyle transitions of households across seasons is displayed in \Figref{lda:fig:sanky_diagram:seasons}. 

We find that the Home for dinner lifestyle comprises a larger proportion of households across seasons compared to the other lifestyles. Such a seasonal phenomenon also matches the previous findings in the observations across the entire year in \Figref{lda:fig:ls_mix_6topics}. Each season contains households with all six lifestyles except for summer which does not contain any households with the Steady going lifestyle. One reason could be that the relatively flatter usage profile of Steady going is particularly uncommon during summer months because thermal comfort-related consumption---such as HVAC usage---tend to be turned on and off for multiple hours across a day, yielding a more volatile daily load shape. Whereas in the winter, many homes in California rely on gas--heating and therefore regulation of thermal comfort could yield a flatter pattern of electricity use. Furthermore, we find that some households stay in a single lifestyle across all seasons, whereas other homes switch between two or more lifestyles over the course of four seasons (see \ref{lda:appx:sec:pop:change} for more details). Such an observation motivates us to investigate the distinctions between those lifestyle-consistent households and lifestyle-changing households.

\subsection{Energy lifestyle analyses}
To understand what determines  different lifestyles, we explore a number of energy use characteristics derived from raw smart meter data from households. Unlike many other studies~\cite{Kwac2018LifestyleSeg, beckel2014revealing, xu2017household}, our energy use characteristics (also known as features) are generated using raw energy data from households and does not rely on the previously generated load shape dictionary or energy attributes. We first illustrate the features associated with corresponding lifestyles and then explore the changes overtime of various features of lifestyles across seasons. This allows us to identify those households that change lifestyles across seasons (Changer) and those who do not (No Changer). 

\subsubsection{Features of energy use}
Once we have constructed our lifestyles, each household is associated with a single lifestyle. Conditioning on these lifestyles, all households that are labeled in the same lifestyle are grouped. We find that each group of households has unique distributions of certain energy use features. These features are extracted from raw electricity use from households, such as mean daily energy use, ratio of morning to whole day energy use, and peak hour frequency (normalized), described in detail in \tableref{lda:tab:appx:features:def}. 

\begin{table}[!htpb]
    \centering
    \begin{small}
    \caption{Description of features of electricity use}
    \label{lda:tab:appx:features:def}
    \begin{tabular}{c|l}
    \toprule
         \textbf{Feature} & \multicolumn{1}{c}{\textbf{Description}} \\
    \midrule
        \rowcolor{cyan!10}
         $E_{day}$ & mean of daily energy use   \\
         \rowcolor{cyan!10}
         $E_{hour}$ & mean of hourly energy use \\
         \rowcolor{cyan!10}
         $E_{peak}$ & mean energy use of peak hour in a day, equivalent to $E_{max}$ \\  
         \rowcolor{cyan!10}
         $E_{base}$ & mean base energy use of a day \\ 
         \rowcolor{cyan!10}
         $E_{min}$ & mean of min energy use of a day \\ 
         \rowcolor{cyan!10}
         $E_{morning}$ & morning energy use between 6am to 10am \\
         \rowcolor{cyan!10}
         $E_{noon}$ & morning energy use between 10am to 2pm \\
         \rowcolor{cyan!10}
         $E_{evening}$ & evening energy use between 6pm to 10pm \\
         \rowcolor{cyan!10}
         $E_{night}$ & night energy use between 10pm to 2am \\
         \rowcolor{cyan!10}
         $E_{wholeday}$ & 24 hour energy use \\
    \midrule
        \rowcolor{purple!10}
         $r_{base}$ & {base load ratio, i.e. mean of $\frac{E_{base}}{E_{day}}$} \\
         \rowcolor{purple!10} 
         & mean ratio of min hourly load divided by max hourly load, \\
         \rowcolor{purple!10}
         \multirow{-2}{*}{$r_{min2max}$} &  i.e. mean of $\frac{E_{min}}{E_{max}}$ \\
         \rowcolor{purple!10}
          & mean of morning energy use divided by whole day energy use, \\
          \rowcolor{purple!10}
          \multirow{-2}{*}{$r_{m2w}$} & i.e. mean of $\frac{E_{morning}}{E_{wholeday}}$ \\
          \rowcolor{purple!10}
          & mean of noon energy use divided by whole day energy use, \\
          \rowcolor{purple!10} 
          \multirow{-2}{*}{$r_{n2w}$} & i.e. mean of  $\frac{E_{noon}}{E_{wholeday}}$ \\
          \rowcolor{purple!10}
          & mean of evening energy use divided by whole day energy use, \\
          \rowcolor{purple!10}
          \multirow{-2}{*}{$r_{e2w}$} & i.e. mean of $\frac{E_{evening}}{E_{wholeday}}$ \\
          \rowcolor{purple!10}
          & mean of night energy use divided by whole day energy use, \\
          \rowcolor{purple!10}
          \multirow{-2}{*}{$r_{ni2w}$} & i.e. mean of $\frac{E_{night}}{E_{wholeday}}$ \\ 
          \midrule
          \rowcolor{green!10} 
           & multinomial distribution over 24 hours showing the normalized\\
          \rowcolor{green!10}
          \multirow{-1}{*}{$\pi_{j}$} & frequency of peak hour occurrence. The $j$ takes  value \\
          \rowcolor{green!10}
          &from 0, 1, ..., 23, indicating $j$-th peak hour in a day \\
    \bottomrule 
    \end{tabular}
    \end{small}
\end{table}

\begin{figure}[!hbpt]
    \centering
    \begin{subfigure}[b]{0.49\columnwidth}
    \includegraphics[width=0.99\textwidth]{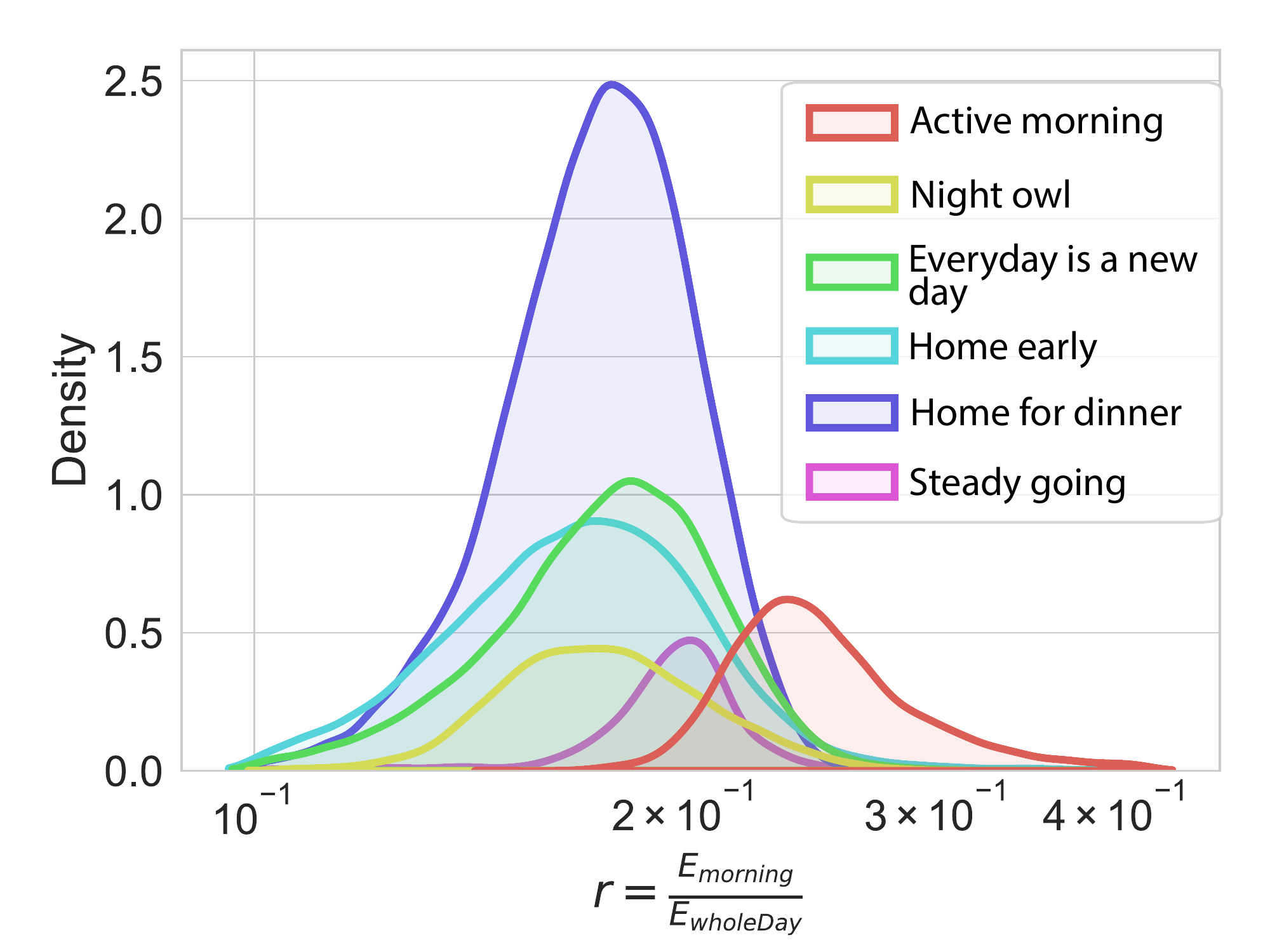}
    \caption{Ratio of morning to whole day energy use}
    \label{lda:fig:year_feat:m2w_ratio}
    \end{subfigure}
    \hfill
    \begin{subfigure}[b]{0.49\columnwidth}
    \includegraphics[width=0.99\textwidth]{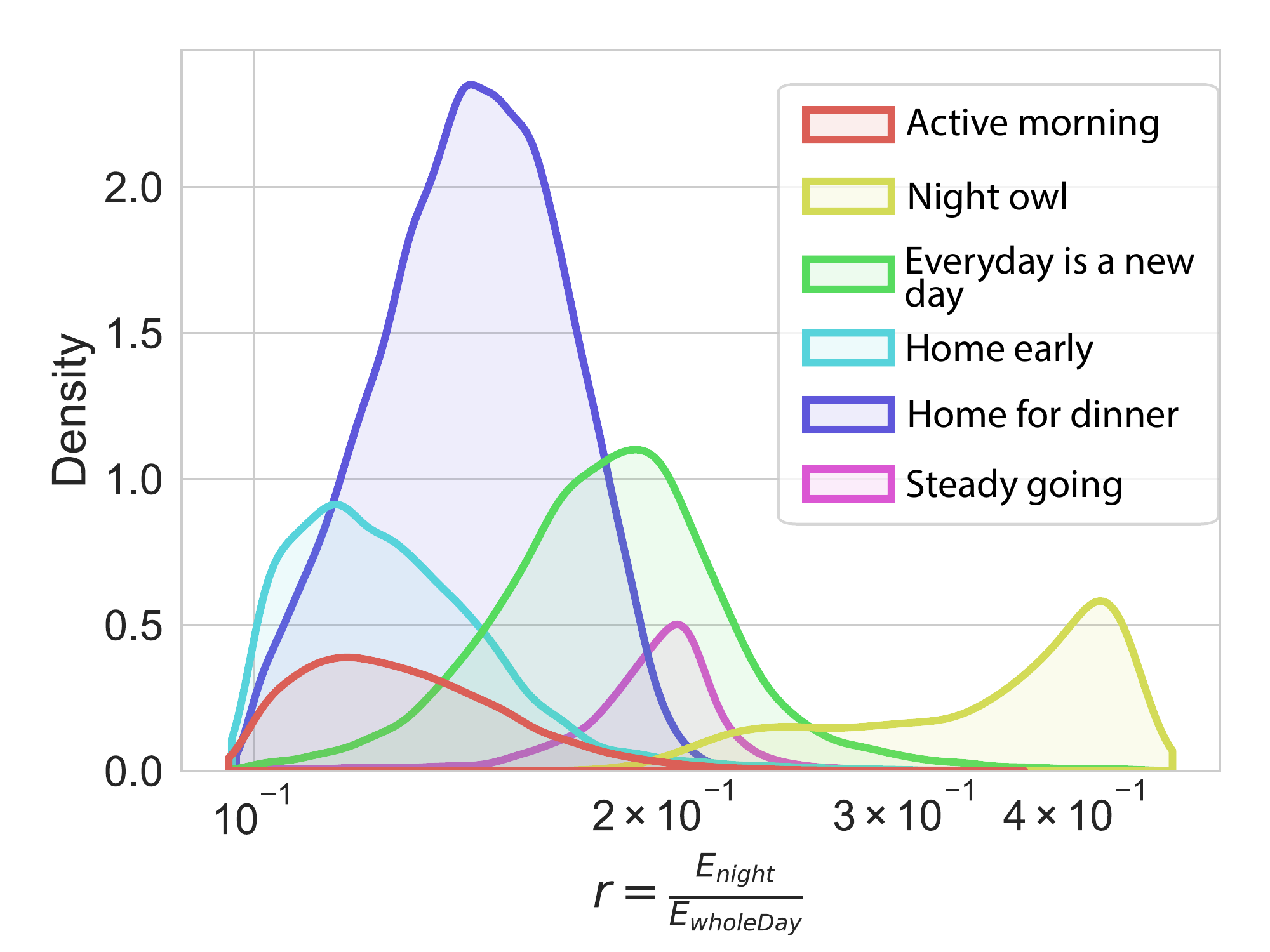}
    \caption{Ratio of night to whole day energy use}
    \label{lda:fig:year_feat:ni2w_ratio}
    \end{subfigure}
    \hfill
    \caption{Load features of different lifestyles. (\subref{lda:fig:year_feat:m2w_ratio}) suggests that Active morning style has a higher ratio of morning to whole day energy than other lifestyles. (\subref{lda:fig:year_feat:ni2w_ratio}) reflects that Night owl has a significantly higher ratio of night to whole day energy than other lifestyles. }
\end{figure}

We present a few examples showing that certain lifestyles can be distinguished from feature distributions (\Figref{lda:fig:year_feat:m2w_ratio} and \Figref{lda:fig:year_feat:ni2w_ratio}). Specifically, \Figref{lda:fig:year_feat:m2w_ratio} indicates that the Active morning group has the highest ratio of morning (6am--10am) to whole day energy use whereas the mean is approximately 0.24. The Home for dinner group has a mean ratio value of approximately 0.18 with a substantial portion of homes having an even lower value of 0.10. Other lifestyles such as Night owl, Everyday is a new day, and Home early have the mean ratio value between 0.16--0.19. In short, these distributions are consistent with initial insights about our lifestyles as Active morning has higher energy use than other lifestyles in the morning time period. The Night owl lifestyle has the highest ratio of night (10pm--2am), suggesting that many homes use energy between 10pm-2am, accounting for approximately 44\% of the whole day use in that lifestyle group. Other lifestyles have mean ratio values below 0.15 except for Everyday is a new day and Steady going lifestyles, both of which have non-trivial energy consumption during the night period. 

\begin{figure}[!htpb]
    \centering
    \begin{subfigure}[t]{0.49\textwidth}
    \includegraphics[width=0.99\columnwidth]{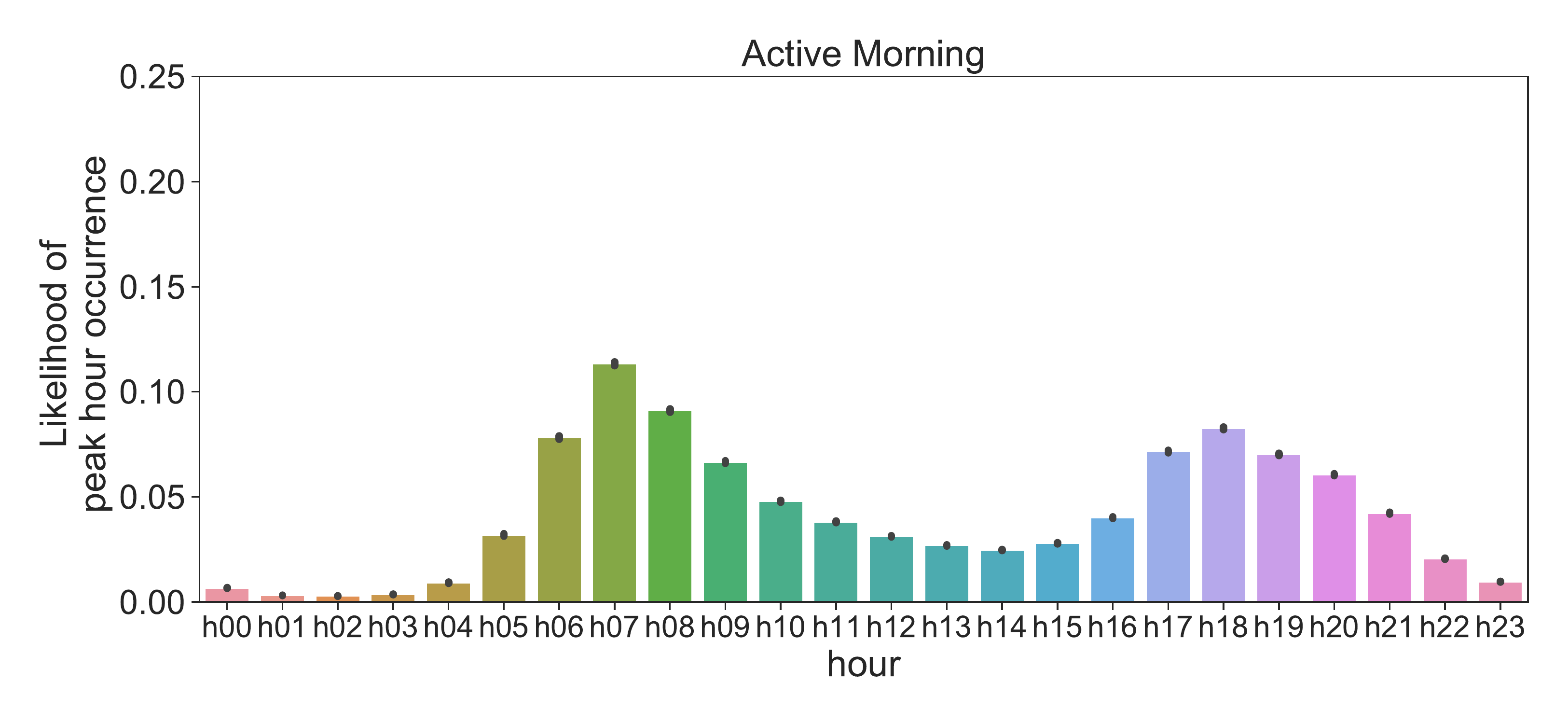}
    \caption{active morning}
    \label{lda:fig:feat:pk_hr:s0}
    \end{subfigure}
    \hfill
    \begin{subfigure}[t]{0.49\textwidth}
    \includegraphics[width=0.99\columnwidth]{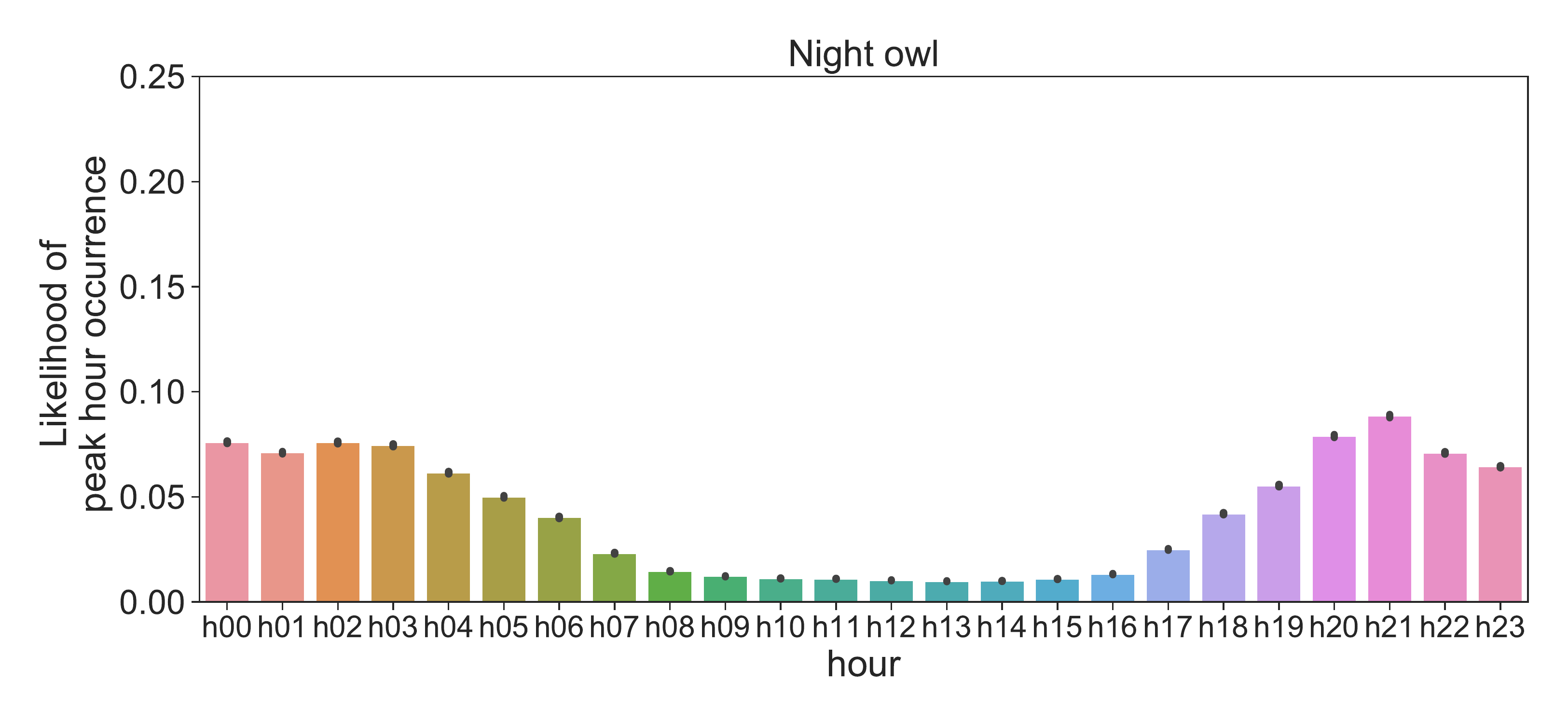}
    \caption{night owl}
    \label{lda:fig:feat:pk_hr:s1}
    \end{subfigure}
    \begin{subfigure}[t]{0.49\textwidth}
    \includegraphics[width=0.99\columnwidth]{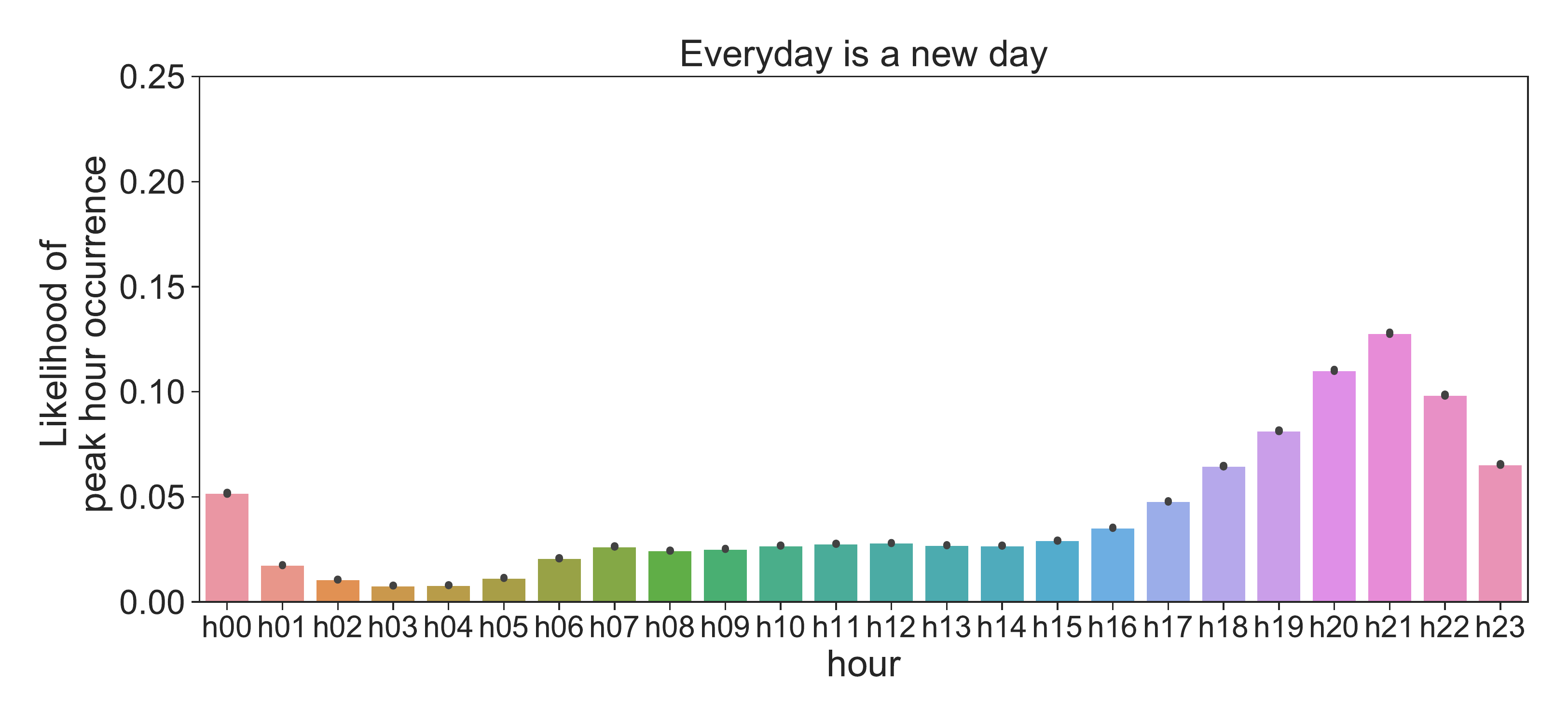}
    \caption{everyday is a new day}
    \label{lda:fig:feat:pk_hr:s2}
    \end{subfigure}
    \hfill
    \begin{subfigure}[t]{0.49\textwidth}
    \includegraphics[width=0.99\columnwidth]{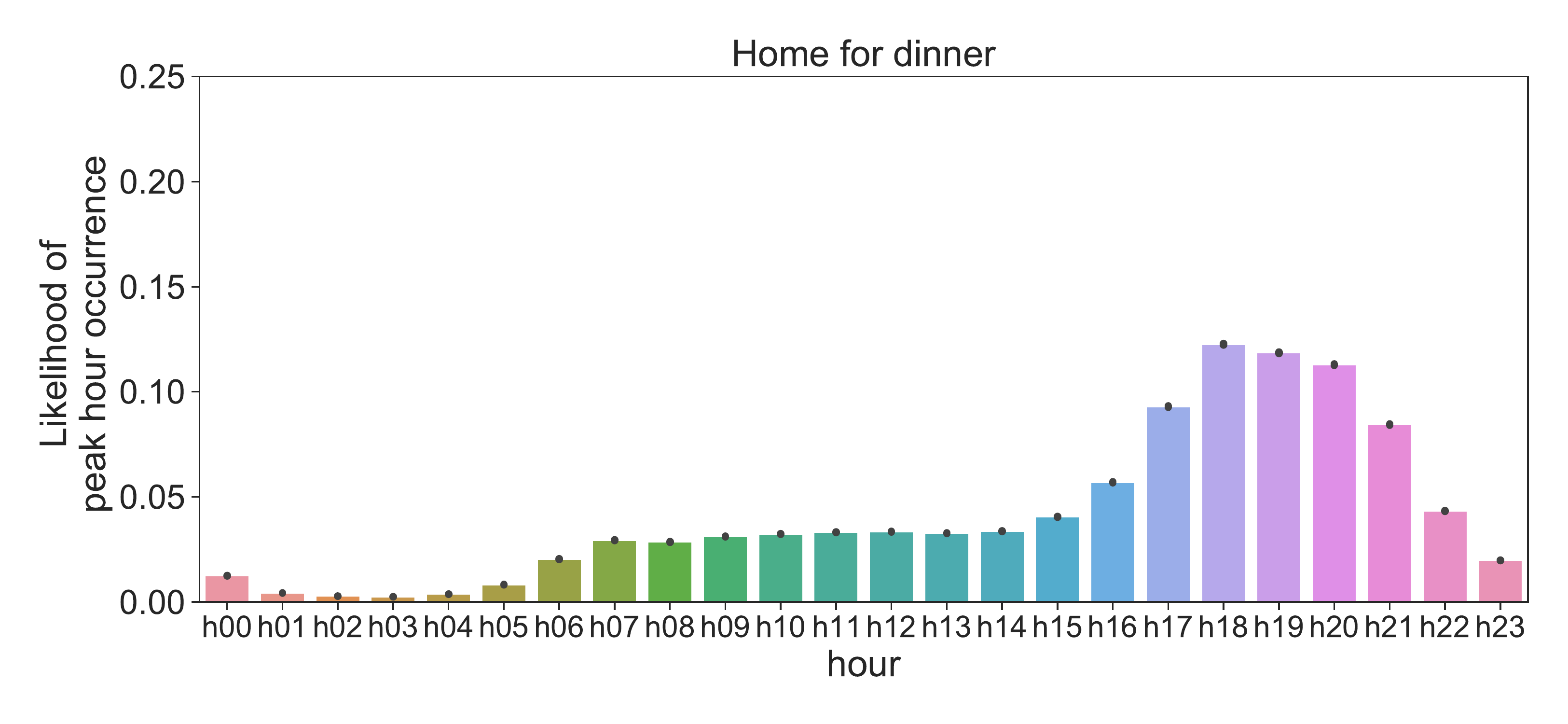}
    \caption{home for dinner}
    \label{lda:fig:feat:pk_hr:s4}
    \end{subfigure}
    \caption{Peak hour distribution per selected lifestyle.  Each sub-figure shows the average frequency of peak hour occurrence for all homes in the corresponding style group.}
    \label{lda:fig:feat:pk_hr:dist:ex:sec4}
\end{figure}

Apart from the intra-day's ratio of energy use, we compare the peak hour occurrence of the different lifestyles. We present the four most prevalent lifestyles among households in \Figref{lda:fig:feat:pk_hr:dist:ex:sec4}.  \Figref{lda:fig:feat:pk_hr:dist:ex:sec4} suggests that the distributions of peak hour frequencies align with lifestyles even though the frequency of occurrences are extracted from raw energy use. For example, the pattern of peak hours frequency for Night owl (\Figref{lda:fig:feat:pk_hr:s1}) closely matches with its lifestyle curve (\Figref{lda:fig:ls_mix_6topics}). Similar matches can also be found in other lifestyles like Active morning, Everyday is a new day, and Home for dinner. Such descriptive cross-validation in peak hours demonstrates the value of our lifestyle framework, while also building an understanding around inductive features for various lifestyles. We present additional summaries  of features in \ref{lda:sec:appx:feat:details}.

As the distributions of features differ substantially among various lifestyles, we expect that lifestyles can be identified by using load-related features. To assess this, we establish a classification problem where the lifestyle of the household $i$ is the label $y_i$ and the features are the observed predictors $\vx_i$. We therefore learn a mapping $f$ such that $y_i=f(\vx_i), \forall i \in 1\dots N$ where $N$ is the number of samples. For interpretability and robustness, we apply random forest (RF) as our classification model with 25 estimators and other default settings from Scikit-learn~\cite{scikit-learn}. After splitting the training, validation, and test sets using the portions of 70\%, 10\%, and 20\%, followed by selecting and calibrating features, we then fit a RF model with a classification accuracy of 68.5\% (for details, see \ref{lda:appx:sec:clf:ls:head}). We find that Night owl is the easiest lifestyle to classify having approximately 82\% accuracy. In contrast, Home early is the most difficult lifestyle to model with 47\% accuracy since a significant portion is miss-classified as Home for dinner. These observations are also supported by the classification results of precision, recall, and F1 score (shown in \ref{lda:appx:sec:clf:ls:head}). 

\begin{figure}[!hbpt]
    \centering
    \begin{subfigure}[t]{0.70\columnwidth}
    \includegraphics[width=1.\textwidth]{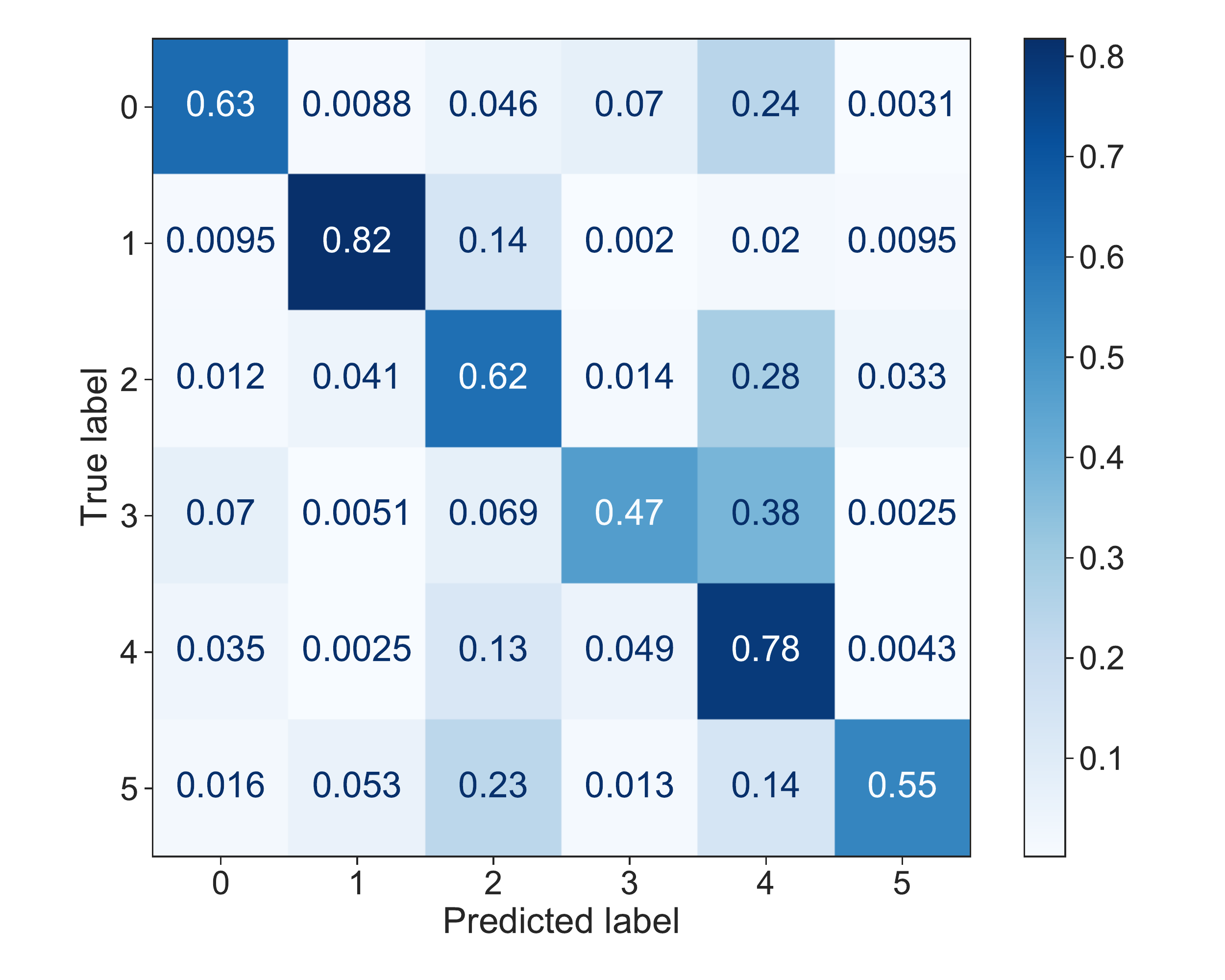}
    \caption{Confusion matrix for classifying lifestyles. The lifestyles are labeled 0 to 5 with 0=Active morning, 1=Night owl, 2=Everyday is a new day, 3=Home early, 4=Home for dinner, 5=Steady going.}
    \label{lda:fig:ls:clf:cm}
    \end{subfigure}
    ~
    \begin{subfigure}[t]{0.70\columnwidth}
    \includegraphics[width=1\textwidth]{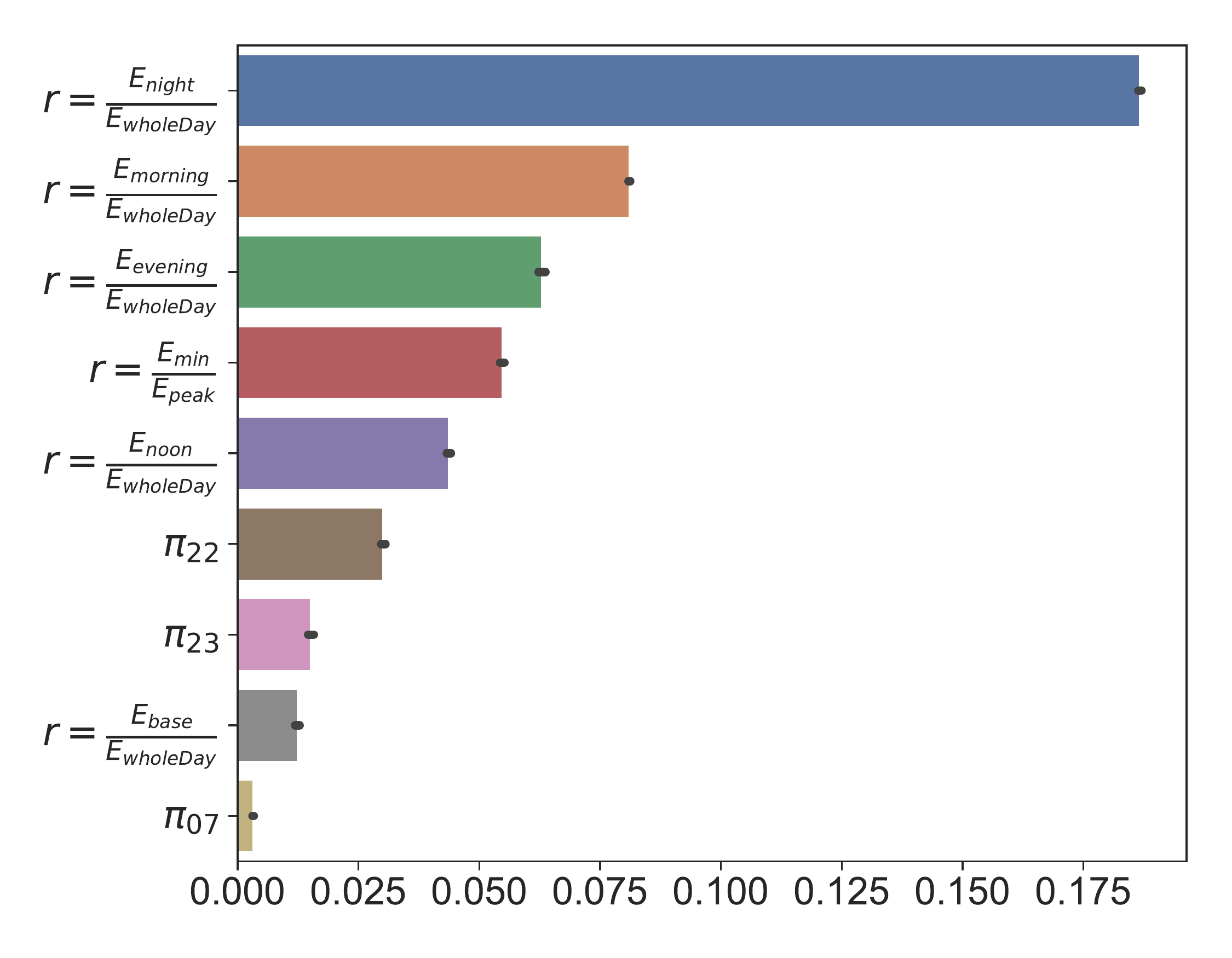}
    \caption{Feature importance}
    \label{lda:fig:ls:clf:featImp}
    \end{subfigure}
    \caption{Classification results. (\subref{lda:fig:ls:clf:cm}) The confusion matrix suggests that Night owl has the highest accuracy at 0.82. In contrast, Home early has the lowest accuracy, 0.47.  (\subref{lda:fig:ls:clf:featImp}) The top nine most important features for classifying a home’s lifestyle.}
    \label{lda:fig:ls:clf:res}
\end{figure}

In addition to comparing the feature distributions of lifestyles and classifying each lifestyle based on household energy consumption, we investigate what features have important roles in determining lifestyles. We use a model-agnostic permutation importance score described in \cite{breiman2001random, fisher2019all} to estimate the importance of the features in our RF model, and discover that the features constructed as various ratios play major roles in identifying lifestyles (\Figref{lda:fig:ls:clf:featImp}). 

We find the mean ratio of night to whole day usage is the most important feature,  contributing to approximately 18\% of additional accuracy compared to a case where the ratio is identically distributed (i.e., random assignment), followed by the mean ratio of morning to whole day that contributes an additional 8\% accuracy. We also observe that the peak hour frequencies at 7th, 22nd, 23rd hour are non-trivial in determining the lifestyle, suggesting that the peak consumption in the night around 10pm-11pm and in the morning around 7am are important features. As an additional robustness step, we verify that these top features are not highly correlated (see \ref{lda:appx:sec:clf:ls:head}).

\clearpage

\subsubsection{Dynamics in energy lifestyles across time}\label{lda:sec:exp:res:chan_nochan:heading}

\begin{figure}[!hbpt]
    \centering
    \includegraphics[width=0.99\textwidth]{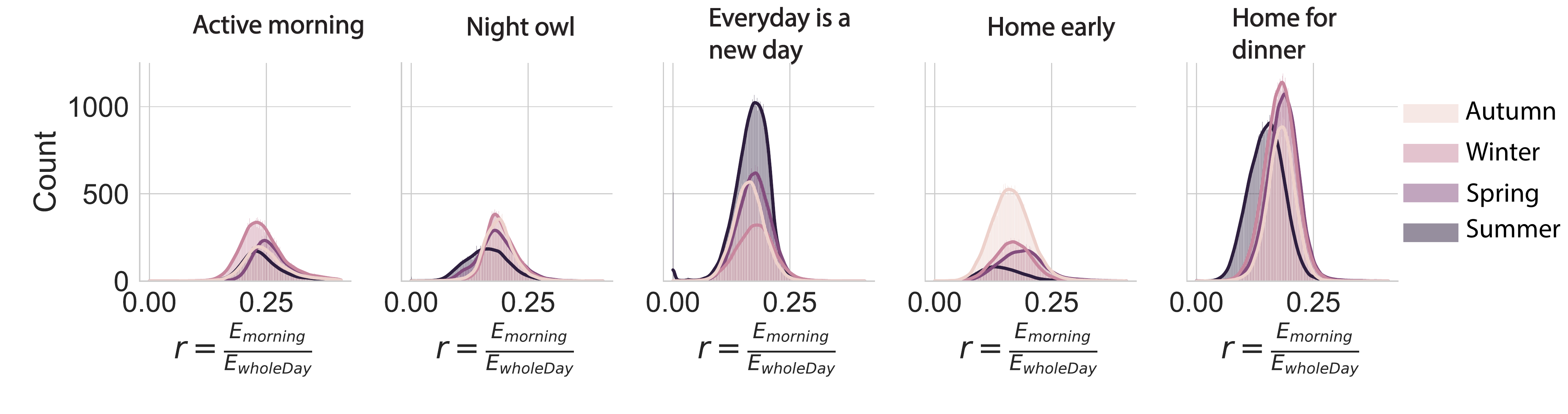}
    \caption{Ratio of morning to whole day of energy displayed seasonally per lifestyle. }
    \label{lda:fig:nochan:season:r_m2w}
\end{figure}

\begin{figure*}[!hbpt]
    \centering
    \includegraphics[width=0.99\textwidth]{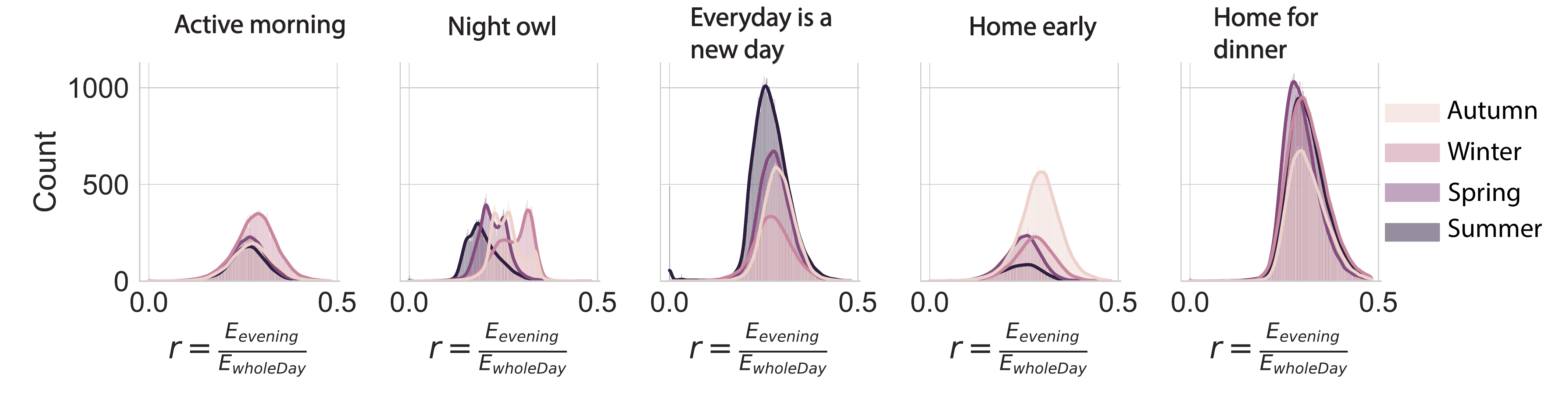}
    \caption{Ratio of evening to whole day of energy displayed seasonally per lifestyle. }
    \label{lda:fig:nochan:season:r_e2w}
\end{figure*}

We next compare the distributions of features at the seasonal-level because certain households may change lifestyles (i.e., Changer, 72.6\% of households) or may not change lifestyles (i.e., No Changer, 27.4\% of households) across a single year period. Since the Steady going lifestyle does not occur in the summer, the following analysis is focused on the remaining lifestyles. First, we assess the characteristics of No Changer households, which comprise 27.4\% of the households in our samples, in terms of load-related features. In particular, we compare both the ratio of morning to whole day usage and the ratio of evening to whole day usage across four seasons to check stability of the feature distributions among various lifestyle groups. We find that Active morning, Everyday is a new day, Home early, and Home for dinner have very stable distributions across four seasons. Consistent with the lifestyle name, Active morning is more influenced by the morning to whole day ratio value (approximately 0.26) compared with any other lifestyle's mean ratio (\Figref{lda:fig:nochan:season:r_m2w}). Although Night owl households tend to keep this lifestyle across multiple seasons, we note that the ratio of morning to whole day usage of the Night owl lifestyle shifts toward smaller values in the summer compared to other seasons, indicating some homes either increase their whole day energy use or reduce their consumption in the morning period during the summer. To confirm the No Changers' stability of load characteristics, we further compared the ratio of evening to whole day usage across the seasons in \Figref{lda:fig:nochan:season:r_e2w}. We observe that all lifestyles have stable distributions of this ratio, with means located between 0.27 to 0.32, except for the Night owl lifestyle that has a mean of 0.19 in summer and 0.34 in winter. Other features, such as mean load and peak load, also demonstrate the stability of No Changer households in various lifestyles (for more information, see~\ref{lda:appx:sec:clf:nochanger}).
%

\begin{figure*}[!hpbt]
    \centering
    \includegraphics[width=0.95\textwidth]{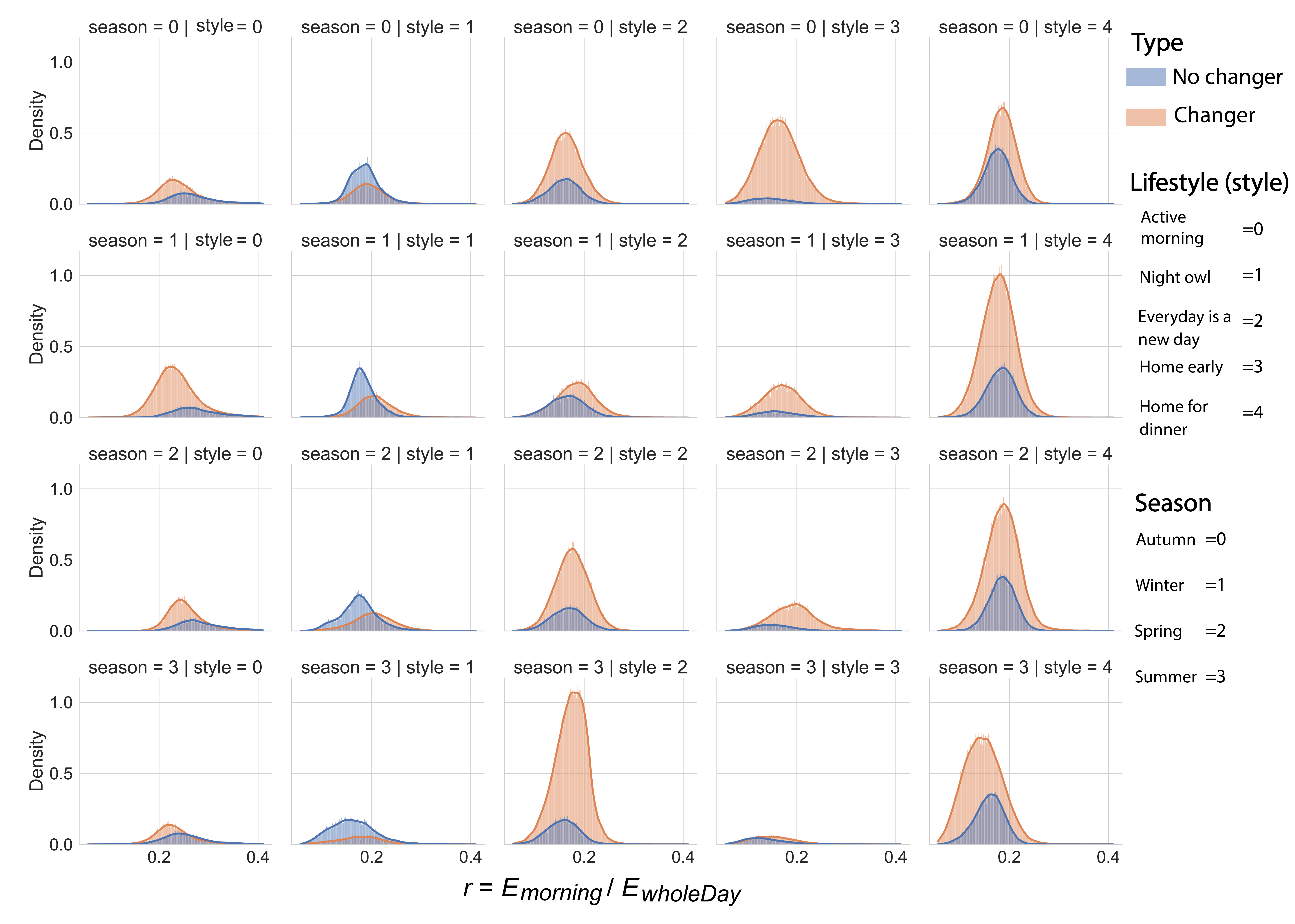}
    \caption{Ratio of morning to whole day of energy displayed seasonally per lifestyle. Lifestyle is abbreviated to ``style'' for visualization purposes.}
    \label{lda:fig:comp:chan:nochan:season:r_m2w}
\end{figure*}

Second, we compare the distributions of load features between Changer and No Changer to understand the difference between these two groups across various lifestyles. Specifically, we evaluate these two groups given a lifestyle and a season, and then expand the evaluation over multiple seasons and lifestyles. For example, the distribution of the ratio of morning to whole day usage is expressed in \Figref{lda:fig:comp:chan:nochan:season:r_m2w}, which suggests three insights. First, in the Active morning lifestyle, the Changers' mean is lower than the No Changers' mean over four seasons. Such a pattern indicates that No Changers tend to consume more in the morning than Changers. Second, overall the No Changers have lower means than Changers for the Night owl, Everyday is a new day, Home early, and Home for dinner lifestyles in four seasons. When comparing the composition of these attributes, No Changers in those lifestyles have higher consumption in the afternoon than in the morning, indicating that morning usage is relatively small. Thus, the Changers could have higher morning usage because they are not restricted to a single lifestyle. Third, the population of Changers (72.6\%) is much larger than that of No Changers (27.4\%). Many Changers switch their lifestyles between Everyday is a new day, Home early, and Home for dinner. In the winter, Changers are mainly concentrated in the Active morning and Home for dinner lifestyles. In contrast, in the summer, Changers are mainly located in Everyday is a new day and Home for dinner. Alternative comparisons using the base--to--peak ratio also suggest that No Changers differ from Changers across seasons (see \ref{lda:appx:sec:clf:nochanger}).

%

\begin{figure}[!hpbt]
    \centering
    \begin{subfigure}[t]{0.73\columnwidth}
    \includegraphics[width=0.85\textwidth]{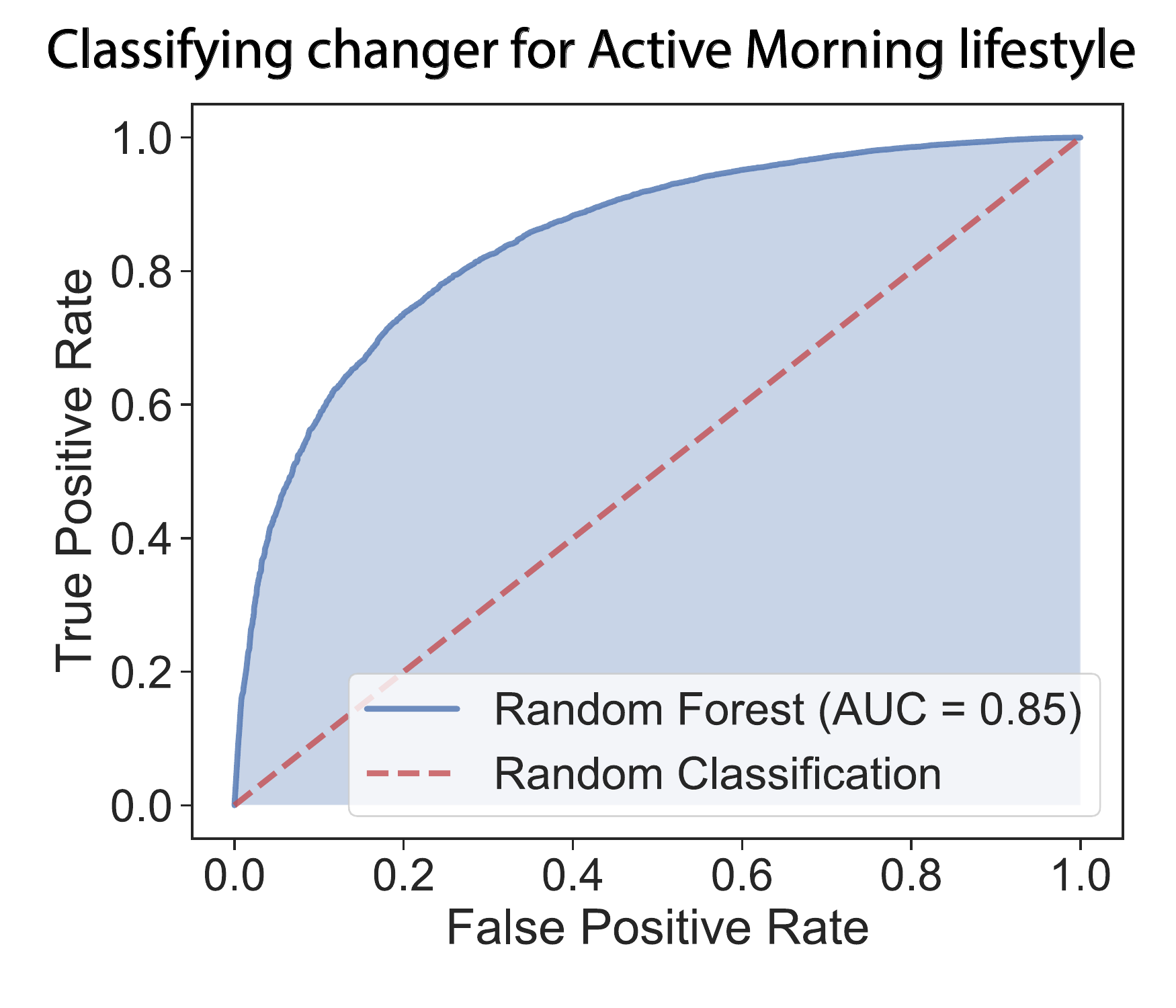}
    \caption{Receiver operating characteristic (ROC) curve displaying the true positive rate (TPR) and false positive rate (FPR) with an area under curve (AUC) of 0.85. }
    \label{lda:fig:ls:clf:chan_nochan:roc}
    \end{subfigure} 
    \hfill
    \begin{subfigure}[t]{0.77\columnwidth}
    \includegraphics[width=0.89\textwidth]{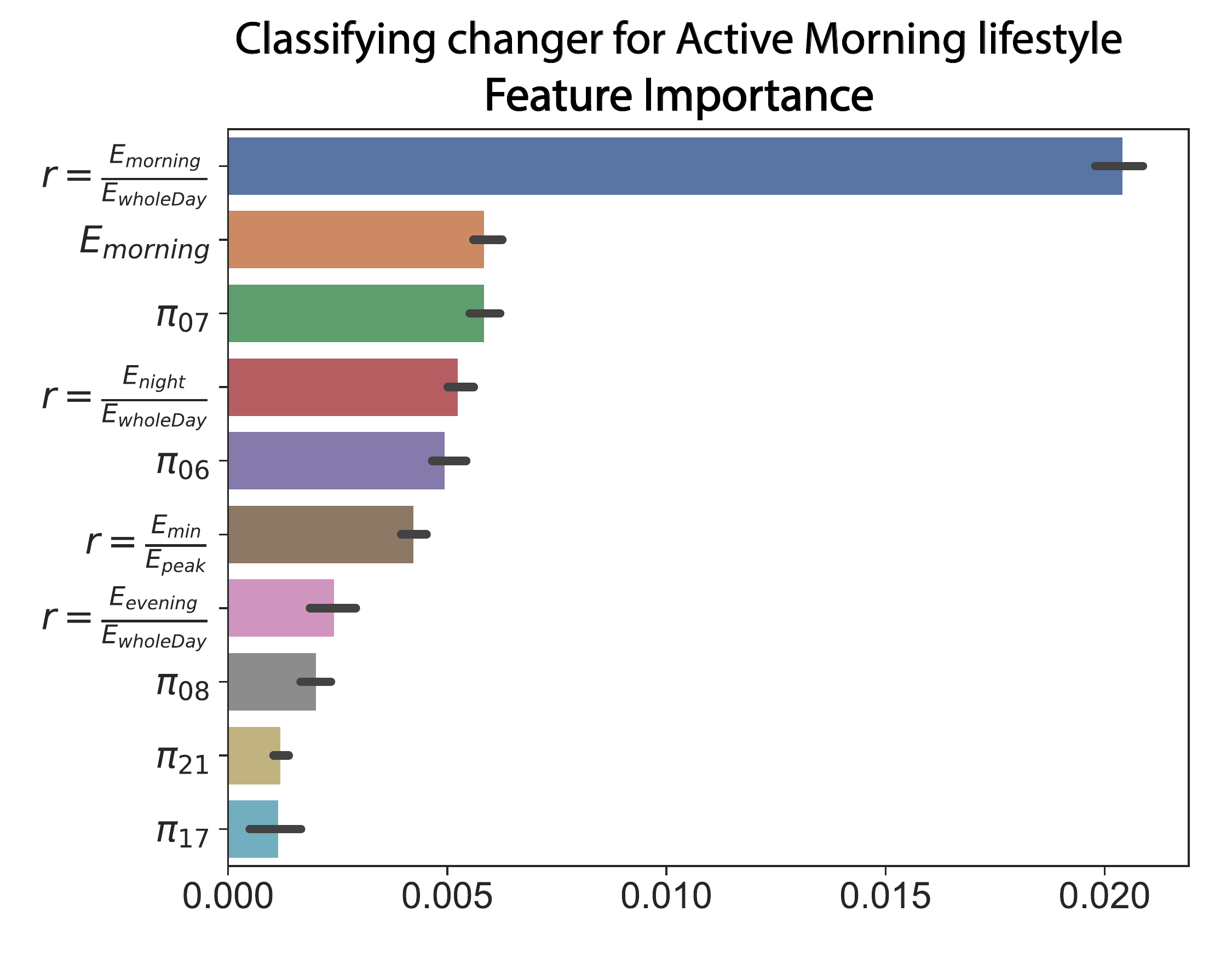}
    \caption{The top 15 features are first selected according to the $F$-value from $\chi^2$ tests between labels (changer or no changer) and features. The top 10 important features are then identified after we permuted the features and fitted a Random Forest model. }
    \label{lda:fig:ls:clf:chan_nochan:featImp}
    \end{subfigure}
    \caption{Identifying Changer vs. No Changer}
    \label{lda:fig:ls:clf:chan:nochan:auc_featImp}
\end{figure}

To accurately classify a Changer or No Changer in each lifestyle, we label the No Changer homes as 0 and label the Changer homes (who possessed the corresponding lifestyle once and then switched to other lifestyles) as 1, and then apply a RF model to this binary classification problem. For example, the lifestyle of Active morning achieves $87.9$\% identification accuracy  (\ref{lda:appx:sec:clf:nochanger}). Because the positive and negative samples are not evenly distributed, the binary decision can be adjusted for a low false positive rate, as shown in the receiver
operating characteristic (ROC) curve in \Figref{lda:fig:ls:clf:chan_nochan:roc} (where the red dotted line denotes  performance of random selection).  The area under the ROC curve (AUC) is 0.85, meaning that a randomly selected positive example (i.e., a Changer household) is more likely to be a Changer than a randomly selected negative example (i.e., a No Changer household) with probability 0.85. 

Once the classifier is fitted to identify a Changer, we evaluate the top determinant features by again using the permutation importance method. \Figref{lda:fig:ls:clf:chan_nochan:featImp} suggests that the ratio of morning to whole day usage, morning energy use, and peak hour frequency at the 7th hour (7am-8am) are among the top three most important features. Such findings indicate that the pattern of energy consumption in the morning period can largely determine whether a household is a Changer or No Changer in the Active morning lifestyle. We also verify that these top importance features are not highly correlated (see the correlation heatmap in~\ref{lda:appx:sec:clf:detail:head}), which demonstrates the robustness of our results regarding important features.       

We assess both the classification performance and feature importance when identifying Changers and No Changers in other lifestyles (\ref{lda:appx:sec:clf:nochanger}). The results show that classifying Changers in the Night owl lifestyle has the highest AUC value of 0.97, and doing so in the Home for dinner lifestyle has the lowest AUC value of 0.77. Such different performances of the AUC metric suggest that identifying Changers in the Night owl group is much easier than identifying Changers in the Home for dinner group.  For feature importance, we find individual lifestyles to have their own prominent features that determine Changers separately, but the features that are characterized by various ratios of energy use play important roles in all lifestyles. In general, features related to certain time spans within a day (such as ratio of evening to wholeday energy use) can be applied to identify whether a household is a Changer or not, and have a higher importance compared to volume-based features (e.g., base load, hourly mean load, etc.).      

\clearpage

\section{Discussion and Conclusion}
\subsection{Contributions and key findings} 
In this research, we present a new approach for constructing dynamic energy lifestyles by applying LDA to residential electricity demand data. Our framework is highly scalable and extensible, while also being flexible enough to accommodate different time intervals and completely new sources of residential energy data from other locations and contexts. Using this dynamic lifestyle approach, we can greatly simplify the interpretation of energy lifestyle patterns by using a method that generates a sparse number of energy attributes that are then used to generate a manageable set of energy lifestyle profiles. We show this process of generating energy lifestyles is robust to multiple load shape dictionary inputs and time intervals. We also demonstrate that these derived energy lifestyles can be associated with certain energy use characteristics, even though these energy use characteristics were not originally applied in constructing the lifestyles themselves. 

We also generate some key insights about residential energy lifestyles. First, our approach reveals that the most frequently occurring of our lifestyles is Home for dinner (40\%), a lifestyle with a 7pm--9pm peak in usage, overlapping a time period where electricity demand system-wide is also highest. We then find that Home early (20\%) and Everyday is a new day (19\%) are the next most frequently occurring lifestyles, both which have very distinct patterns of usage compared to Home for dinner. We also find that when we examine lifestyle membership across seasons, many households manifest multiple lifestyles, and only 27.4\% of households do not change lifestyles throughout the year. This suggests that static representations of energy use patterns across extended time horizons, such as a year or longer, may not be capturing important heterogeneity related to lifestyle change. Finally, we find that our lifestyle approach is related to volume-based characteristics of energy consumption, presenting opportunities for new applications and analysis.

We then apply these energy use characteristics to further interpret these lifestyles and provide insight into how such an approach toward lifestyle analysis could be used in practice. This energy lifestyle analysis approach can also be applied across different time horizons, allowing for applications at varying time intervals to examine temporal dynamics. While in our experiment we applied a seasonal time interval, shorter (e.g., monthly) or longer (e.g., yearly) time horizons for energy lifestyles can also be estimated---dependent on data availability. Such an approach provides the ability to generate meaningful insights that can be applied to a wide variety of energy program designs and use cases. This approach may be particularly useful for utilities who need to understand household energy lifestyles and lifestyle change patterns across time, such as identifying the demand flexibility for households to time-of-use programs~\cite{stelmach2020exploring, di2017nudging, afzalan2019residential}. We describe some potential applications of the energy lifestyle framework in detail below.

\subsection{Potential applications of the dynamic lifestyles framework}

We have identified three potential applications for this lifestyle analysis approach or energy services stakeholders, each considering a different aspect of energy program design. These stakeholders may be as diverse as state and municipal governments, utilities and energy service providers, and other third-parties interested in promoting consumer energy efficiency and deploying clean energy technology, such as solar installers. The first application is in identifying households with lifestyle patterns that are most appropriate for installation of behind--the--meter resources, such as residential solar and battery storage systems, an area of increasing interest due to its potential for power outage resiliency~\cite{zanocco2021lights}. Taking the example of households with the energy lifestyle Home early, these households may be particularly well--suited for rooftop residential solar as they have a pattern of usage that begins in midday, when solar energy potential is higher. A household where usage tends to peak later in the day and during evening hours with less solar energy potential, such as Home for dinner, would be less suitable for targeting residential solar unless it was combined with battery storage (i.e., a solar plus storage system)~\cite{pena2019optimized}. 

For demand response programs for utilities, certain energy lifestyles we derived from our experimental data suggest differing demand flexibility for households, especially when considering demand responsiveness to time-of-use pricing, which typically occurs during weekdays when system-level demand is high, such as in the late afternoon and early evening.  For households in Everyday is a new day, their daily energy use is highly varied. This suggests that these households could be more able to change their daily energy use patterns, making them flexible in their demand because their energy use patterns are less structured compared to other energy lifestyles, such as Home early and Home for dinner. Energy lifestyles that are less flexible, such as Home early and Home for dinner, however, may be better suited for energy efficiency programs, because their lifestyle patterns indicate stable electricity patterns with little day-to-day variation.

While both of these examples are related to households that display relatively static energy lifestyle patterns, how these patterns differ across time is also important for potential applications in practice. First, if a household always displays a particular energy lifestyle pattern, this suggests that the household has a higher affinity toward the pattern of energy use within this lifestyle compared to a household that displays a change in lifestyle across time. Next, the number of lifestyle changes that a household undergoes on a seasonal basis, and the variety of these lifestyles, imparts important information about the household. Households that are constantly undergoing change will likely be difficult to target for demand response programs~\cite{todd2019spillover} given the instability of their daily usage patterns. However, these households may be better candidates for energy efficiency programs such as smart thermostat or smart A/C programs being deployed by utility companies~\cite{dusparic2017residential}.

There are also opportunities to use this energy lifestyle analysis framework to inform energy intervention design used by stakeholders ranging from utilities to consumer advocacy and educational organizations, where households attempt to change their lifestyles to promote energy use patterns that save them money while also lessening their burden on the grid and carbon emissions. To do so, households that have a particular lifestyle with peak demand that corresponds with system demand, such as Home for dinner, could attempt to change their usage to a different lifestyle pattern, such as Steady going or Active morning, with less usage concentrated during peak periods of system demand. This energy lifestyle approach could then be used to determine if there is a shift in lifestyles, and also could become the basis in which to assess whether the household had successfully implemented this change. Moreover, such an approach may be used for households to quickly monitor their own energy lifestyle and make adjustments based on changes in the home or other new activity patterns~\cite{albert2013smart}. In this respect, communicating information about energy use to customers via their lifestyle profile may be more impactful than other forms of more traditional energy use informational summaries (e.g., monthly kWh or energy cost).

Given the wide applications of this dynamic lifestyle approach to a variety of stakeholders, for which we have only provided a few examples, as well as the ability of iteratively updating energy lifestyles, we see great potential for building and extending this framework. However,  our approach has some limitations. First, while we are able to verify these energy lifestyles using other energy use characteristics that were not included in the formulation of the lifestyles, we do not have an additional external measure to verify the presence or absence of this lifestyle based on other, non--energy--use information about household characteristics~\cite{beckel2014revealing}, such as number and age of occupants and patterns of activities in the home. Incorporating such data, if available, would be an important addition to this work and would bolster our framework's ability to provide insights about energy lifestyles. Additionally, the data that we applied in experiments to generate these energy lifestyles is from the early 2010s and therefore does not include recent trends in electricity use patterns within households related to smart home appliances, electric vehicles, and behind--the--meter resources~\cite{carvallo2021framework}, because these technologies were not yet widespread during this time period. To the extent the adoption of these technologies will impact the formulation of these lifestyles themselves is not directly known, but we expect that recent trends in the deployment of solar, storage, and electric vehicles will have some discernible influence.

As our approach to constructing energy lifestyles is data driven, we do not anticipate that household energy producing or consuming technologies that were not prevalent in our original data will present any substantial challenges or bottlenecks when introduced in our energy lifestyle analysis workflow. However, as the deployment of such technologies, as well as demand response due to variable pricing programs (e.g., time-of-use), can represent changes to energy lifestyles, new lifestyles--with new proportions and dynamics--could emerge in more recent household energy data (e.g., from the early 2020s). Our workflow is designed to be scalable to new data as it can adopt different clustering methods (or distance metrics) for generating load shape dictionaries (see \ref{lda:appx:sec:cluster:details:all}), can extract energy attributes using LDA to form energy lifestyles without additional fine-tuning (see \ref{lda:sec:appx:gen:ene:attr} and \ref{lda:appx:sec:n:ls}), and can be applied to much larger datasets without substantial computational burdens (see \ref{lda:sec:appx:compute:time}).

\subsection{Conclusion and next steps}

We conceptualized and implemented a new approach for understanding energy lifestyles that can simplify interpretations about household energy use, has a high potential for applicability and scalability, and can measure changes in energy use across time. There are four immediate directions for future research as an extension to this work. First, this dynamic lifestyle approach can address a cold start problem in identifying patterns of use for new residential customers. Because this lifestyle approach can identify lifestyles using very sparse data inputs, energy providers could recommend energy program enrollment based on lifestyles after only the first few months of meter activation. Second, this dynamic lifestyle approach can be applied to additional residential datasets spanning different time periods and geographies to explore intra- and inter-yearly patterns in lifestyles as well as the influence of context and climate. Third, some steps of our lifestyles approach can incorporate privacy preserving methods, such as differential privacy~\cite{dwork2008differential, dwork2014algorithmic} or generative adversarial privacy~\cite{huang2017context,chen2020generating}, to alleviate the concerns of revealing  sensitive information of an individual household~\cite{chen2018understanding,chen2020energy}, which is an important direction for future work. Lastly, using information about residential electricity data coupled with demographic and household characteristics, our method can further validate and provide new insights about lifestyles by identifying the characteristics related to different lifestyles and their dynamics across time.

\section{Acknowledgements}
The authors would like to thank their colleagues in the Stanford Sustainable Systems Lab (S3L) for their feedback and support of this work. The authors would also like to thank Pacific Gas \& Electricity for providing the data used in this study. The work presented herein was funded in part by TotalEnergies in a research agreement with Stanford University. The views and opinions of authors expressed herein do not necessarily state or reflect those of the funding source. 



\bibliographystyle{elsarticle-num} 
\bibliography{LDA-Lifestyles-refs}

\clearpage
\appendix

\renewcommand{\thefigure}{A-\arabic{figure}}

\setcounter{figure}{0}
\renewcommand{\thetable}{A-T\arabic{table}}
\setcounter{table}{0}

\section{}
\subsection{Related work in topic modeling}\label{lda:appx:sec:related:topic:models}
Traditional topic modeling methods include Latent Semantic Analysis (LSA), probabilistic Latent Semantic Analysis (pLSA), and Latent Dirichlet Allocation (LDA). LSA takes a matrix of observations---households (documents) and load shapes (words)---and decomposes it into a separate household--attribute (topic) matrix and an attribute-loadshape matrix. This dimensionality reduction can be performed using truncated Singular Value Decomposition (SVD)~\cite{landauer1998introduction}. The core idea of pLSA is to find a probabilistic model with latent topics that can generate the data in our households (document)--loadshape (word) matrix. Instead of using SVD deterministically, pLSA finds the latent factors of energy attributes (topics) using an Expectation-Maximization approach~\cite{hofmann2001unsupervised}. Yet, pLSA is not flexible enough for our purposes because it cannot assign probabilities to new households. To overcome this issue, LDA is commonly adopted as the standard approach to find latent attributes (topics). LDA is a Bayesian approach of pLSA, which uses Dirichlet priors for the household(document)-attribute(topic) and attribute--loadshape(word) distributions. This has been shown to have better generalization in topic modeling~\cite{blei2003latent, blei2006dynamic}. This motivates us to use LDA in our energy lifestyle analysis workflow.

\subsection{Description of Latent Dirichlet Allocation}\label{lda:appx:deferred:LDA:detail:explain}
In this section, we describe details of Latent Dirichlet Allocation (LDA) and its application in constructing lifestyles. We use the notation listed in \tableref{lda:tab:appx:LDA:notation:desc}.

\begin{table}[!bpht]
    \centering
    \begin{tabular}{ll}
    \toprule
         Notation & Description  \\
         \midrule
         $k$ & Index of attributes (topics) \\ 
         $K$ & Number of attributes \\
         $i$ & Index of shapes \\
         $j$ & Index of homes or users \\
         $\alpha$ & Dirichlet prior on the attributes in a home \\
         $\beta$ & Dirichlet prior weight of shapes in a attribute \\
         $\theta_j$ & Attribute distribution of home $j$\\
         $\theta_{jk}$ & Proportion of attribute $k$ in home $j$  \\
         $\psi_k$ & Shape distribution of attribute $k$  \\
         $\psi_{ki}$ & Probability of word $i$ occurring in attribute $k$ \\
         $s_j$ & Shape collection of home $j$ \\
         $s_{ji}$ & Shape $i$ in $s_j$ \\
         $z_{ji}$ & Attribute assignment for shape $s_{ji}$ from home $j$ \\
         $M$ & Number of homes \\
         $N_j$ & Number of shapes in home $j$ \\
         \bottomrule
    \end{tabular}
    \caption{LDA symbol description}
    \label{lda:tab:appx:LDA:notation:desc}
\end{table}

The LDA model first prescribes $K$ attributes, with each attribute $k$ associated with a distribution $\psi_k$ over shapes in the dictionary. In particular, $\psi_k$ is sampled from a Dirichlet distribution $Dir(\beta)$. Based on these created attributes, a home $j$ (namely a collection of shapes $s_j$) is generated by first sampling a distribution $\theta_j$ over $K$ attributes from another Dirichlet distribution $Dir(\alpha)$, which determines attribute assignment for each shape in $s_j$, and then choosing each shape $s_{ji}$ based on $\theta_j$. In generating each shape $s_{ji}$, LDA first samples a particular attribute $z_{ji} \in \{1,..., K\}$ from multinomial distribution $\theta_j$, and then the shape $s_{ji}$ is selected from a multinomial distribution $\psi_{z_{ji}}$. This process can be summarized in the following steps:

\begin{tcolorbox}[title=Steps in Latent Dirichlet Allocation]
\begin{tabular}{rl}
step1: & Pick shape distribution of each attribute $k$ by $\psi_k \sim Dir(\beta)$ \\
step2: & Pick attribute distribution for each home $j$ by $\theta_j \sim Dir(\alpha)$ \\
step3: & For each home $j$, for each shape $s_{ji}$ in $j$:\\
 & $\qquad$ Pick an attribute $z_{ji} \sim \theta_j$; \\
 & $\qquad$ Pick a shape $s_{ji} \sim \psi_{z_{ji}}$ \\
\end{tabular}
\end{tcolorbox}

The model fitting can be completed by using variational expectation-maximization (EM)~\cite{blei2006dynamic,blei2012probabilistic} or Markov Chain Monte Carlo methods (e.g., Gibbs sampling \cite{griffiths2004finding}). Both methods can infer the posterior of attribute distribution $\theta$ and attribute-shape distribution $\psi$ efficiently. In our experiment, we use sklearn~\cite{scikit-learn} with variational EM algorithm\footnote{\url{https://scikit-learn.org/stable/modules/generated/sklearn.decomposition.LatentDirichletAllocation.html}} to perform the computation.

\subsection{Description of datasets}\label{lda:appx:sec:des:dataset}
\begin{figure}[!hbpt]
    \centering
    \includegraphics[width=0.35\textwidth]{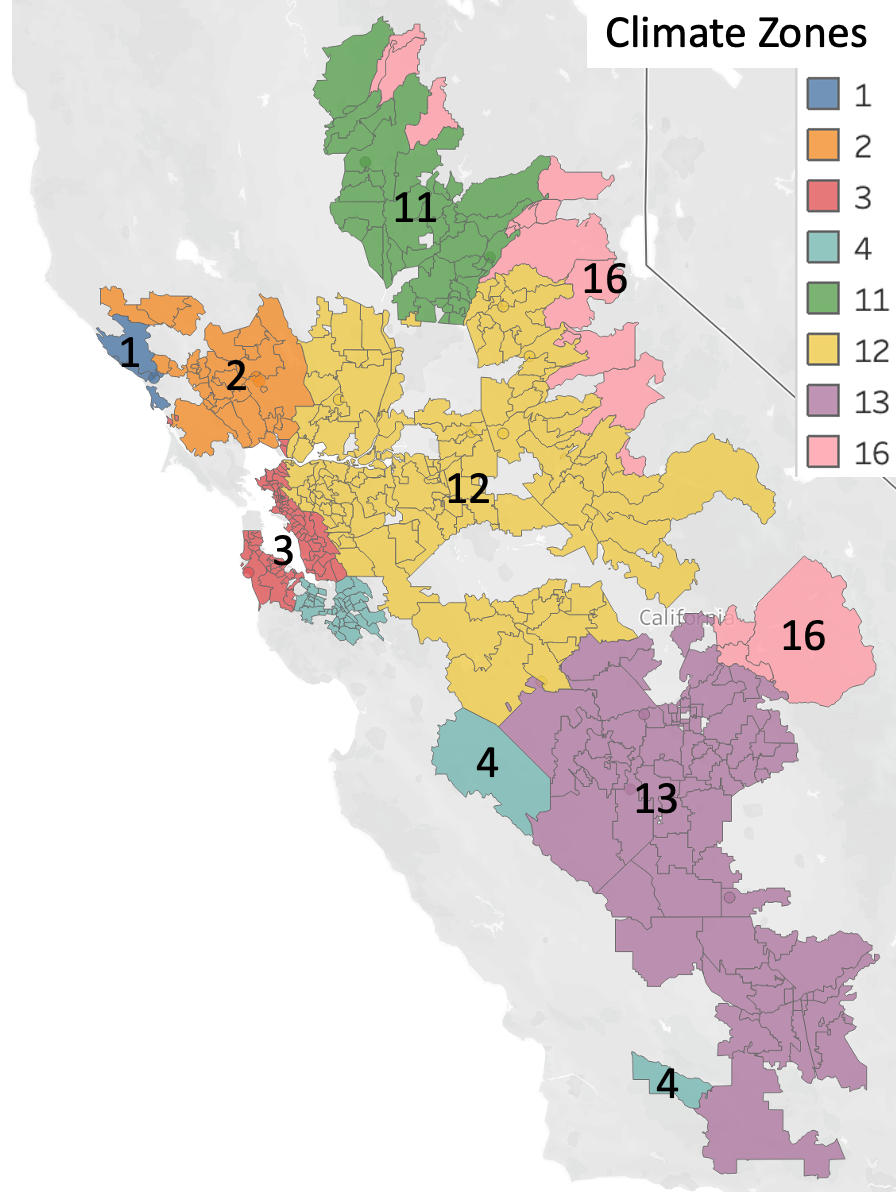}
    \caption{Households are located in eight climate zones in California, USA.}
    \label{lda:fig:homes:climatezones}
\end{figure}

Our sampled households covers 436 ZIP codes and eight different climate zones in California shown in \Figref{lda:fig:homes:climatezones}.

\subsection{Details of clustering load shapes}\label{lda:appx:sec:cluster:details:all}

\subsubsection{Evaluating different distances}
\label{lda:appx:sec:cluster:detail:distance}
To generate a robust and meaningful dictionary of load shapes, we compare several different distances such as cosine distance, $\normlone$ distance (Manhattan distance), $\normltwo$ distance (Euclidean distance), and dynamic time warping (DTW). For simplicity of explanations, we consider two vectors $\va$ and $\vb \in \R_{+}^m$ (e.g. $m=24$) in the following context.

\textbf{Cosine distance} is a measure of similarity between two non-zero vectors of an inner product space. The distance is expressed as 
\begin{align}
    d_{cos}(\va, \vb) = 1 - \frac{\va^T \vb}{\|\va\|_2 \|\vb\|_2 } \quad.
\end{align}

\textbf{\boldsymbol{$\normlone$} distance} is a measure of the element-wise absolute difference between two vectors. The expression is 
\begin{align}
    d_{\ell_1}(\va, \vb) = \|\va - \vb\|_1 = \sum_{k=1}^{m} \big|\va[k] - \vb[k]\big| ,
\end{align}
where $\va[k]$ is the $k$-th dimension in vector $\va$.

\textbf{\boldsymbol{$\normltwo$} distance} is a measure of element-wise squared gap between two vectors. The expression is 
\begin{align}
    d_{euc}(\va, \vb) = \|\va - \vb\|_2 = \sqrt{\sum_{k=1}^{m} \big(\va[k] - \vb[k]\big)^2} .
\end{align}

\textbf{Dynamic Time Warping distance} (DTW) is a method that calculates an optimal match between two given sequences~\cite{sakurai2005ftw}. We adopt a popular implementation that is based on dynamic programming:
\begin{align}
\begin{footnotesize}
    d_{DTW}(\va, \vb) = D(m, m), \text{ when }  D(i, j) = \min \begin{cases} 
    D(i-1, j) + \nu(i,j) \\
    D(i-1, j-1) + \nu(i, j) \\
    D(i, j-1) + \nu(i,j) 
    \end{cases},
    1 \leq i, j \leq m
\end{footnotesize}
\end{align}
where $D$ is a matrix that records the optimal warping value between the two vectors $\va$ and $\vb$; the $\nu(i,j)$ computes the cost between $\va[i]$ and $\vb[j]$ (the cost is Euclidean distance in this example); and the base case is $D(0, 0) = (\va[0] - \vb[0])^2$. 

\textbf{Hybrid distance}: we additionally apply a mixture of the $\normltwo$ and DTW distances to compute the distance between two vectors: 
\begin{align}
    d_{hybrid}(\va, \vb) = \gamma d_{euc}(\va, \vb) + (1-\gamma) d_{DTW}(\va, \vb),
\end{align}
where the $\gamma \in [0, 1]$ is the parameter to weigh the trade-off between two the distance metrics.

\subsubsection{Clustering methods}
\label{lda:appx:sec:cluster:detail:meth}
We apply several clustering methods including $k$-means, $k$-medians, hierarchical clustering, and DBSCAN for a thorough evaluation. 
For the center based methods (i.e., $k$-means and $k$-medians), we minimize the following objective (also known as distortion):
\begin{equation}
    \min \sum_{i=1}^N d\big(\vx_i, \phi(\vx_i, C)\big), 
\end{equation}
where $N$ is the sample size, $d$ is the distance metric, and $\phi(\vx_i, C)$ returns the nearest cluster center $c \in C$ to $\vx_i$. When $d$ is the Euclidean distance and $\phi$ finds the nearest center using Euclidean distance, we have 
\begin{align}
    \min \sum_{i=1}^N ||\vx_i - \phi(\vx_i, C)||_2^2 = \min \sum_{i=1}^N ||\vx_i - \mu_{c^{(i)}}||_2^2 \quad,
\end{align}
where $c^{(i)}$ is the cluster label for $i$-th data point. To express  the function $\phi$ more specifically, the $k$-means method updates the cluster centers by the following iterations until convergence: 
\begin{align}
    c^{(i)} = \argmin_{j} ||\vx_i - \mu_j||_2,\quad \mu_j = \frac{\sum_{i=1}^N \1\{c^{(i)} = j\}\vx_i }{\sum_{i=1}^N \1\{c^{(i)} = j\}} \quad.  \label{lda:appx:cluster:method:kmeans:eq2}
\end{align}
The $k$-medians method differs from the previous $k$-means clustering when calculating the cluster center. Instead of taking the mean $\mu_j$ in \eqrefp{lda:appx:cluster:method:kmeans:eq2}, we compute the median as the center $\tilde{\mu}_j$ so that 
\begin{align}
    \tilde{\mu}_j = median\big\{\vx_{i=\{1...N\}}\big\}, \text{if } c^{(i)}=j, \forall i \in {1 \dots N} \quad. 
\end{align} 

Hierarchical clustering is an agglomerative (hierarchical) approach, from the bottom individual point to up-level the whole dataset, that builds nested clusters in a successive manner \cite{johnson1967hierarchical,Jure2014MiningMass}. It has three popular implementations by minimizing different distances (objectives): Ward linkage~\cite{ward1963hierarchical}, average linkage~\cite{sokal1962comparison}, and complete linkage~\cite{defays1977efficient}. The Ward's linkage method measures the distance between two clusters, $A$ and $B$, which is how much the sum of squares will increase when we merge them:
\begin{align}
\begin{split}
    \Delta(A, B) = \sum_{i \in A \bigcup B} ||x_i - c_{A\cup B} ||^2 -  \sum_{i \in A} ||x_i - c_{A} ||^2 - \sum_{i \in B} ||x_i - c_{B} ||^2 = \frac{n_A n_B}{ n_A + n_B} ||c_A - c_B ||^2
\end{split}
\end{align}
where $c_A, c_B$ are the centers of clusters $A$ and $B$, and $n_A, n_B$ are the number of points in clusters $A$ and $B$. $\Delta$ denotes the merging cost of putting $A$ and $B$ together. The average linkage calculates the mean distance of all possible pairs of points in two clusters. The complete linkage method calculates the farthest distance of two points allocated in two clusters. 
We pick Ward linkage because it gives a more stable result compared with other two types of linkages. 

DBSCAN~\cite{ester1996density}, known as density-based spatial clustering of applications with noise, does not need to specify the number of clusters beforehand. It requires two key parameters, $\eps$ and $n_{\min}$, which define the neighborhood's distance and the minimum number of points to form a cluster. Higher $n_{\min}$ or lower $\eps$ indicate higher density to form a cluster. Choosing $\eps$ and $n_{\min}$ depends on domain knowledge of the data; hence, we evaluate multiple combinations and find it does not scale well for our use case. 

Both hierarchical and DBSCAN clustering do not compute cluster centers during iterations; therefore, we add an additional step to calculate a barycenter~\cite{petitjean2011global} of the points in each cluster to obtain a representative center. The barycenter is similar to the notion of a center in convex clusters, so we use the sequential averaging technique to compute the cluster center in the context of dealing with time series trajectories~\cite{Petitjean2014-ICDM-2}. 


\subsubsection{Evaluating clustering performances}\label{lda:appx:sec:eval:clustr:perform}
To compare multiple clustering methods with different distances, we mainly use two evaluation metrics: \emph{Calinski-Harabaz Index}~\cite{calinski1974dendrite} and \emph{Davies-Bouldin Index}~\cite{davies1979cluster}. Both metrics are widely adopted to evaluate clustering models. A higher \emph{Calinski-Harabasz Index} (CHI) relates to a model with better defined clusters, whereas a lower \emph{Davies-Bouldin Index} (DBI) is suggested for a model with a better separation between the clusters. To compare different clustering methods with various distances, we randomly draw 1000 data samples and record the cluster labels that yields the highest CHI and DBI scores when we search the number of clusters from $\{2, 4, 6, 8, 10, 12, 14, 16\}$. We repeat this exercise five times and present the results of the means of CHI and DBI in \tableref{lda:tab:appx:cluster:perform:compare}.    

\begin{table}[!hbpt]
\centering
\begin{footnotesize}
\begin{threeparttable}
    \centering
    \caption{Clustering method comparison. We report the means of both \emph{Calinksi-Harabaz Index} (CHI) and \emph{Davies-Bouldin Index} (DBI) after 5 rounds of random tests. A higher CHI indicates a model can yield a better separation of clusters. In contrast, a lower DBI suggests a better separation between the clusters. We see that $k$-medians with the hybrid distance  gives the best clustering performance. }
    \label{lda:tab:appx:cluster:perform:compare}
    \begin{tabular}{c|c|c|c}
    \toprule
         method & distance &  $CHI \uparrow$ & $DBI \downarrow$ \\
        \midrule
            & Euclidean & 107.42 & 4.53 \\
            & cosine & 102.31 & 3.87 \\
        $k$-means & $\ell_1$ & 99.51 & 4.14\\
            & DTW & 113.93 & 3.89 \\ 
            & $d_{hybrid}(\gamma=0.5)$ & 116.76 & 3.67 \\
        \midrule
             & Euclidean & 109.53 & 4.50 \\
             & cosine & 108.11 & 4.05 \\
        \textbf{$k$-median} & $\ell_1$ & 102.40 & 4.19\\
             & DTW & 115.84 & 3.82 \\
             & \boldsymbol{$d_{hybrid}(\gamma=0.5)$} & \textbf{118.31} & \textbf{3.54}\\
        \midrule
             & Euclidean & 93.21 & 4.99\\
             & cosine & 92.18 & 4.81 \\
        Hierarchical (Ward) & $\ell_1$ & 90.53 & 5.16 \\
             & DTW & 98.65 & 4.87\\
             & $d_{hybrid}(\gamma=0.5)$ & 101.32 & 4.58\\
        \midrule
             & Euclidean & 82.44 & 5.17 \\
             & cosine & 85.37 & 5.29 \\
        DBSCAN ($\eps=0.1$) & $\ell_1$ & 80.15 & 5.18 \\
             & DTW & 88.03 & 5.25\\
             & $d_{hybrid}(\gamma=0.5)$ & 89.75 & 5.07 \\
        \bottomrule
    \end{tabular}
\end{threeparttable}
\end{footnotesize}
\end{table}

\begin{figure*}[!hbpt]
    \centering
    \includegraphics[width=0.72\textwidth]{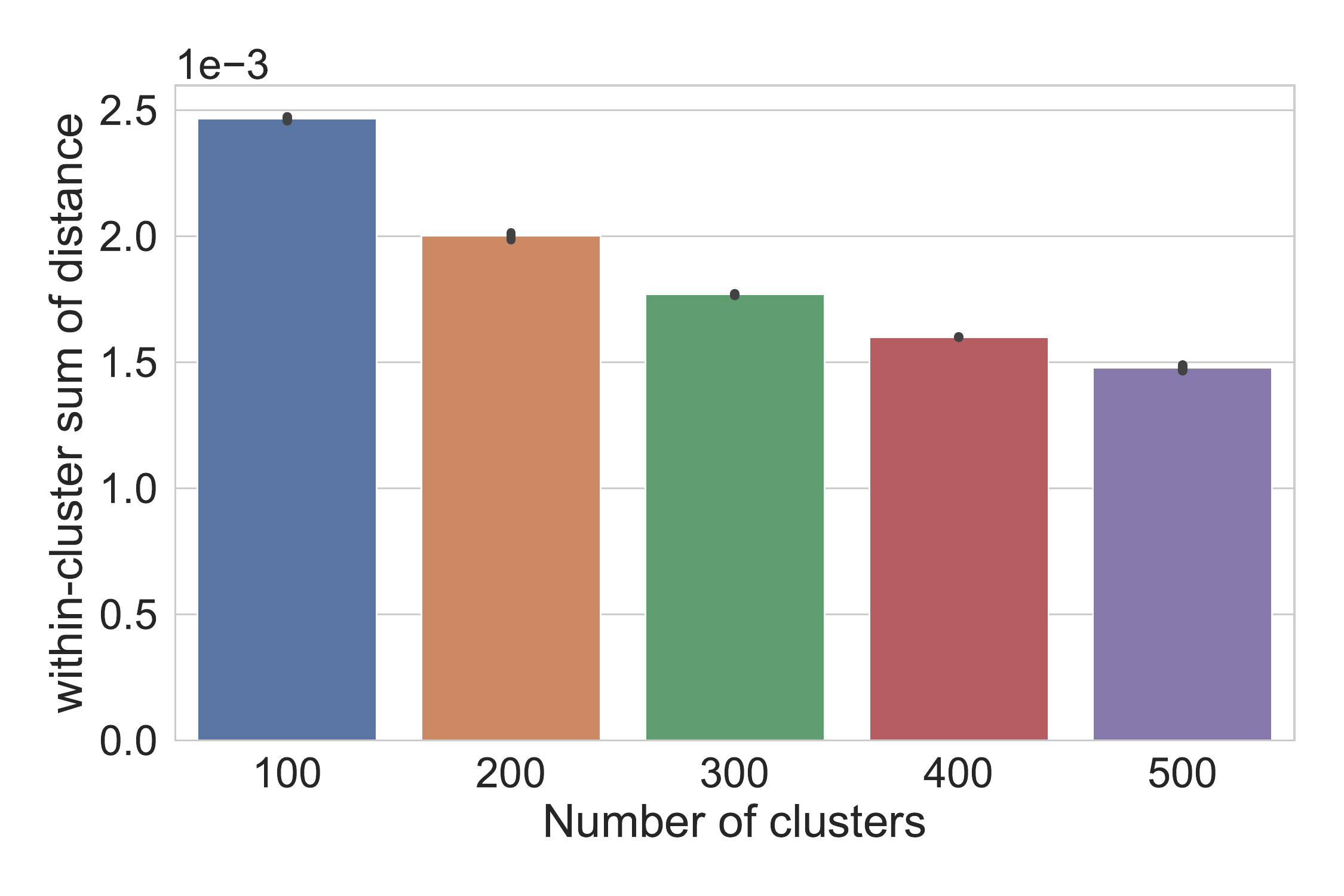}
    \caption{Choosing the size of the load shape dictionary. When increasing the number of clusters above 200, we have limited marginal gain of reducing the within cluster sum of distances. Thus, we choose 200 as an appropriate dictionary size.}
    \label{lda:fig:appx:dict:size:compare}
\end{figure*}

{\subsubsection{Determining the dictionary size}\label{lda:appx:sec:cluster:detail:dic:size}}
Once the $k$-median method with the $d_{hybrid}$ is chosen, we explore the appropriate size of the load shape dictionary. In particular, we tested the size of 100, 200, 300, 400, and 500 load shapes. Such a comparison involves two stages of clustering processes: 1) we randomly partition 60,000 homes into 600 bins where each bin has 100 homes, and then we run clustering on these 100 $\times$ 365 data points for each bin to create 100 cluster centers. 2) Having these 100 clustered load shapes times the 600 bins, we run another round of clustering on $100 \times 600$ data points to yield the cluster centers with the size ranging from 100 to 500. \Figref{lda:fig:appx:dict:size:compare} suggests that a size of 200 reduces the within-cluster distortion dramatically around 20\%, which is much more prominent than at other sizes. Thus, we pick 200 clusters as the size of the load shape dictionary.       

%
\clearpage
\subsection{Generating distinct attributes}
\label{lda:sec:appx:gen:ene:attr}
\begin{figure*}[!hpbt]
    \centering
    \includegraphics[width=0.9\textwidth]{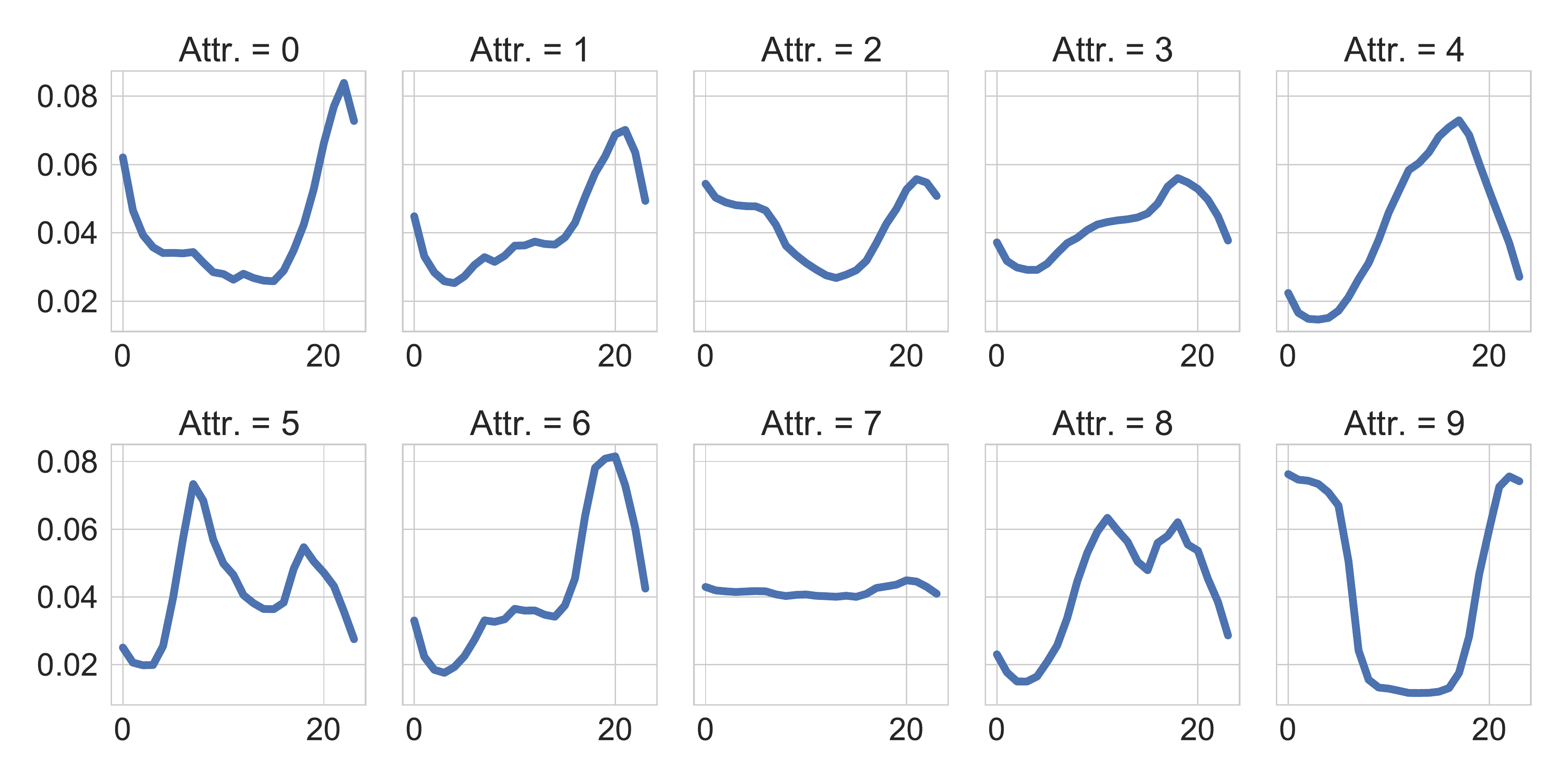}
    \caption{Ten attributes are obtained after applying LDA initially. A few center curves are similar, such as Attribute 1 and Attribute 6. We then construct a projection matrix according to correlation distance to reduce the number of attributes (a.k.a, number of topics) down to six.}
    \label{lda:fig:appx:init:topic10}
\end{figure*}

\begin{figure*}[!hbpt]
    \centering
    \includegraphics[width=0.75\textwidth]{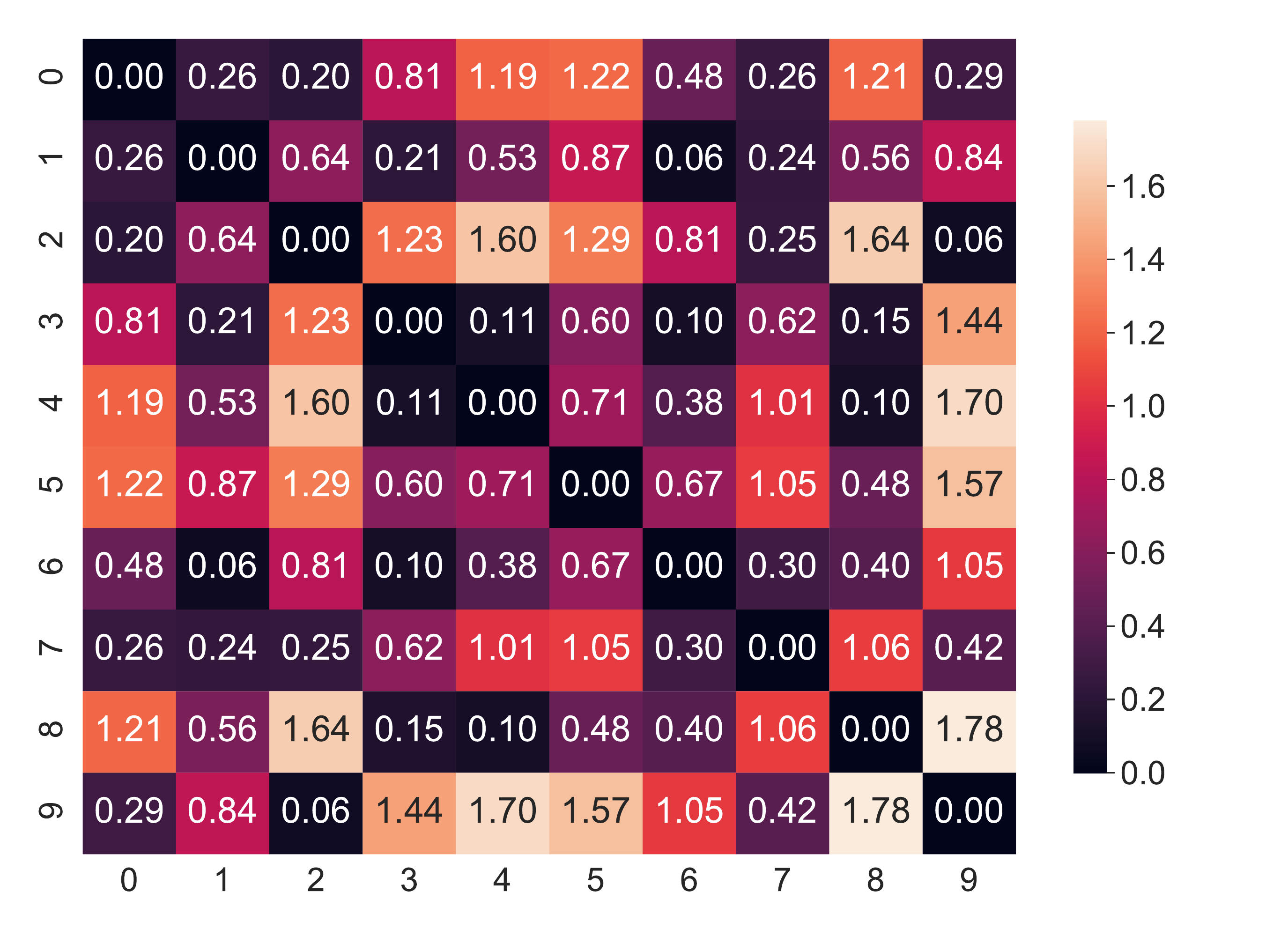}
    \caption{Correlation distance heatmap of ten attributes that are obtained from \Figref{lda:fig:appx:init:topic10}}
    \label{lda:fig:appx:topic10:corr:dist}
\end{figure*}

\begin{figure*}[!hbpt]
    \centering
    \includegraphics[width=0.90\textwidth]{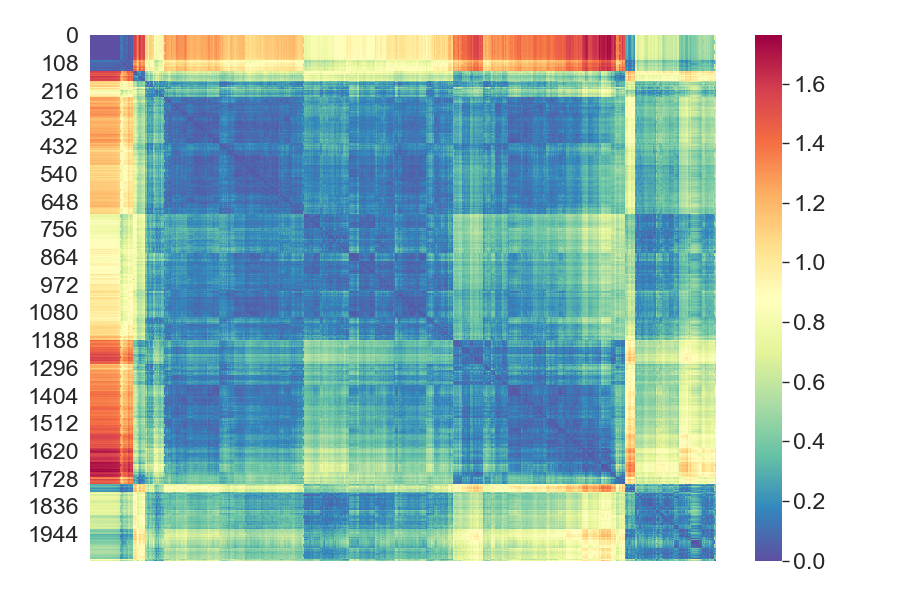}
    \caption{Distance heatmap. 2000 homes are randomly sampled and their pairwise correlation distances appear to be segmented into six main blocks along the diagonal.}
    \label{lda:fig:appx:example:homes:distances}
\end{figure*}

Before synthesizing the energy lifestyles of households, we need to find the representative attributes that compose the multiple load patterns for households. Thus, teasing out distinct latent attributes of energy usage is a crucial building block. We apply LDA with a prescribed $K=10$ number of attributes (topics), displayed in \Figref{lda:fig:appx:init:topic10}. After fitting the LDA model, we find several attributes are very similar to each other such as \emph{attribute 1} and \emph{attribute 6} in \Figref{lda:fig:appx:init:topic10}. 
A further calculation of the correlation distances between attributes (normalized 24-dimensional vectors) also demonstrates that some attributes are very close and can be merged together (\Figref{lda:fig:appx:topic10:corr:dist}), where the correlation distances between two vectors $\va$ and $\vb$ with their associated elements means $\mu_{\va}$ and $\mu_{\vb}$ can be expressed as 
\begin{align}
    d_{corr} = 1 - \frac{(\va-\mu_{\va})(\vb-\mu_{\vb})}{\|\va-\mu_{\va}\|_2 \|\vb-\mu_{\vb}\|_2} . 
    \label{lda:eq:appx:corr:distance}
\end{align}
We set the threshold as 0.1 to indicate that two attributes are very similar, and then find the nearest neighbors of the energy attributes based on that criterion. Once the neighbors are settled, we merge similar shapes together by 1) constructing a projection matrix $A^TA$ where $A = D_{corr} + I$ and where $D_{corr}$ consists of either zeros or ones, where ones mean when $d_{corr}$ is less than 0.1 in entries, mentioned in \eqrefp{lda:eq:appx:corr:distance} and $I$ is the identity matrix; 2) scanning through columns and pruning the $A^TA$ once the corresponding rows are located. In our experiment, we prune down to six dimensions, because each dimension has its distinct attribute shape (\Figref{lda:fig:6topicshapes}). Additionally, we qualitatively verify that six attributes are robust for a large population by randomly sampling 2000 homes and comparing their correlation distances on the attribute spaces prior to projection (\Figref{lda:fig:appx:example:homes:distances}). We observe that homes are nested mainly into 5 to 6 diagonal blocks, which supports our previous merge operation of simplifying the energy attributes.   

\clearpage

\subsection{Determining the number of lifestyles}\label{lda:appx:sec:n:ls}
\begin{figure}[!hbpt]
    \centering
    \includegraphics[width=0.5\textwidth]{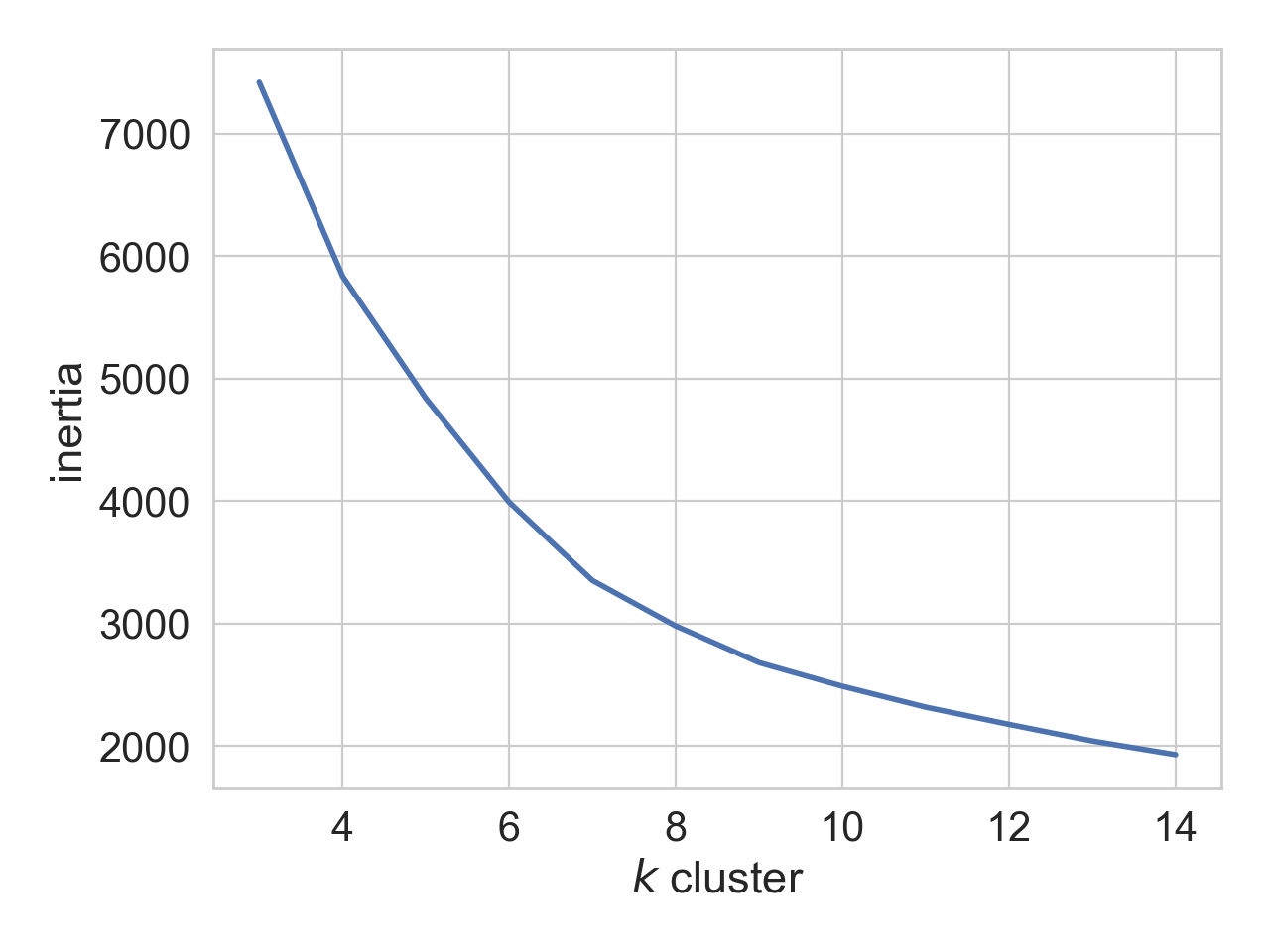}
    \caption{Inertia vs. number of clusters. Inertia is the sum of squared distances of samples to their closest cluster center.}
    \label{lda:fig:appx:ls:cluster:n:elbo}
\end{figure}

\begin{figure*}[!hbpt]
    \centering
    \includegraphics[width=1.09\textwidth]{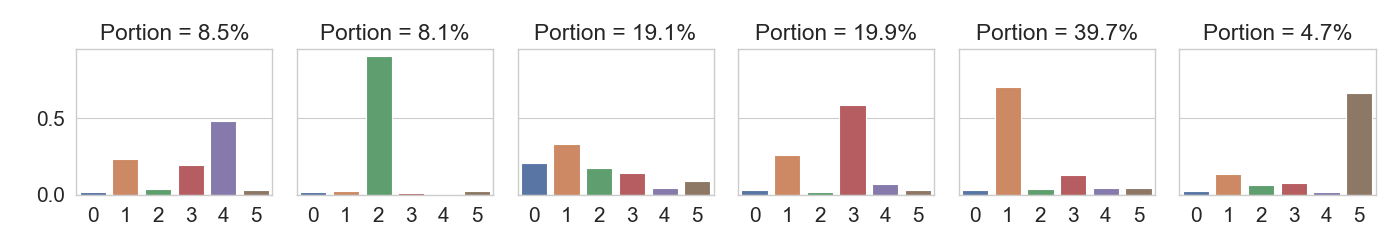}
    \caption{Weights of energy attributes}
    \label{lda:fig:appx:ls_mixWeights_6topics}
\end{figure*}

Having determined the energy attributes, we use a six-dimensional vector to represent each home. In order to obtain prototypical attributes distribution of these homes, we need to segment all the homes using another round of clustering. We use $k$-means with $K=6$, because this setting gives a distinguishable and meaningful result, according to the inertia heuristic (the sum of squared distances of samples to their closest cluster center).
The elbow plot of the inertia is displayed in \Figref{lda:fig:appx:ls:cluster:n:elbo}. 
The corresponding centers of the attribute weights are shown in \Figref{lda:fig:appx:ls_mixWeights_6topics}. 

\clearpage
\subsection{Population change over seasons}\label{lda:appx:sec:pop:change}
We provide detailed population splits of six lifestyles over four seasons in \tableref{lda:tab:appx:population:split:ls:seasons}. The numbers are the counts, and percentage values in the parentheses are proportions of the population. From the table, we find that the lifestyle of Home for dinner is the most frequently occurring, usually accounting for about 40\% of the households except in the autumn when it accounts for 33.02\% of the samples. 
\begin{table}[!hbpt]
    \centering
    \begin{footnotesize}
    \begin{threeparttable}
    \caption{Population splits of lifestyles in seasons}
    \label{lda:tab:appx:population:split:ls:seasons}
    \begin{tabular}{c|c|c|c|c}
    \toprule
         &  \textbf{Autumn} N (\%) & \textbf{Winter} N (\%) & \textbf{Spring} N (\%) & \textbf{Summer} N (\%) \\
        \midrule
        Active morning & 4027 (6.71\%) & 7836 (13.06\%) & 4713 (7.85\%) & 3968 (6.61\%) \\
        Night owl & 5174 (8.62\%) & 4844 (8.07\%) & 5504 (9.17\%) & 4557 (7.60\%) \\
        Everyday is a new day & 8632 (14.39\%) & 5509 (9.18\%) & 10750 (17.92\%) & 21311 (35.52\%) \\
        Home early &  18963 (31.61\%) & 11975 (19.96\%) & 10254 (17.09\%) & 4420 (7.37\%) \\
        Home for dinner & 19813 (33.02\%) & 26084 (43.47\%) & 24458 (40.76\%) & 25744 (42.91\%) \\
        Steady going & 3391 (5.65\%) & 3752 (6.25\%)& 4221 (7.04\%) & 0 $(0\%)$\tnote{$\dagger$} \\
        \bottomrule
    \end{tabular}
    \begin{tablenotes}
    \item[$\dagger$] We do not observe that households in our samples have a flat pattern of energy use (i.e., steady going lifestyle) across many days in the summer. 
    \end{tablenotes}
    \end{threeparttable}
    \end{footnotesize}
\end{table}

\subsection{Features of energy usage}
\label{lda:sec:appx:feat:details}
We show distributions of additional features associated with different lifestyles. The definitions of features are provided in \tableref{lda:tab:appx:features:def}.

\begin{figure}[!htpb]
    \centering
    \begin{subfigure}[t]{0.49\textwidth}
    \includegraphics[width=0.99\columnwidth]{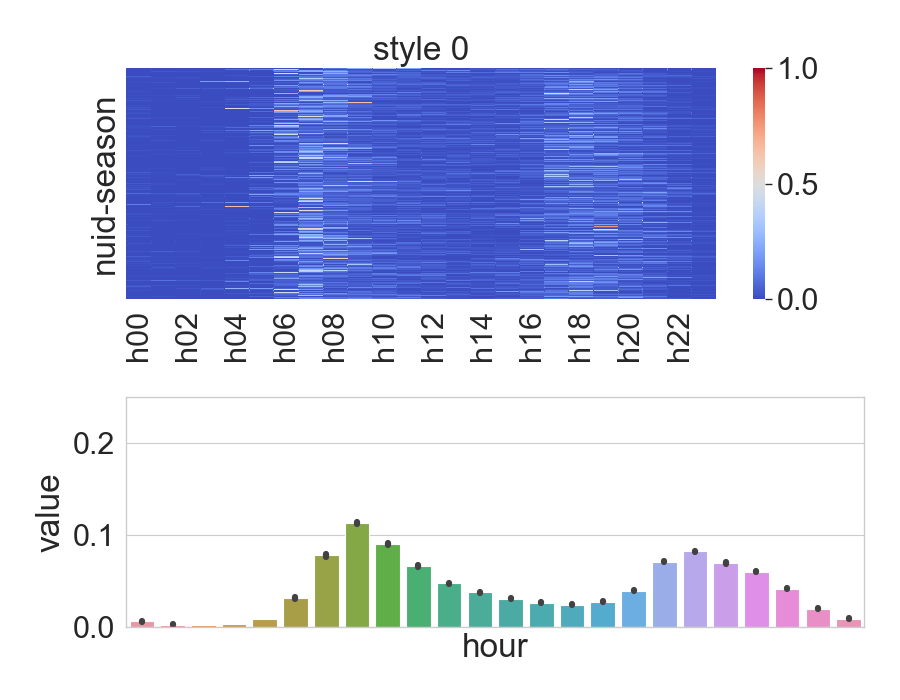}
    \caption{active morning}
    \label{lda:fig:pk_hr:s0}
    \end{subfigure}
    \hfill
    \begin{subfigure}[t]{0.49\textwidth}
    \includegraphics[width=0.99\columnwidth]{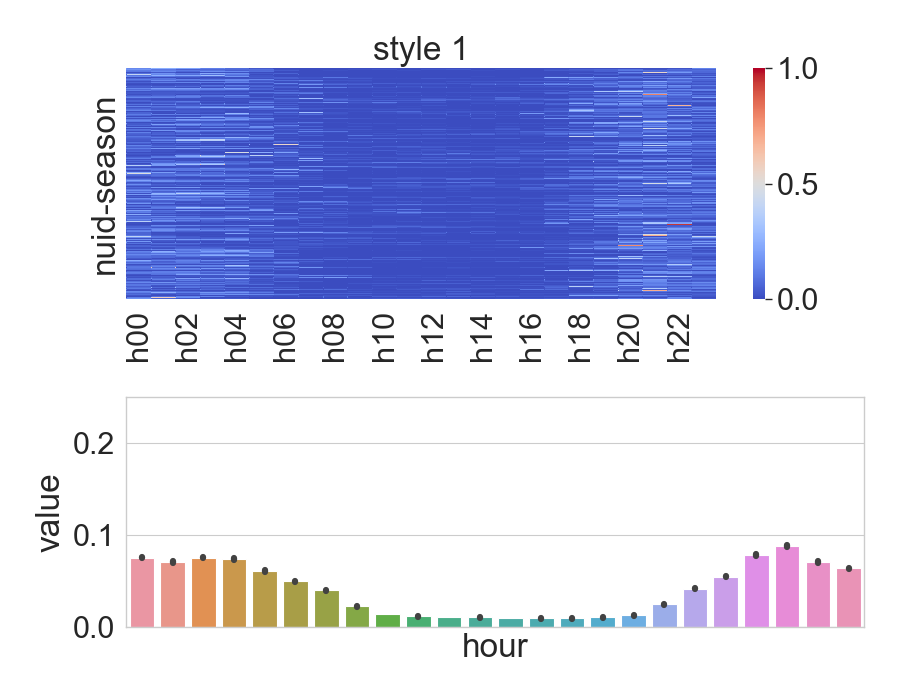}
    \caption{night owl}
    \label{lda:fig:pk_hr:s1}
    \end{subfigure}
    \begin{subfigure}[t]{0.49\textwidth}
    \includegraphics[width=0.99\columnwidth]{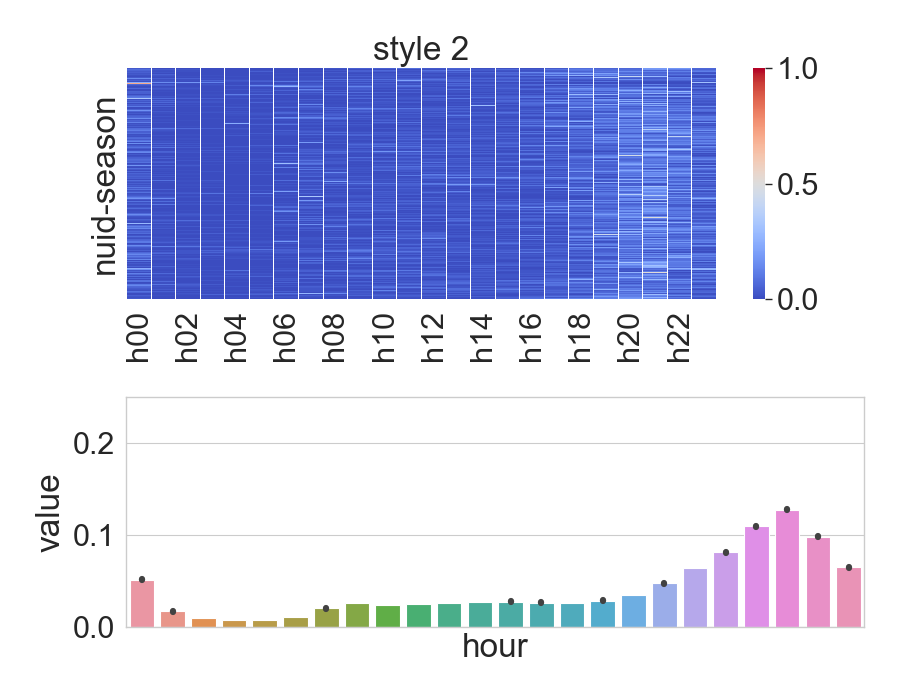}
    \caption{everyday is a new day}
    \label{lda:fig:pk_hr:s2}
    \end{subfigure}
    \hfill
    \begin{subfigure}[t]{0.49\textwidth}
    \includegraphics[width=0.98\textwidth]{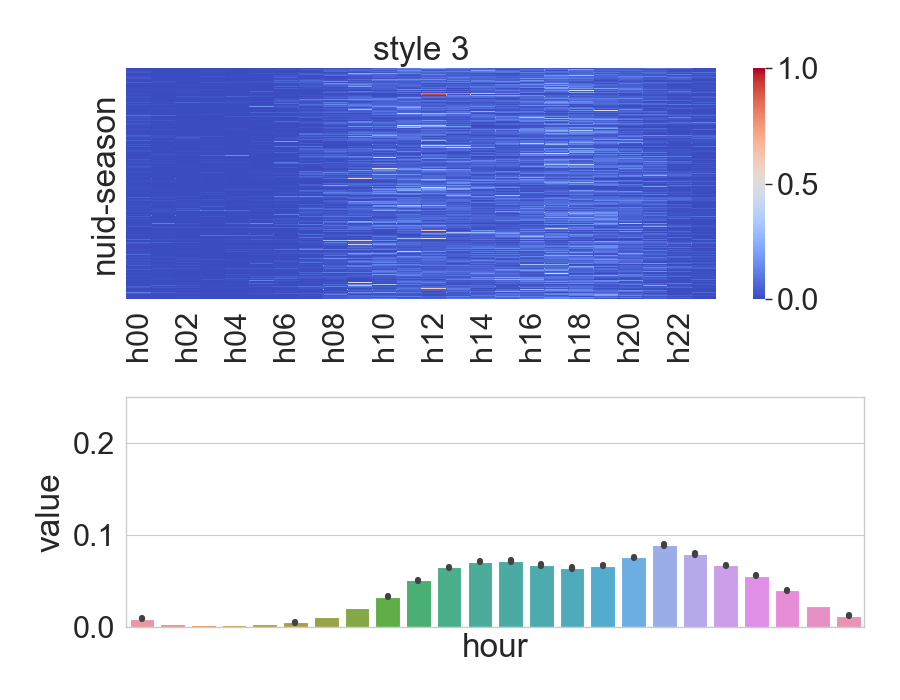}
    \caption{Home early lifestyle.}
    \label{lda:fig:appx:pk_hr:s3}
    \end{subfigure}
    \begin{subfigure}[t]{0.49\textwidth}
    \includegraphics[width=0.99\columnwidth]{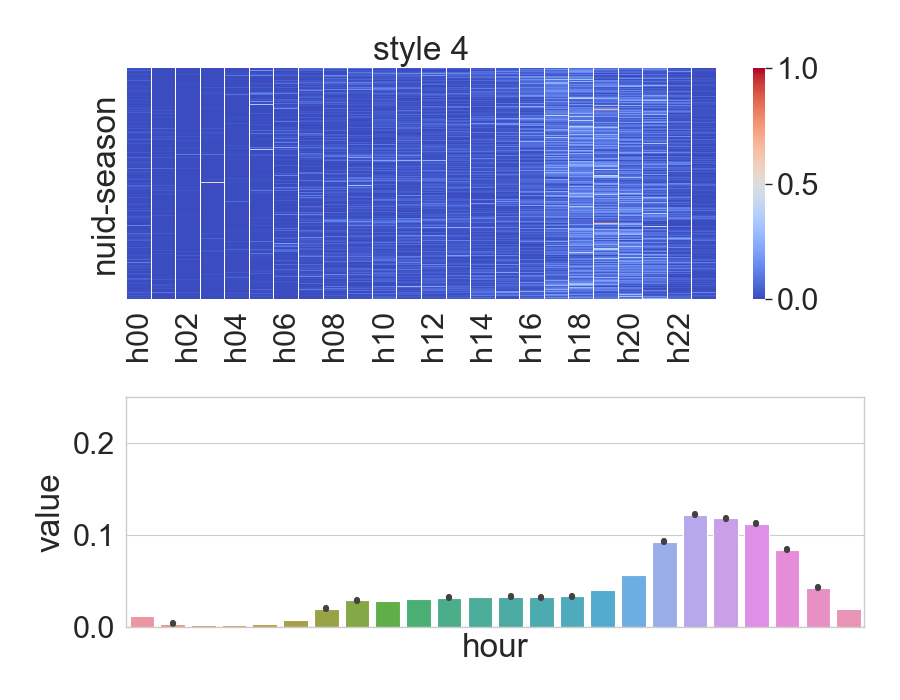}
    \caption{home for dinner}
    \label{lda:fig:pk_hr:s4}
    \end{subfigure}
    \caption{Peak hour distribution over a day (hour 0 -- hour 23). In each sub-figure, the upper panel shows the heatmap of peak hour frequency when each home in a season is represented by each row stacked by seasons. The lower panel is the averaged frequency of peak hour occurrence for all homes in the corresponding lifestyle group.}
    \label{lda:fig:pk_hr:dist:ex:sec4}
\end{figure}

\begin{figure}[!hpbt]
    \centering
    \begin{subfigure}[t]{0.49\textwidth}
    \includegraphics[width=0.99\columnwidth]{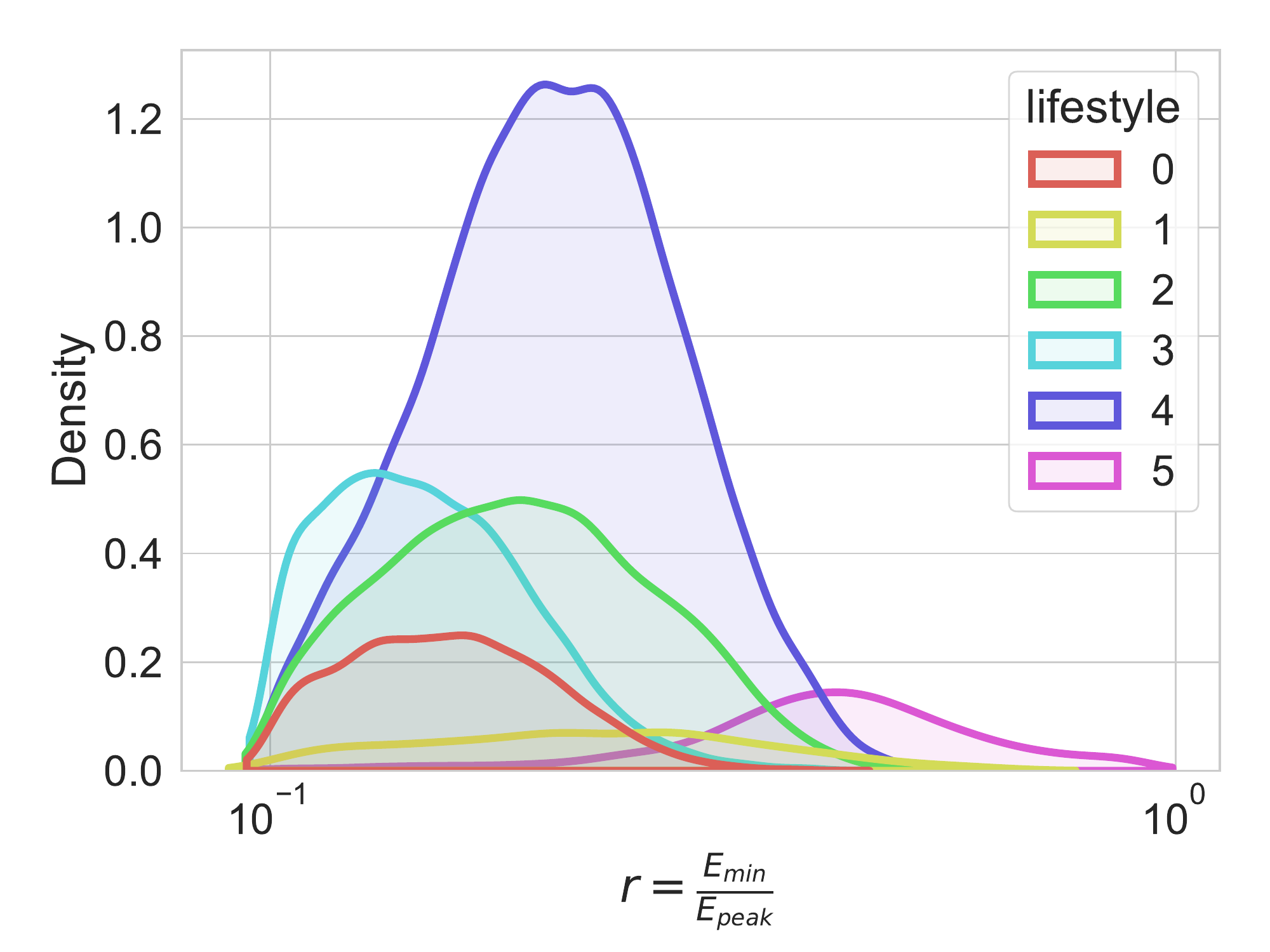}
    \caption{Ratio of hourly min- over max-energy in 24 hours}
    \label{lda:fig:appx:bp_r}
    \end{subfigure}
    \hfill
    \begin{subfigure}[t]{0.49\textwidth}
    \includegraphics[width=0.99\columnwidth]{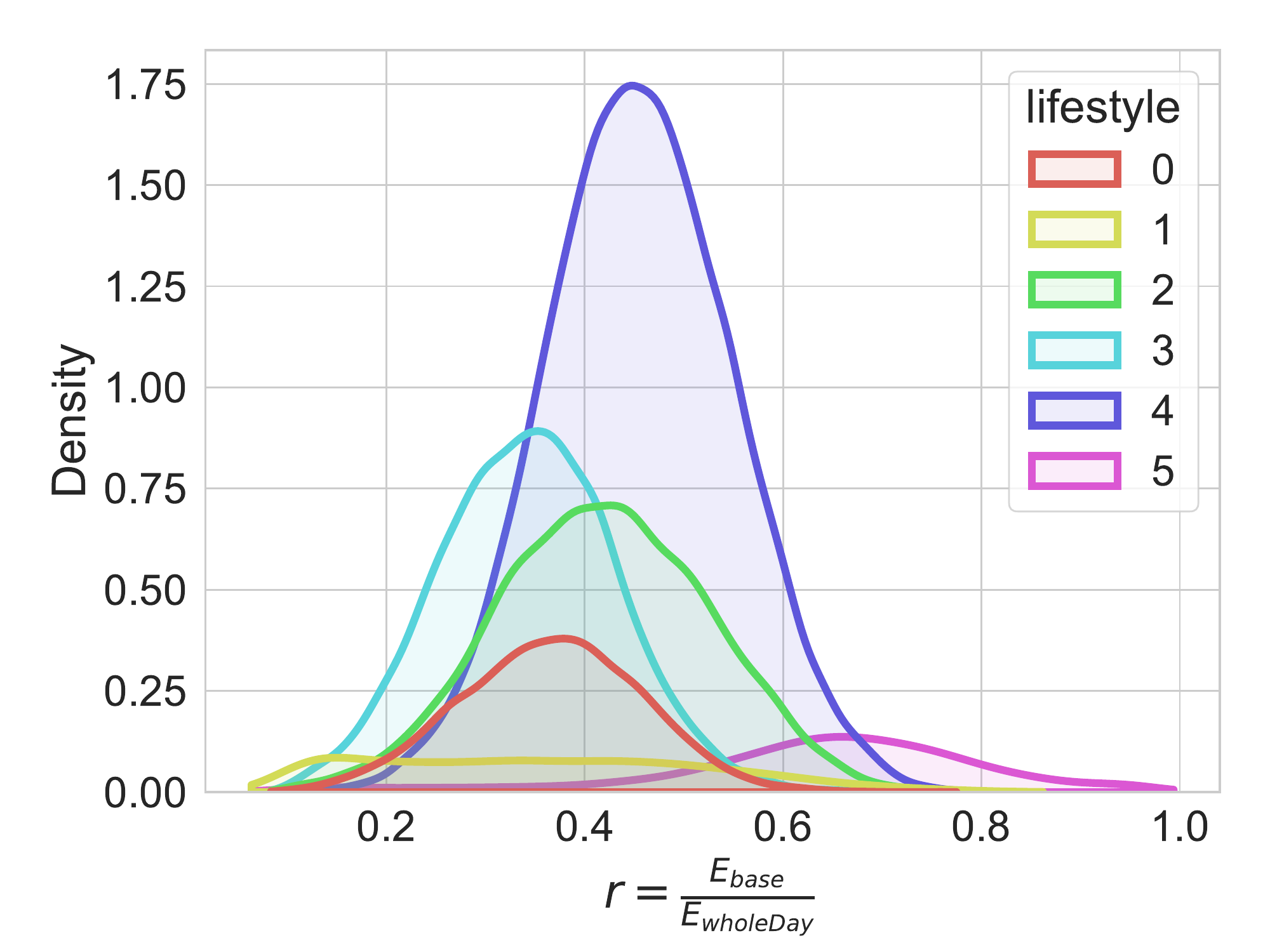}
    \caption{Ratio of base over whole day energy use in 24 hours}
    \label{lda:fig:appx:base_p}
    \end{subfigure}
    \hfill
    \begin{subfigure}[t]{0.49\textwidth}
    \includegraphics[width=0.99\columnwidth]{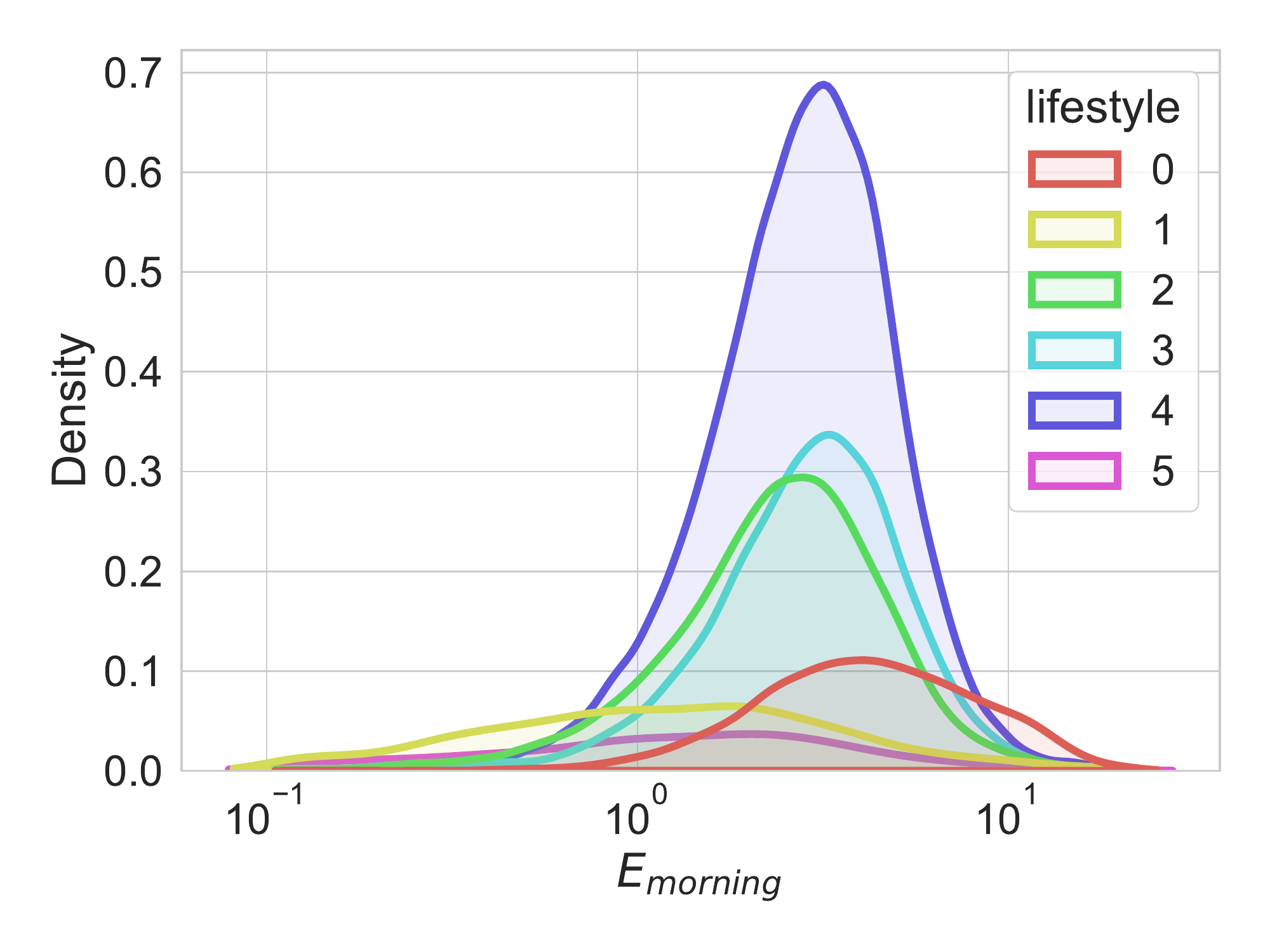}
    \caption{Morning energy use (6am - 10am)}
    \label{lda:fig:appx:c_morning}
    \end{subfigure}
    \hfill
    \begin{subfigure}[t]{0.49\textwidth}
    \includegraphics[width=0.99\columnwidth]{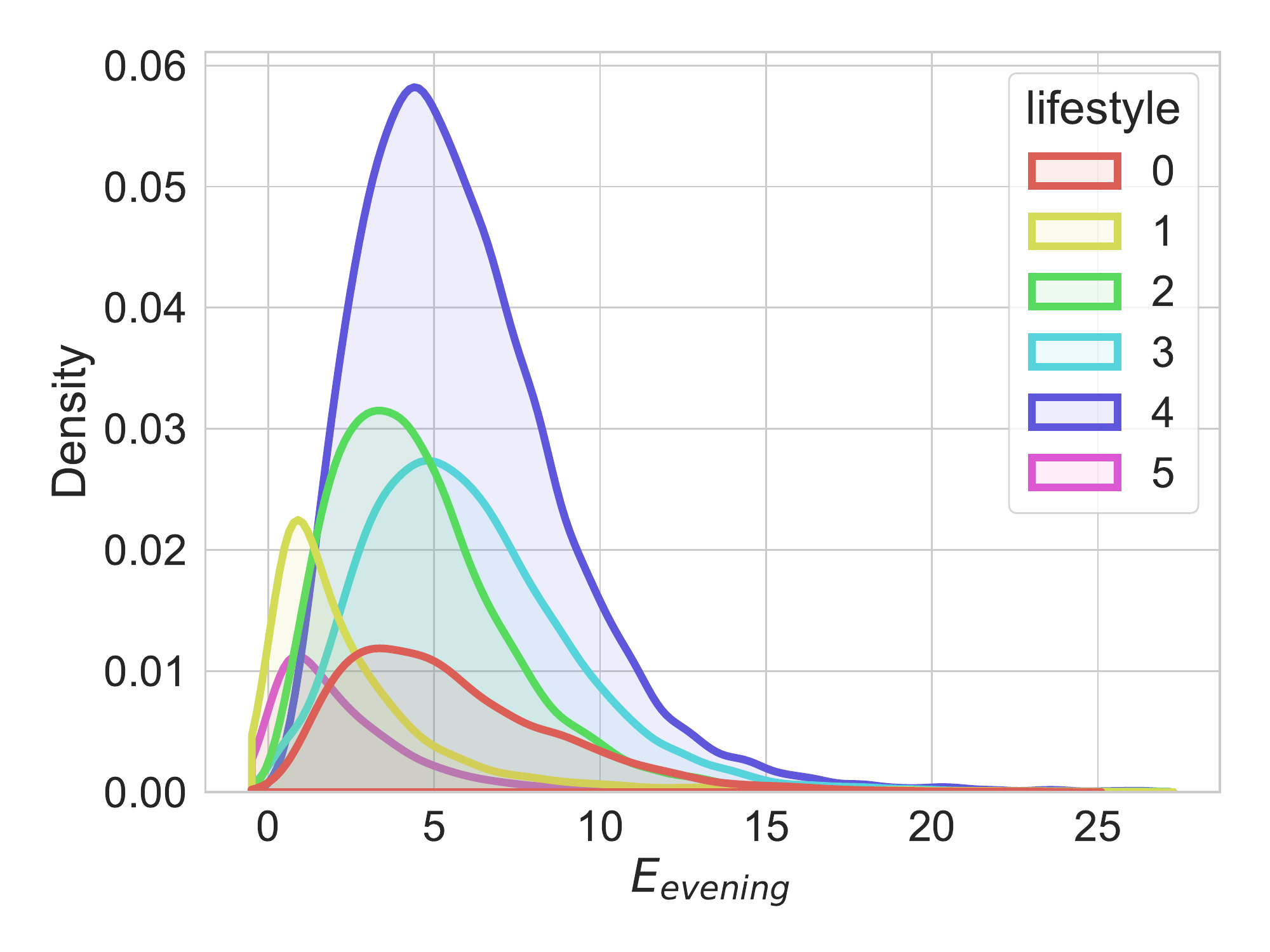}
    \caption{Evening energy use (6pm-10pm)}
    \label{lda:fig:appx:c_eve}
    \end{subfigure}
    \hfill
    \begin{subfigure}[t]{0.49\textwidth}
    \includegraphics[width=0.99\columnwidth]{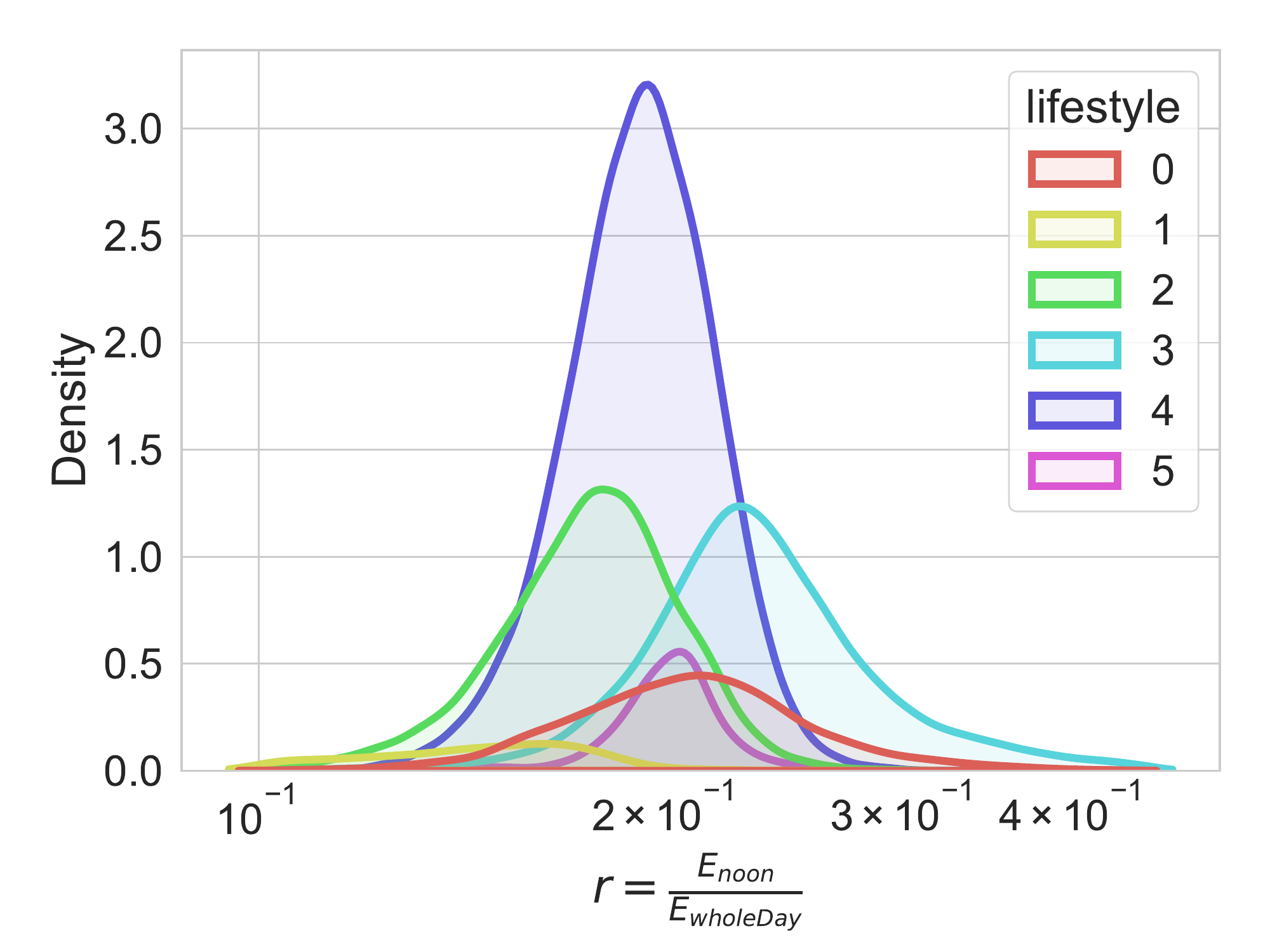}
    \caption{Ratio of noon over whole day energy use}
    \label{lda:fig:appx:r_no2w}
    \end{subfigure}
    \hfill
    \begin{subfigure}[t]{0.49\textwidth}
    \includegraphics[width=0.99\columnwidth]{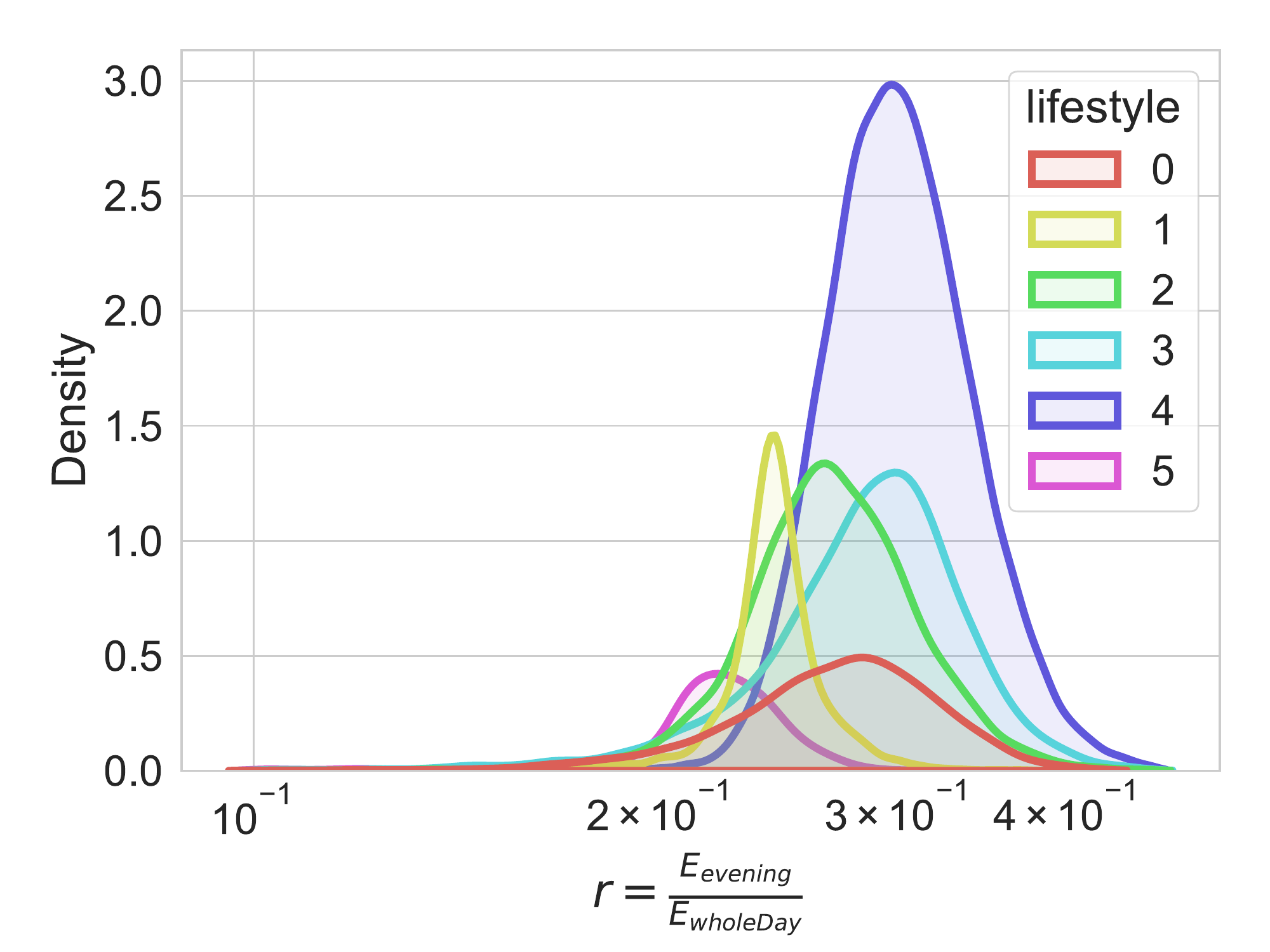}
    \caption{Ratio of evening over whole day energy use}
    \label{lda:fig:appx:r_e2w}
    \end{subfigure}
    \caption{Distributions of different energy usage features characterizing the distinctions between lifestyles.}
    \label{lda:fig:appx:ls:feat:annual}
\end{figure}

First, we provide the peak hour distribution for the Home early lifestyle in addition to the other lifestyles mentioned in \Figref{lda:fig:pk_hr:dist:ex:sec4}. Second, multiple year-specific features are displayed in \Figref{lda:fig:appx:ls:feat:annual} over six lifestyles.

\clearpage

\subsection{Classification details}\label{lda:appx:sec:clf:detail:head}
\subsubsection{Identifying lifestyles}\label{lda:appx:sec:clf:ls:head}

We provide the performance details of classifying lifestyles using Random Forest (RF) fed with normalized load features. With 70\%/10\%/20\% train/validation/test split of the data, the RF model is calibrated with 25 estimators and other default settings from Scikit-learn\footnote{\url{https://scikit-learn.org/stable/modules/generated/sklearn.ensemble.RandomForestClassifier.html}}.

\begin{figure}[!hbpt]
    \centering
    \includegraphics[width=0.99\textwidth]{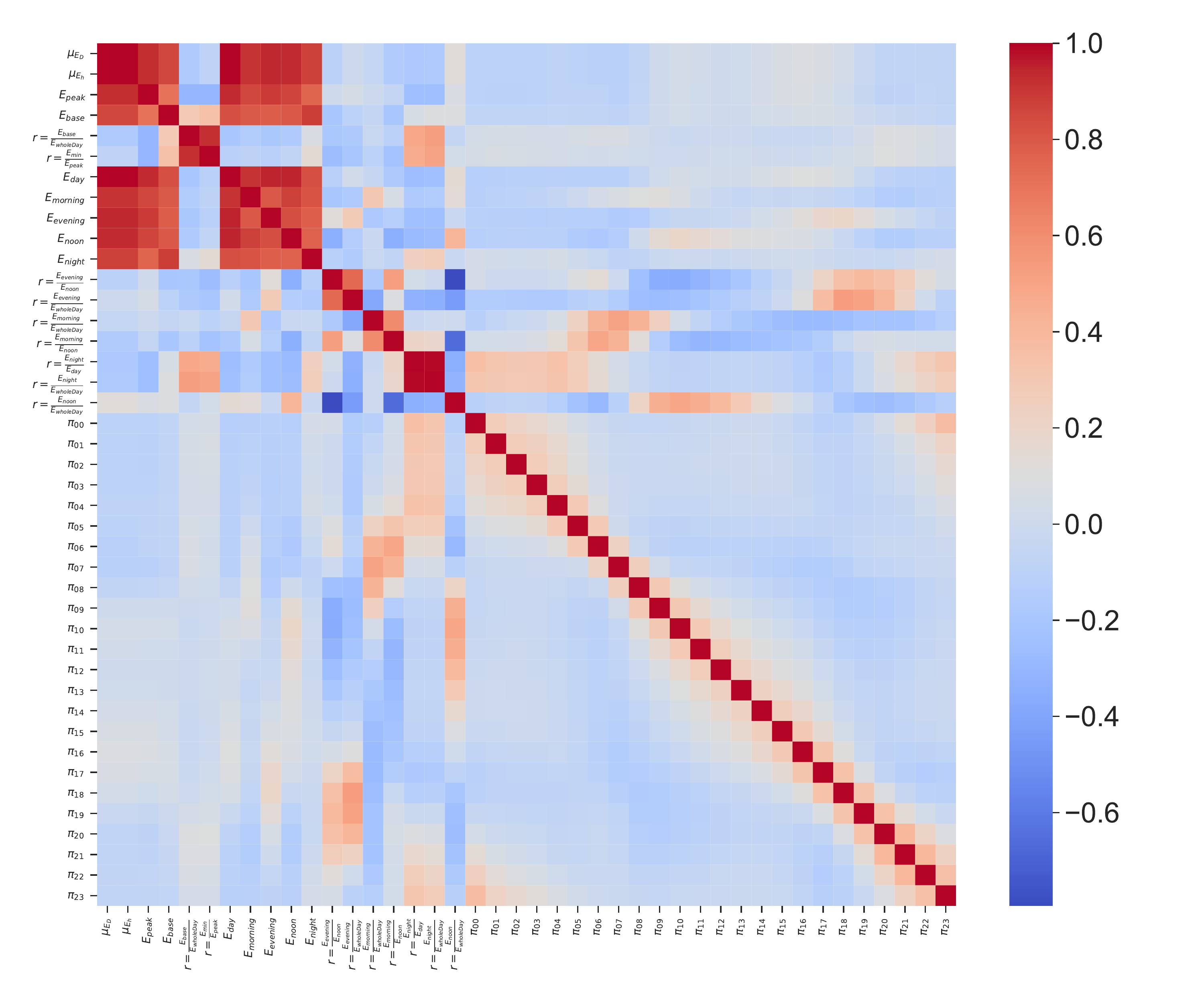}
    \caption{Correlation heatmap of features.}
    \label{lda:fig:chan_nochan:feat:corr:heatmap}
\end{figure}

As shown in \tableref{lda:tab:appx:ls:clf:metrics}, Night owl has the highest F1 score around 0.84. In contrast, the Home early lifestyle has the lowest F1 score about 0.55, indicating this is a difficult lifestyle to identify. 
The feature correlation is displayed in \Figref{lda:fig:chan_nochan:feat:corr:heatmap}.

\begin{table}[!hbpt]
    \caption{Lifestyle classification performance}
    \label{lda:tab:appx:ls:clf:metrics}
    \centering
    \begin{tabular}{r|l|c|c|c}
    \toprule
        style index & lifestyle & precision & recall & F1 score \\
        \midrule
        0 & active morning  & 0.7164 & 0.6315 & 0.6713  \\
        1 & night owl & 0.8657 & 0.8176 & 0.8409 \\
        2 & everyday is a new day & 0.6411 & 0.6191 & 0.6299 \\
        3 & home early & 0.6551 & 0.4685 & 0.5463 \\
        4 & home for dinner & 0.6652 & 0.7841 & 0.7198\\
        5 & steady going & 0.6417 & 0.5493 & 0.5919 \\
        \midrule 
        \multicolumn{5}{r}{average acc = 0.685} \\
        \bottomrule
    \end{tabular}
\end{table}

\clearpage

\subsubsection{Identifying No Changer}\label{lda:appx:sec:clf:nochanger}
\begin{figure}[!hpbt]
    \centering
    \includegraphics[width=0.99\textwidth]{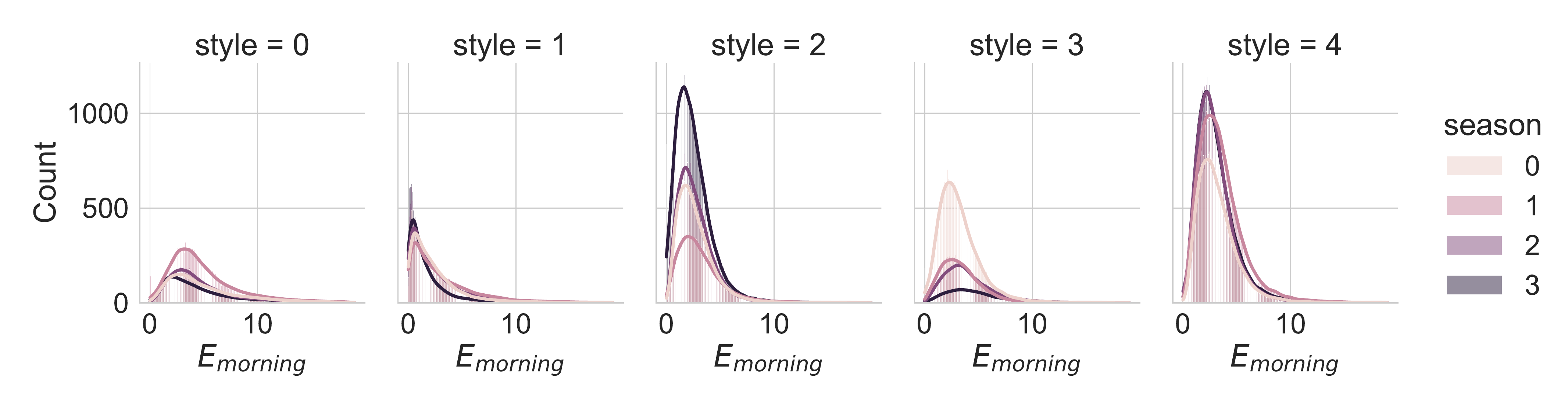}
    \caption{Distribution of morning energy use (in KWh) over four seasons for lifestyles}
    \label{lda:fig:appx:feat:season:c_morning}
\end{figure}

\begin{figure}[!hpbt]
    \centering
    \includegraphics[width=0.99\textwidth]{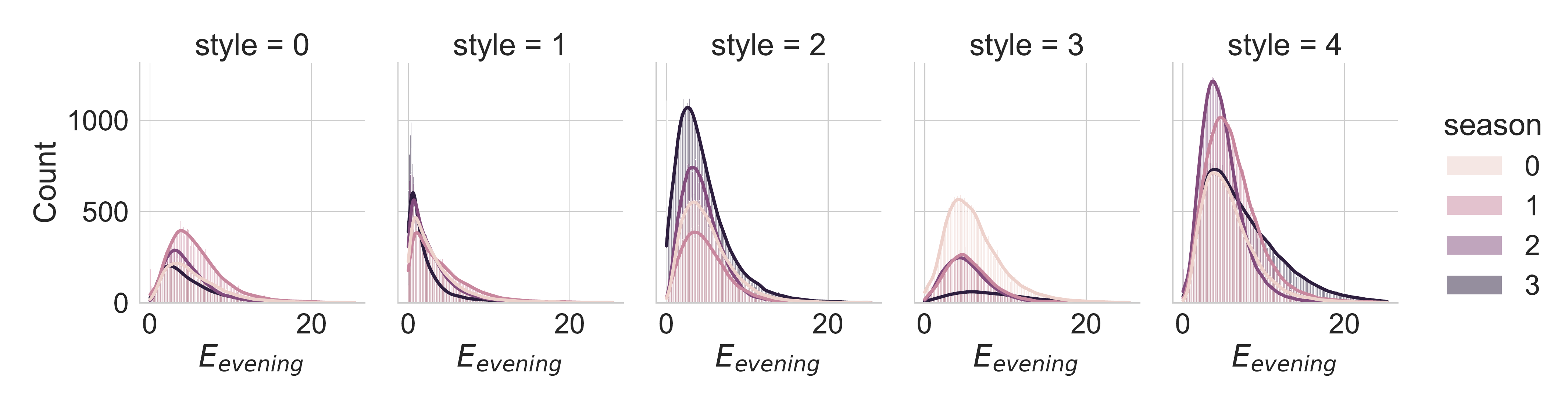}
    \caption{Distribution of evening energy use (in KWh) over four seasons for lifestyles}
    \label{lda:fig:appx:feat:season:c_eve}
\end{figure}

\begin{figure}[!hpbt]
    \centering
    \includegraphics[width=0.99\textwidth]{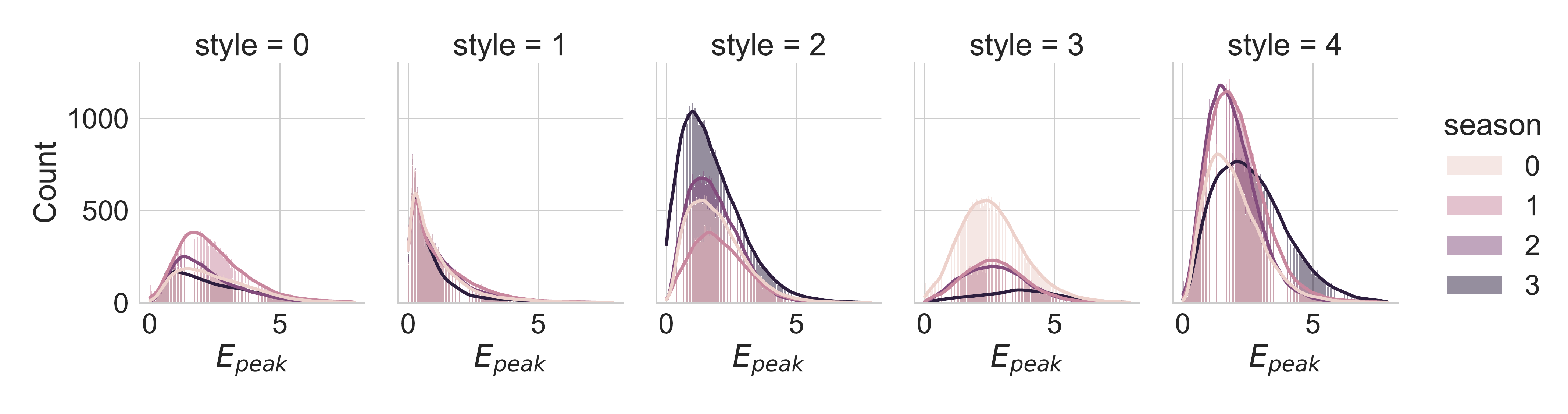}
    \caption{Distribution of daily peak energy (in KWh) over four seasons for lifestyles}
    \label{lda:fig:appx:feat:season:pk_ene}
\end{figure}

\begin{figure}[!hpbt]
    \centering
    \includegraphics[width=0.99\textwidth]{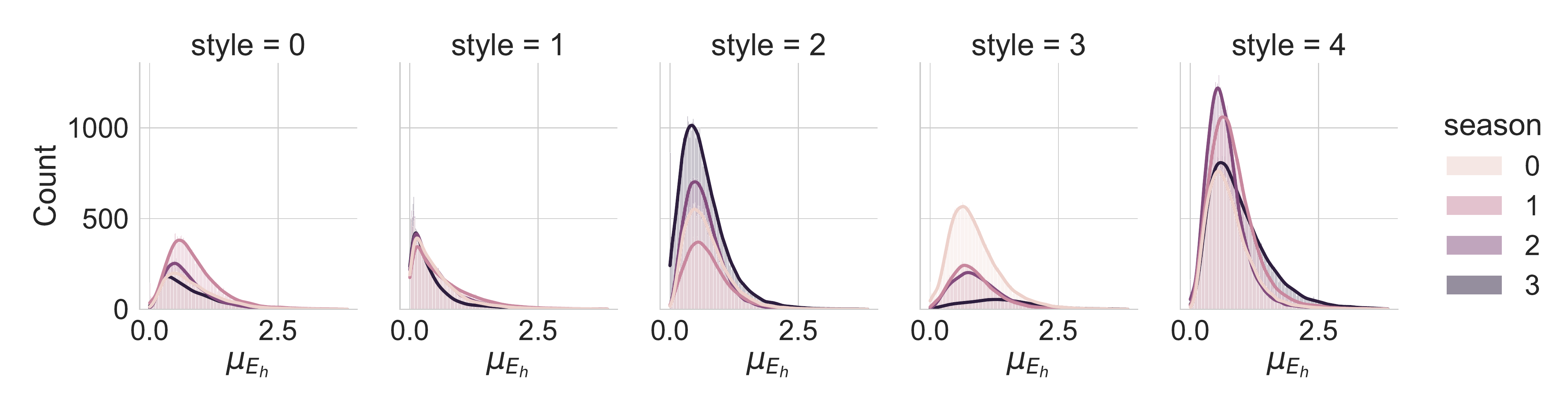}
    \caption{Distribution of hourly mean energy (in KWh) over four seasons for lifestyles}
    \label{lda:fig:appx:feat:season:mu_hr}
\end{figure}

We first present additional plots describing the seasonal features.
Figure~\ref{lda:fig:appx:feat:season:c_morning} to \ref{lda:fig:appx:feat:season:mu_hr} demonstrate the stability of the group of No Changers. These figures 
cover the different distributions of No Changers across four seasons including the features of morning energy use, evening energy use, peak energy use, and hourly average energy use.

\begin{figure}[!hbpt]
    \centering
    \includegraphics[width=0.99\textwidth]{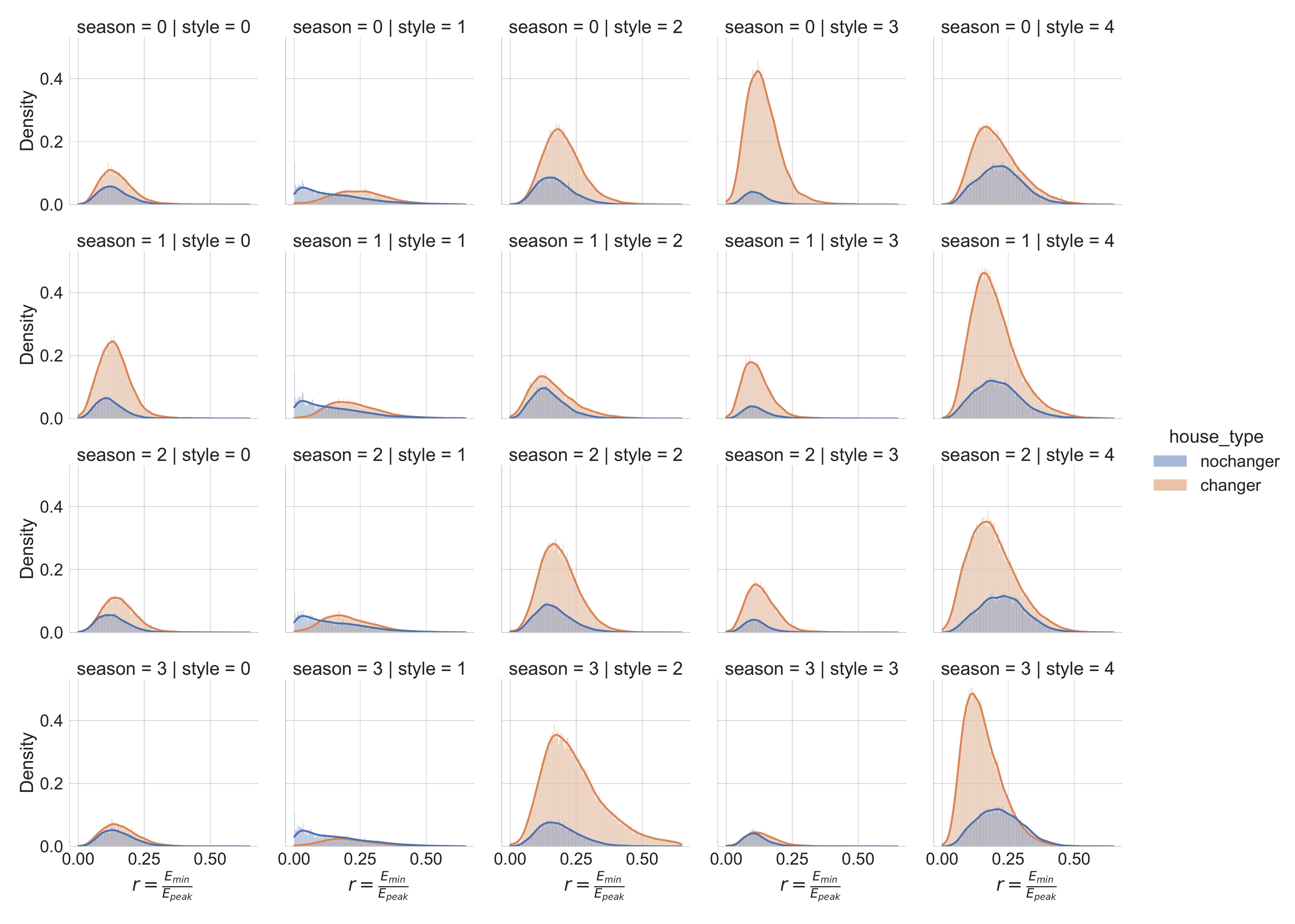}
    \caption{Distributions of min (base) to peak energy ratio for Changers and No Changers of five lifestyles over four seasons.}
    \label{lda:fig:appx:feat:ls:season:ch_noch:bp_r}
\end{figure}

\begin{figure}[!hbpt]
    \centering
    \includegraphics[width=0.99\textwidth]{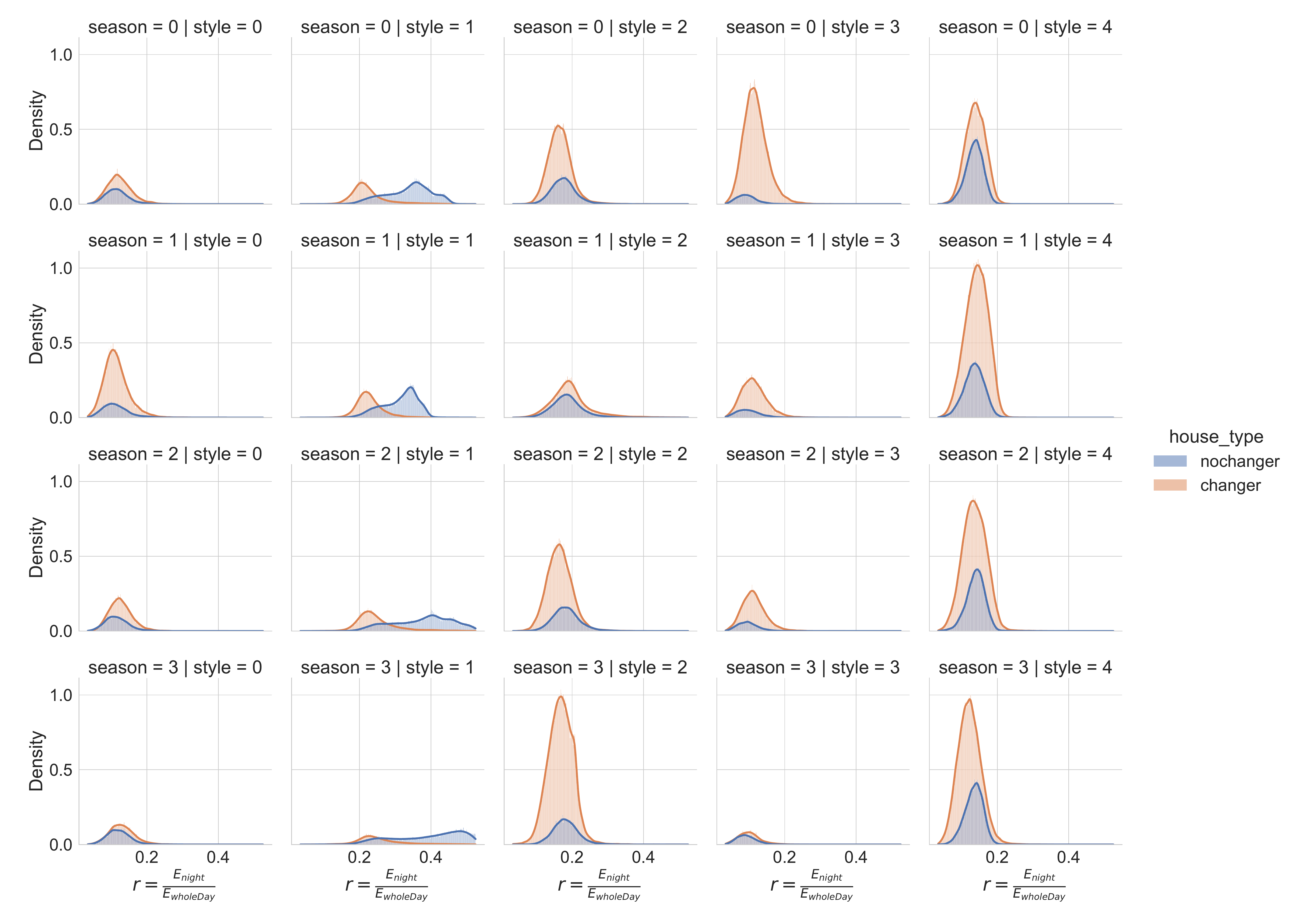}
    \caption{Distributions of night to whole day energy ratio for Changers and No Changers of five lifestyles over four seasons.}
    \label{lda:fig:appx:feat:ls:season:ch_noch:ni2w}
\end{figure}

To provide detailed comparisons between Changers and No Changers, we show the ratio of night to whole day usage and the ratio of noon to whole day usage in \Figref{lda:fig:appx:feat:ls:season:ch_noch:bp_r} and \Figref{lda:fig:appx:feat:ls:season:ch_noch:ni2w} because the distributions of those two features  significantly reveal the seasonal variations for the Changer group. 

\begin{figure}[!bpht]
    \centering
    \begin{subfigure}[t]{0.49\columnwidth}
    \includegraphics[width=0.99\textwidth]{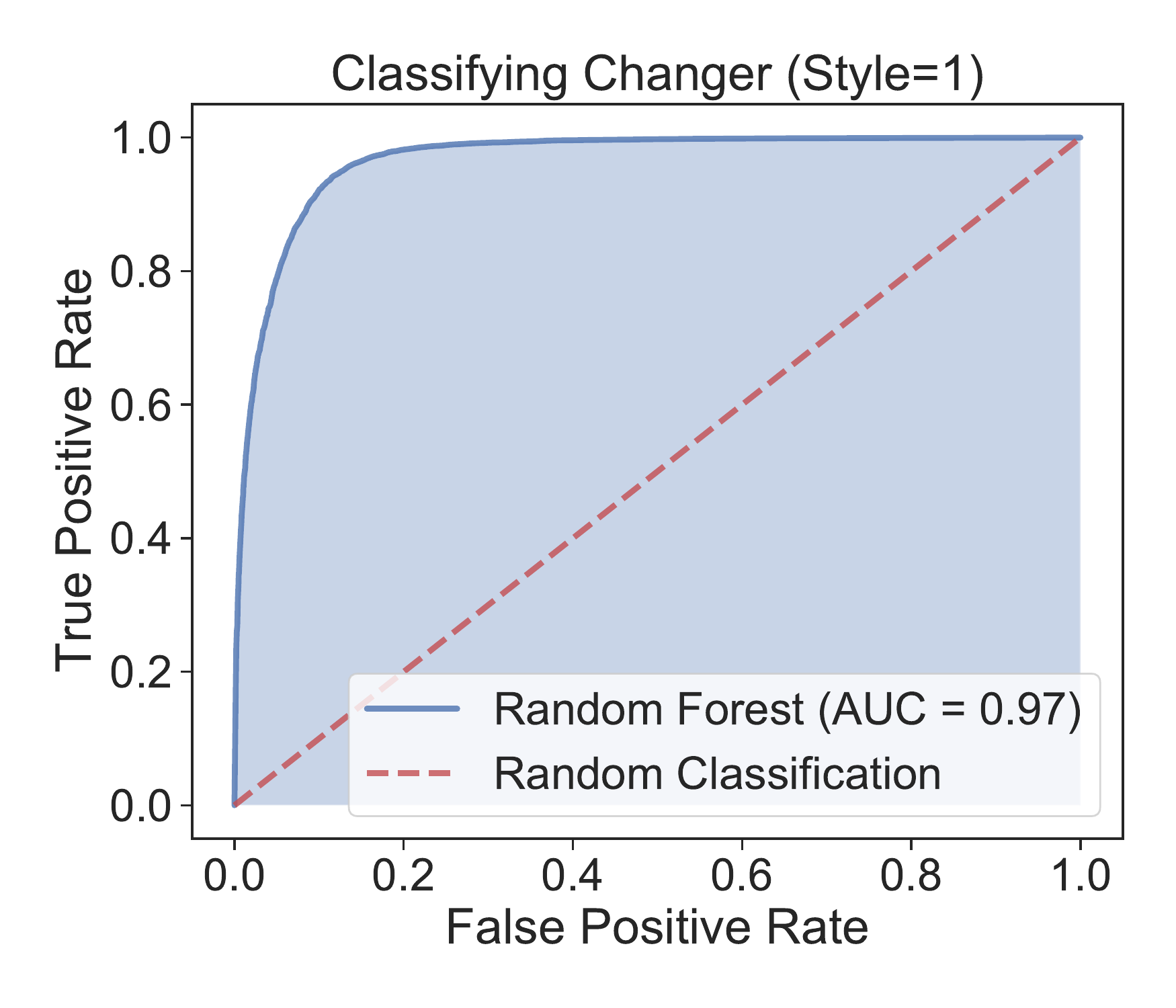}
    \caption{Night owl style}
    \label{lda:fig:appendix:chan_nochan:style1:auc}
    \end{subfigure}
    \hfill
    \begin{subfigure}[t]{0.49\columnwidth}
    \includegraphics[width=0.99\textwidth]{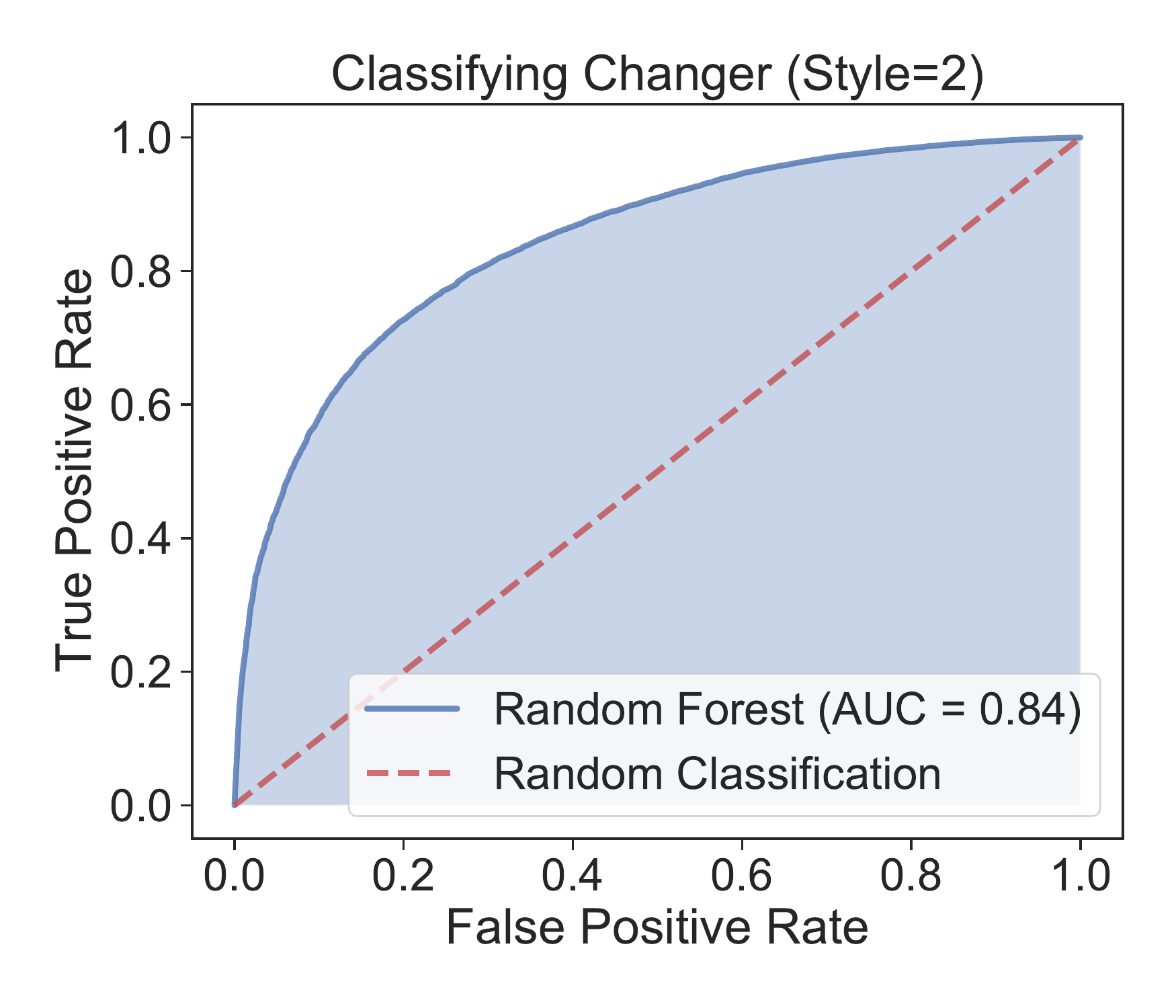}
    \caption{Everyday is a new day style}
    \label{lda:fig:appendix:chan_nochan:style2:auc}
    \end{subfigure}
    \hfill
    \begin{subfigure}[t]{0.49\columnwidth}
    \includegraphics[width=0.99\textwidth]{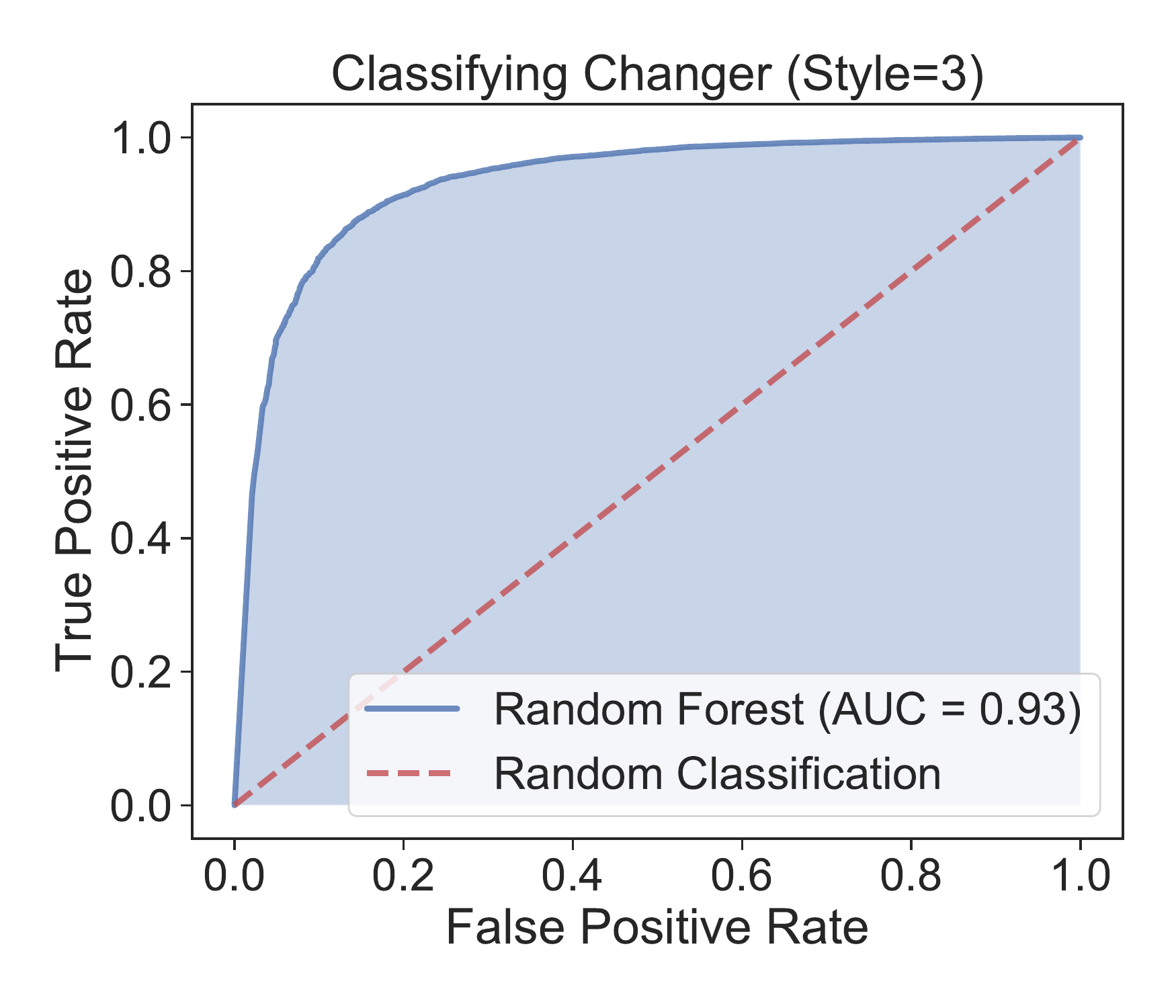}
    \caption{Home early style}
    \label{lda:fig:appendix:chan_nochan:style3:auc}
    \end{subfigure}
    \begin{subfigure}[t]{0.49\columnwidth}
    \includegraphics[width=0.99\textwidth]{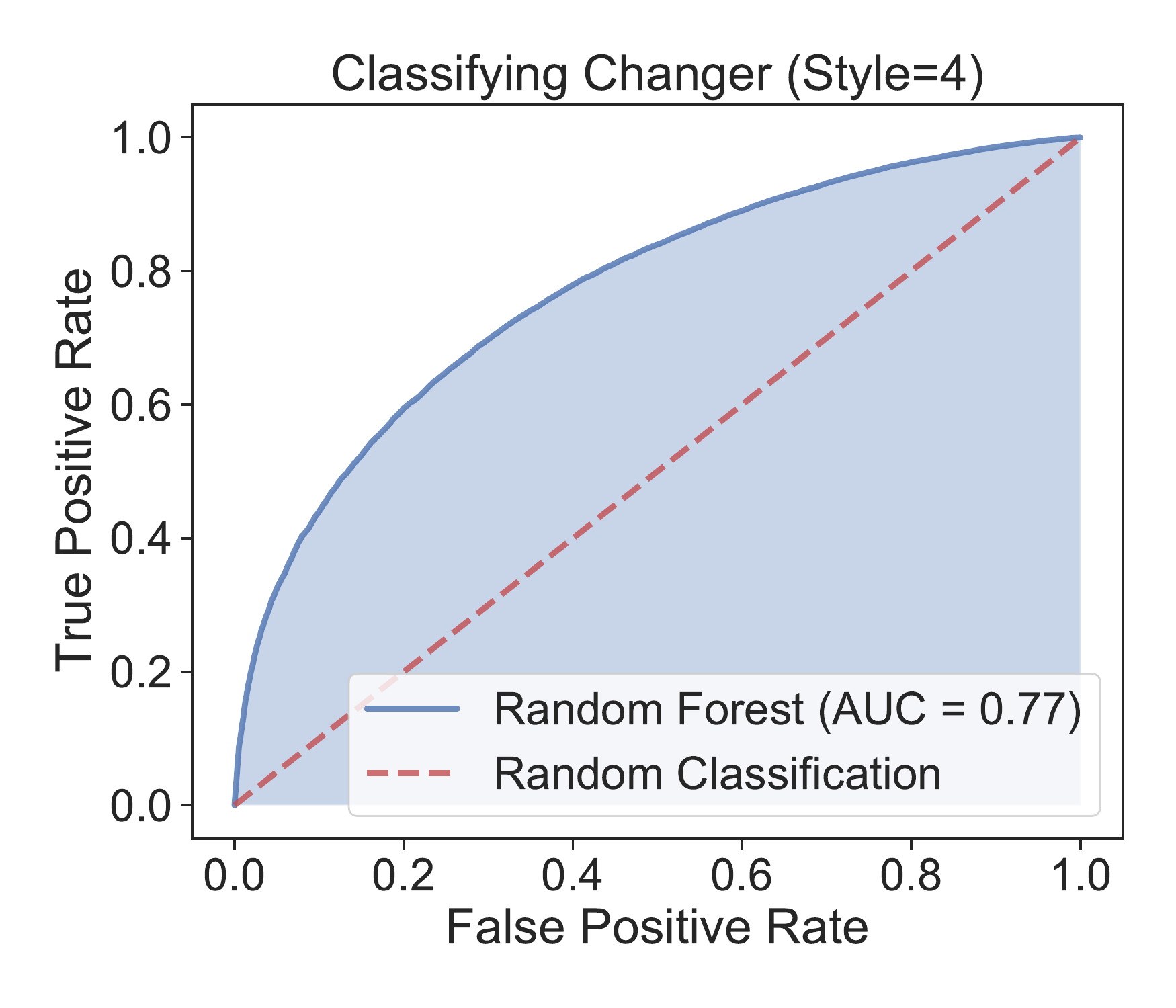}
    \caption{Home for dinner style}
    \label{lda:fig:appendix:chan_nochan:style4:auc}
    \end{subfigure}
    \hfill
    \caption{AUC of classifying Changer vs. No Changer.}
    \label{lda:fig:appendix:chan_nochan:clf:auc}
\end{figure}

We classify Changer vs. No Changer in each lifestyle using RF model with the same setting as previously described. Because the summer season does not have a Steady going lifestyle group, we show the other five lifestyles when each of them has a group of No Changers across four seasons (\tableref{lda:tab:clf:chan_nochan:style0:metrics}, \tableref{lda:tab:clf:chan_nochan:style1:metrics}, \tableref{lda:tab:clf:chan_nochan:style2:metrics}, \tableref{lda:tab:clf:chan_nochan:style3:metrics}, and \tableref{lda:tab:clf:chan_nochan:style4:metrics}). We observe that identifying a Changer is generally easier than identifying a No Changer because of the higher F1 scores. One exception is the Night owl lifestyle, which has similar performance in identifying Changers and No Changers given their relatively similar F1 scores. In addition, we show the Area Under the Receiver Operating Characteristic Curve (AUC) plots identifying Changers vs. No Changers for those five lifestyles (\Figref{lda:fig:appendix:chan_nochan:clf:auc}).   

\begin{figure}[!hbpt]
    \centering
    \begin{subfigure}[t]{0.49\columnwidth}
    \includegraphics[width=0.99\textwidth]{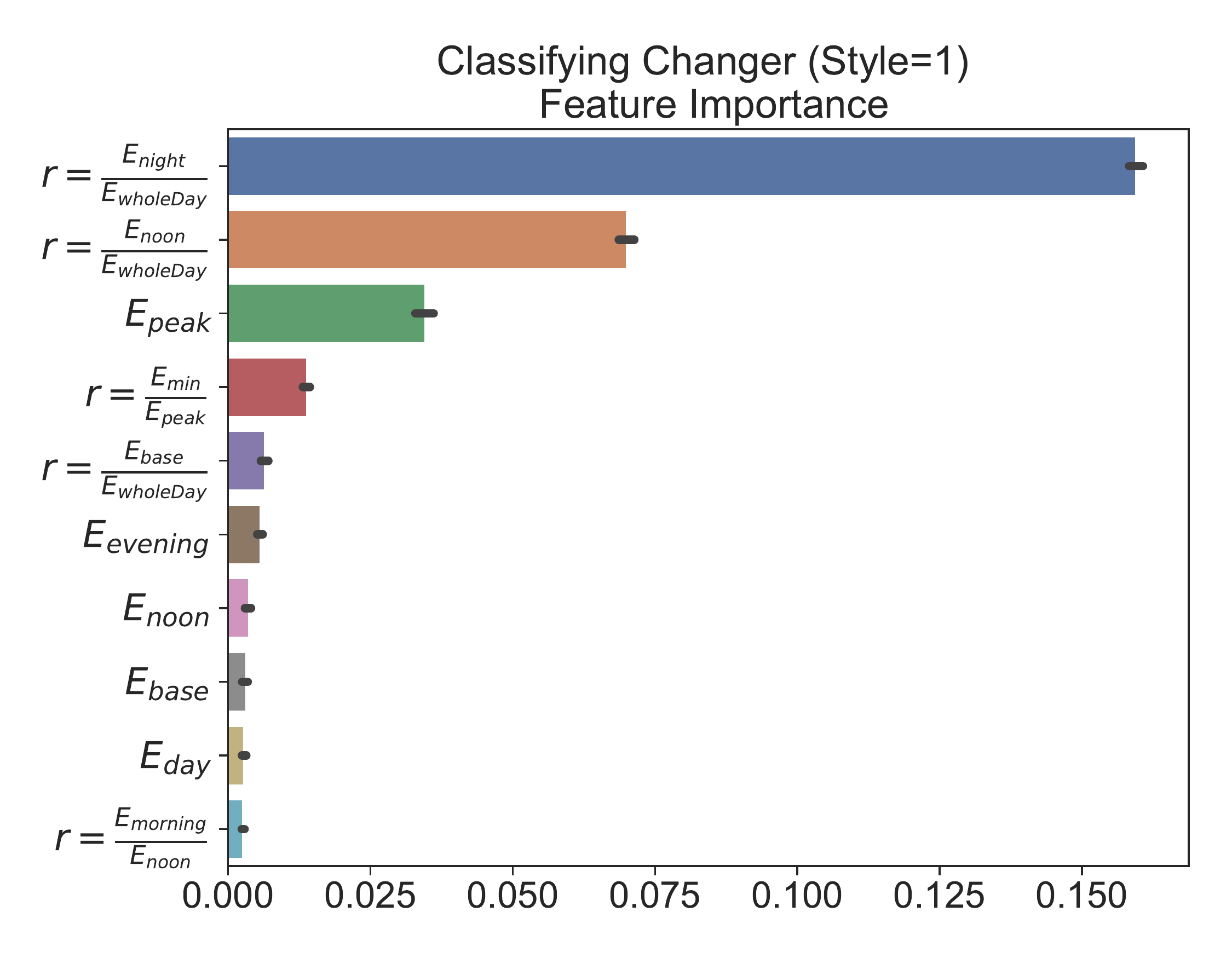}
    \end{subfigure}
    \hfill
    \begin{subfigure}[t]{0.49\columnwidth}
    \includegraphics[width=0.99\textwidth]{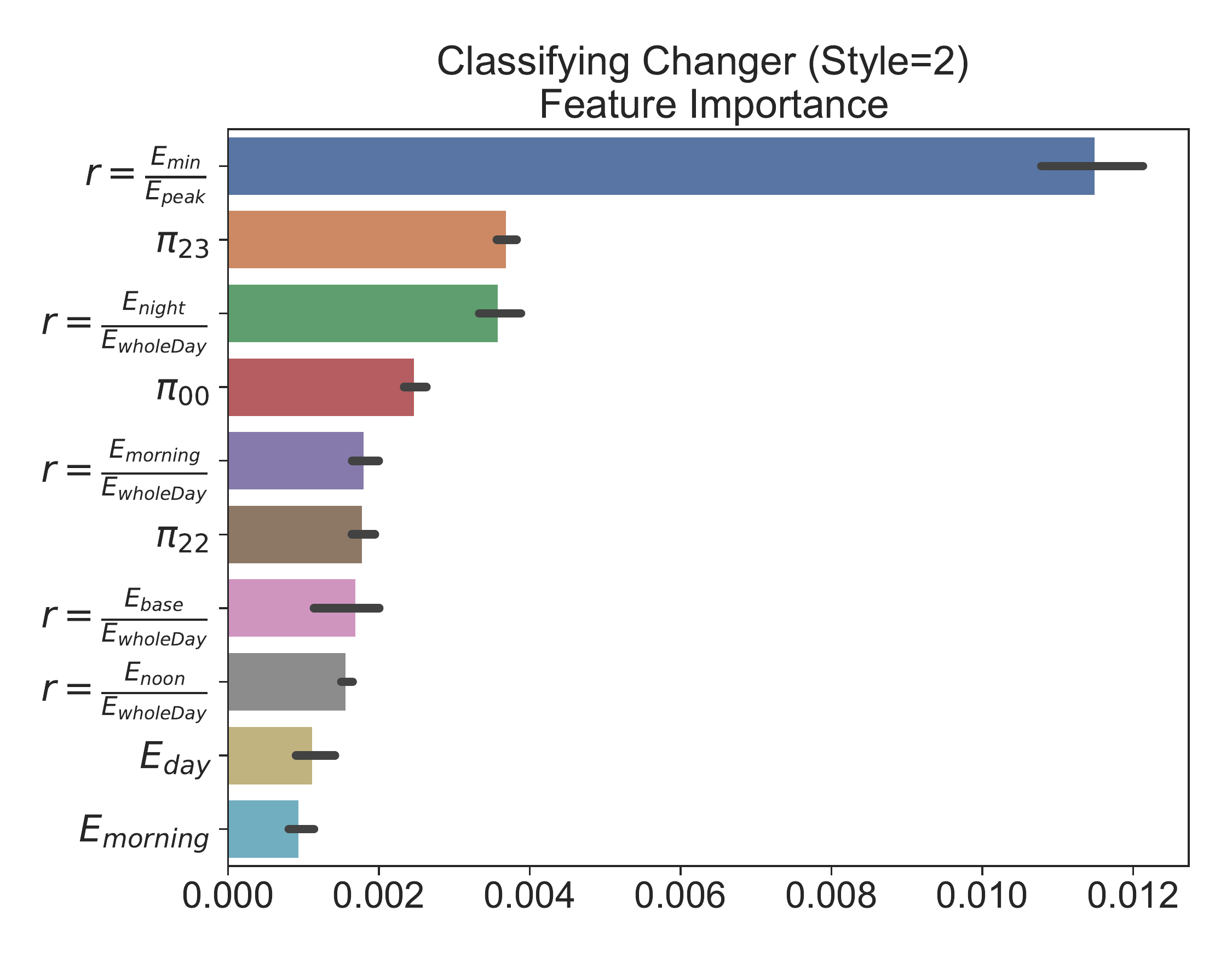}
    \end{subfigure}
    \hfill
    \begin{subfigure}[t]{0.49\columnwidth}
    \includegraphics[width=0.99\textwidth]{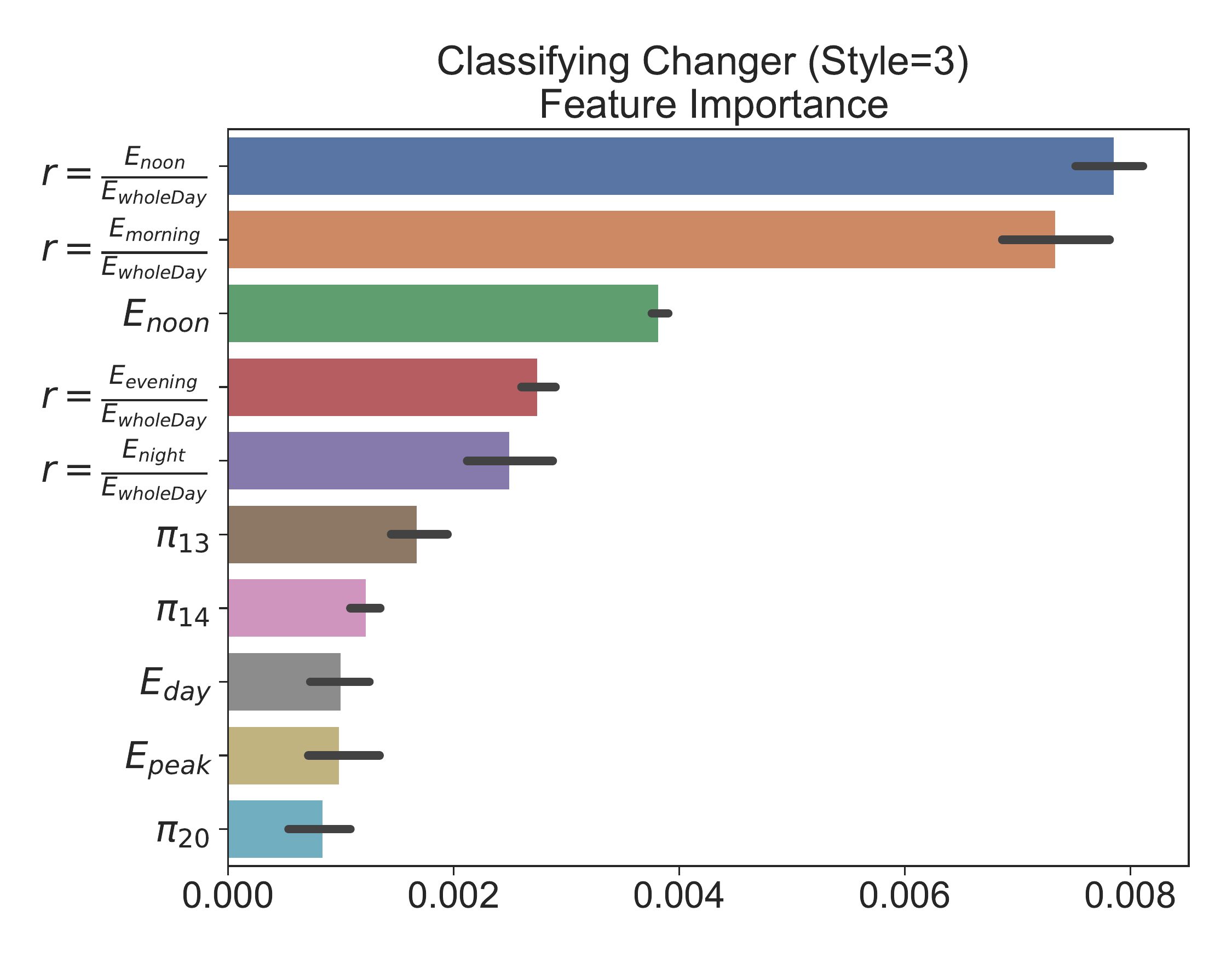}
    \end{subfigure}
    \begin{subfigure}[t]{0.49\columnwidth}
    \includegraphics[width=0.99\textwidth]{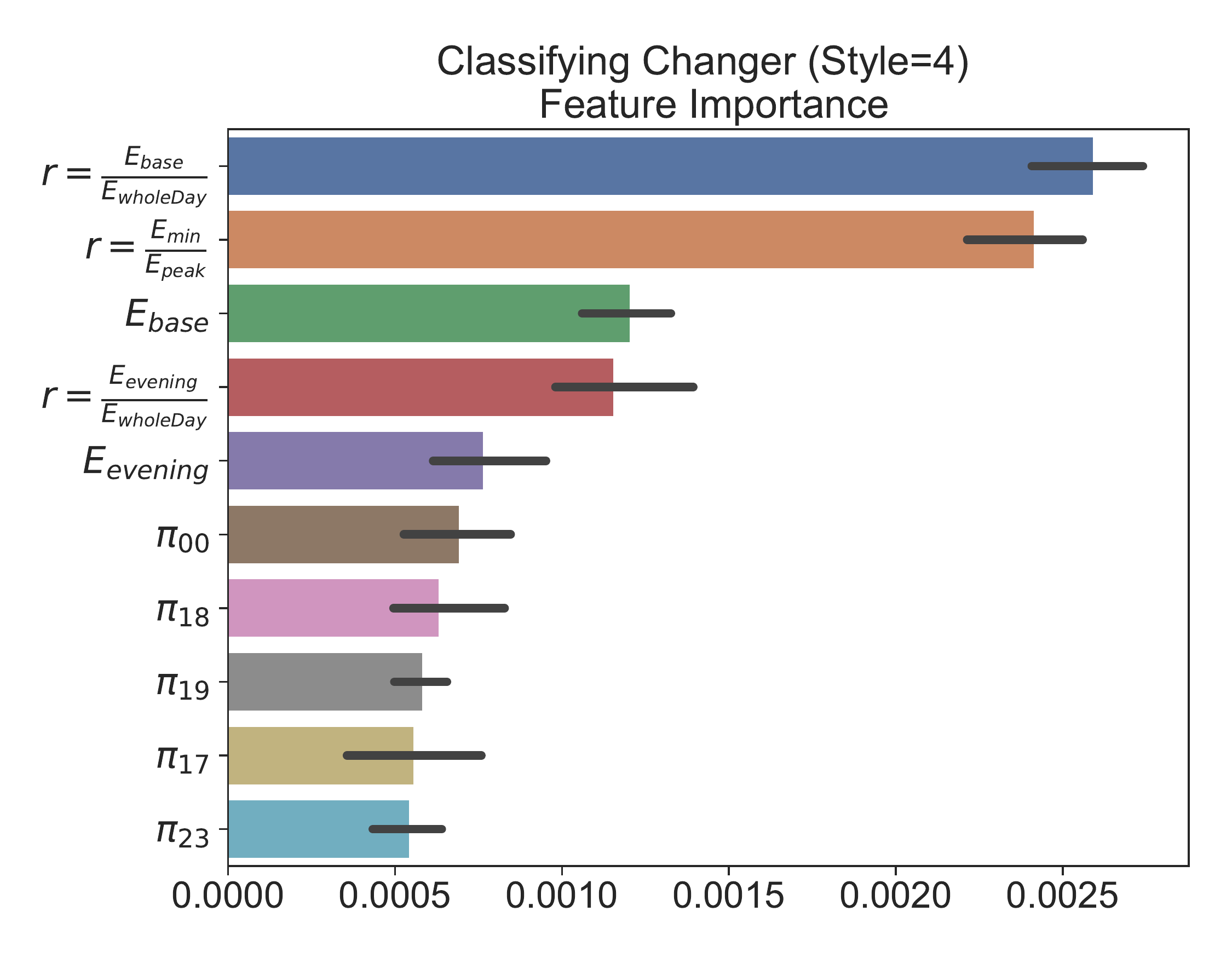}
    \end{subfigure}
    \hfill
    \caption{Feature importance (Changer vs. No changer)}
    \label{lda:fig:appendix:chan_nochan:feat:importance}
\end{figure}

The most important determinants of identifying Changers or No changers are different across the five lifestyles. We present their corresponding top 10 important features in \Figref{lda:fig:appendix:chan_nochan:feat:importance}.  

\begin{table}[!hpbt]
    \caption{Active morning lifestyle}
    \label{lda:tab:clf:chan_nochan:style0:metrics}
    \centering
    \begin{tabular}{r|c|c|c}
    \toprule
        label & precision & recall & F1 score \\
        \midrule
        No Changer (0) & 0.6481 & 0.2502 & 0.3611 \\
        Changer (1) & 0.8922 & 0.9786 & 0.9334 \\
        \midrule
        \multicolumn{4}{r}{average acc = 0.879} \\
        \bottomrule
    \end{tabular}
\end{table}

%
\begin{table}[!hpbt]
    \caption{Night owl lifestyle}
    \label{lda:tab:clf:chan_nochan:style1:metrics}
    \centering
    \begin{tabular}{r|c|c|c}
    \toprule
        label & precision & recall & F1 score \\
        \midrule
        No Changer (0) & 0.9301 & 0.8706 & 0.8994 \\
        Changer (1) & 0.9073 & 0.9508 & 0.9285  \\
        \midrule
        \multicolumn{4}{r}{average acc = 0.917} \\
        \bottomrule
    \end{tabular}
\end{table}
\begin{table}[!hpbt]
    \caption{Everyday is a new day lifestyle}
    \label{lda:tab:clf:chan_nochan:style2:metrics}
    \centering
    \begin{tabular}{r|c|c|c}
    \toprule
        label & precision & recall & F1 score \\
        \midrule
        No Changer (0) & 0.6476 & 0.1533 & 0.2479 \\
        Changer (1) & 0.9083 & 0.9902 & 0.9475 \\
        \midrule
        \multicolumn{4}{r}{average acc = 0.902} \\
        \bottomrule
    \end{tabular}
\end{table}
\begin{table}[!hpbt]
    \caption{Home early lifestyle}
    \label{lda:tab:clf:chan_nochan:style3:metrics}
    \centering
    \begin{tabular}{r|c|c|c}
    \toprule
        label & precision & recall & F1 score \\
        \midrule
        No Changer (0) & 0.6966 & 0.4028 & 0.5104 \\
        Changer (1) & 0.9657 & 0.9897 & 0.9775  \\
        \midrule
        \multicolumn{4}{r}{average acc = 0.957} \\
        \bottomrule
    \end{tabular}
\end{table}

\begin{table}[!hpbt]
    \caption{Home for dinner lifestyle}
    \label{lda:tab:clf:chan_nochan:style4:metrics}
    \centering
    \begin{tabular}{r|c|c|c}
    \toprule
        label & precision & recall & F1 score \\
        \midrule
        No Changer (0) & 0.5919 & 0.1734 & 0.2682  \\
        Changer (1) & 0.9657 & 0.9897 & 0.9775  \\
        \midrule
        \multicolumn{4}{r}{average acc = 0.856} \\
        \bottomrule
    \end{tabular}
\end{table}

\clearpage

\begin{figure}[!bpht]
    \centering
    \includegraphics[width=0.75\textwidth]{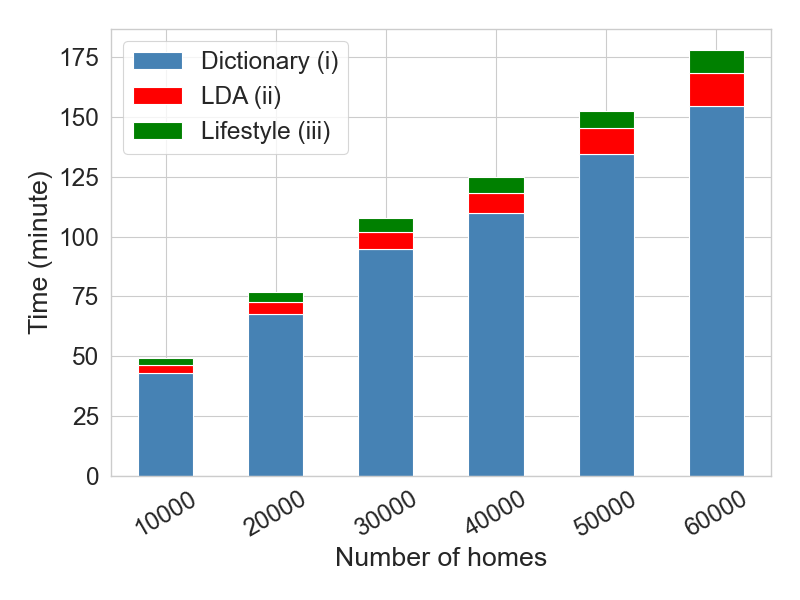}
    \caption{Computation time at different size of samples.}
    \label{lda:fig:appx:runtime:compare}
\end{figure}

{\subsection{Computation efficiency}\label{lda:sec:appx:compute:time}}
The initial input of our method is $N$ homes with 24 hour daily electricity record spanning over 365 days, so the size of the input is $N\times 24 \times 365$. The output is $K$ number of lifestyles. We set $K=6$ and run six different sizes of the dataset when $N=10000, 20000, 30000, 40000, 50000,$ and $60000$. Three steps are involved in our computation process:  (i) generating a dictionary of daily load shapes; (ii) running Latent Dirichlet Allocation to obtain energy attributes; (iii) using clustering to construct energy lifestyles. The computation time is displayed in~\Figref{lda:fig:appx:runtime:compare}.  

\end{document}